\documentclass[lettersize,journal]{style/IEEEtran}
\usepackage{amsmath,amsfonts}
\usepackage[caption=false,font=normalsize,labelfont=sf,textfont=sf]{subfig}
\usepackage{textcomp}

\usepackage[utf8]{inputenc}

\usepackage{todonotes}
\usepackage[acronym]{glossaries}
\glsenableentrycount

\usepackage[hidelinks]{hyperref}
\usepackage{cleveref}
\usepackage{babel}
\usepackage{url}
\usepackage[perpage]{footmisc}

\usepackage{graphicx}
\usepackage[binary-units]{siunitx}
\usepackage{overpic}

\usepackage{ifdraft}
\usepackage{booktabs}

\usepackage[normalem]{ulem}
\usepackage{scalerel}
\usepackage[switch]{lineno}

\usepackage{subfiles}

\newcommand\newreplacement[2]{\def#1/{#2}}
\newreplacement{\BSS}{BrainScaleS}
\newreplacement{\BSSS}{BrainScaleS system}
\newreplacement{\BSSWSS}{BrainScaleS wafer-scale system}
\newreplacement{\BSSWM}{BrainScaleS wafer module}

\newacronym{ananas}{AnaNAS}{Analog Network-Attached Sampling}
\newacronym{auxpwr}{AuxPwr}{Auxiliary Power Supply PCB}
\newacronym{bss1}{\mbox{BSS-1}}{Brain\mbox{ScaleS-1}}
\newacronym{bss2}{\mbox{BSS-2}}{Brain\mbox{ScaleS-2}}
\newacronym{cure}{CURe}{Control Unit for Reticles}
\newacronym{fpga}{FPGA}{field-programmable gate array}
\newacronym{jtag}{JTAG}{Joint Test Action Group}
\newacronym{fg}{FG}{Single-Poly Floating Gate}
\newacronym{hicann}{HICANN}{High Input Count Analog Neural Network}
\newacronym{i2c}{I\textsuperscript{2}C}{Inter-Integrated Circuit}
\newacronym{lvds}{LVDS}{low-voltage differential signaling}
\newacronym{psp}{PSP}{post synaptic potential}
\newacronym{isi}{ISI}{inter spike interval}
\newacronym{powerit}{PowerIt}{Main Power Supply}
\newacronym{sfc}{synfire chain}{synchronous firing chain}
\newacronym{wsi}{WSI}{wafer-scale integration}
\newacronym{adex}{AdEx}{adaptive exponential integrate-and-fire}
\newacronym{lif}{LIF}{leaky integrate-and-fire}
\newacronym{ota}{OTA}{operational transconductance amplifier}
\newacronym{macu}{MaCU}{Main Control Unit}
\newacronym{anab}{AnaB}{Analog Breakout PCB}
\newacronym{anarm}{AnaRM}{Analog Readout Module}
\newacronym{esd}{ESD}{Electrostatic Discharge}
\newacronym{pmbus}{PMBus}{Power Management Bus}
\newacronym{gpib}{GPIB}{General Purpose Interface Bus}
\newacronym{wio}{WIO}{Wafer I/O PCB}

\begin{document}

\title{From Clean Room to Machine Room: Commissioning of the First-Generation BrainScaleS Wafer-Scale Neuromorphic System}

\DeclareRobustCommand{\enumauthorrefmark}[1]{\smash{\textsuperscript{\footnotesize #1}}}

\newcommand{\contributedSymbol}{\IEEEauthorrefmark{1}}
\newcommand{\uheiSymbol}{\enumauthorrefmark{1}}
\newcommand{\ugoeSymbol}{\enumauthorrefmark{2}}

\author{
	\IEEEauthorblockN{%
		Hartmut Schmidt\contributedSymbol\uheiSymbol,
		José Montes\contributedSymbol\uheiSymbol,
		Andreas Grübl\uheiSymbol,
		Maurice Güttler\uheiSymbol,
		Dan Husmann\uheiSymbol,
		Joscha Ilmberger\uheiSymbol,\\
		Jakob Kaiser\uheiSymbol,
		Christian Mauch\uheiSymbol,
		Eric Müller\uheiSymbol,
		Lars Sterzenbach\uheiSymbol,
		Johannes Schemmel\uheiSymbol,
		Sebastian Schmitt\ugoeSymbol\\
	}

	\thanks{

		\IEEEauthorblockA{%
			\contributedSymbol%
			Contributed equally\\
			Email: %
				hschmidt@kip.uni-heidelberg.de, jmontes@kip.uni-heidelberg.de
			Corresponding author: Johannes Schemmel\\
			Email: schemmel@kip.uni-heidelberg.de
		}\newline
		\IEEEauthorblockA{%
			\uheiSymbol%
			Kirchhoff-Institute for Physics,
			Heidelberg University, Germany\\
		}
		\IEEEauthorblockA{%
			\ugoeSymbol%
			Department for Neuro- and Sensory Physiology,
			University Medical Center Göttingen, Germany\\
		}
	}
	\thanks{This work has received funding from the EU ([FP7/2007-2013], [H2020/2014-2020]) under grant agreements 604102 (HBP), 269921 (BrainScaleS), 243914 (Brain-i-Nets), 720270 (HBP SGA1), 785907 (HBP SGA2) and 945539 (HBP SGA3), the \foreignlanguage{ngerman}{Deutsche Forschungsgemeinschaft} (DFG, German Research Foundation) under Germany’s Excellence Strategy EXC 2181/1-390900948 (the Heidelberg STRUCTURES Excellence Cluster), the Helmholtz Association Initiative and Networking Fund (ACA, Advanced Computing Architectures) under Project SO-092, as well as from the Manfred Stärk Foundation.}
}

\maketitle

\thispagestyle{plain}
\pagestyle{plain}

\begin{abstract}
	The first-generation of BrainScaleS, also referred to as \acrlong{bss1}, is a neuromorphic system for emulating large-scale networks of spiking neurons.
Following a ``physical modeling'' principle, its VLSI circuits are designed to emulate the dynamics of biological examples: analog circuits implement neurons and synapses with time constants that arise from their electronic components' intrinsic properties.
It operates in continuous time, with dynamics typically matching an acceleration factor of \num{10000} compared to the biological regime.
A fault-tolerant design allows it to achieve wafer-scale integration despite unavoidable analog variability and component failures.
In this paper, we present the commissioning process of a \acrlong{bss1} wafer module, providing a short description of the system's physical components, illustrating the steps taken during its assembly and the measures taken to operate it.
Furthermore, we reflect on the system's development process and the lessons learned to conclude with a demonstration of its functionality by emulating a wafer-scale \acrlong{sfc}, the largest spiking network emulation ran with analog components and individual synapses to date.

\end{abstract}


\section{Introduction}
Simulating the dynamic properties of large-scale spiking neural networks is challenging due to the massively parallel interactions of their neurons and synapses.
The BrainScaleS neuromorphic architecture proposes a solution to this dilemma by providing inherently parallel computation at nodes operating as neurons and synapses and communicating through asynchronous spikes.
It thereby achieves a constant emulation speed with increasing network sizes~\cite{bruederle10simulator}. %

BrainScaleS implements physical models of neurons and synapses on a CMOS substrate with analog circuits, while the spike communication is digital.
On the one hand, the physical models inherently provide solutions to neuron and synapse dynamics in continuous time, in contrast to the time-discretized and numerically integrated solutions of digital systems and software simulations.
On the other hand, the programmable digital communication of action potentials allows for flexible network topologies and the possibility of using digital logic to feed and read spike events from outside the system.
Furthermore, circuits are operated in strong inversion, targeting dynamics with a typical speedup factor of \num{10000} compared to biological real-time.

The \acrlong{bss1} system utilizes wafer-scale integration to achieve large ASIC counts with energy efficiency and high communication bandwidth.
The structure of its underlying neuromorphic chip and the technology to achieve its wafer-scale integration are introduced in~\cite{schemmel_ijcnn2008,fieres_ijcnn2008,schemmel2010iscas,millner2010vlsi}.
Turning the silicon wafer into a ready-to-use system, though, implicates bringing several additional components, shown in \cref{fig:wafer_module}, to work hand in hand.
For that cause, a commissioning chain is established, which is this paper's focus.

\begin{figure}
  \centering
  \includegraphics[width=0.8\linewidth]{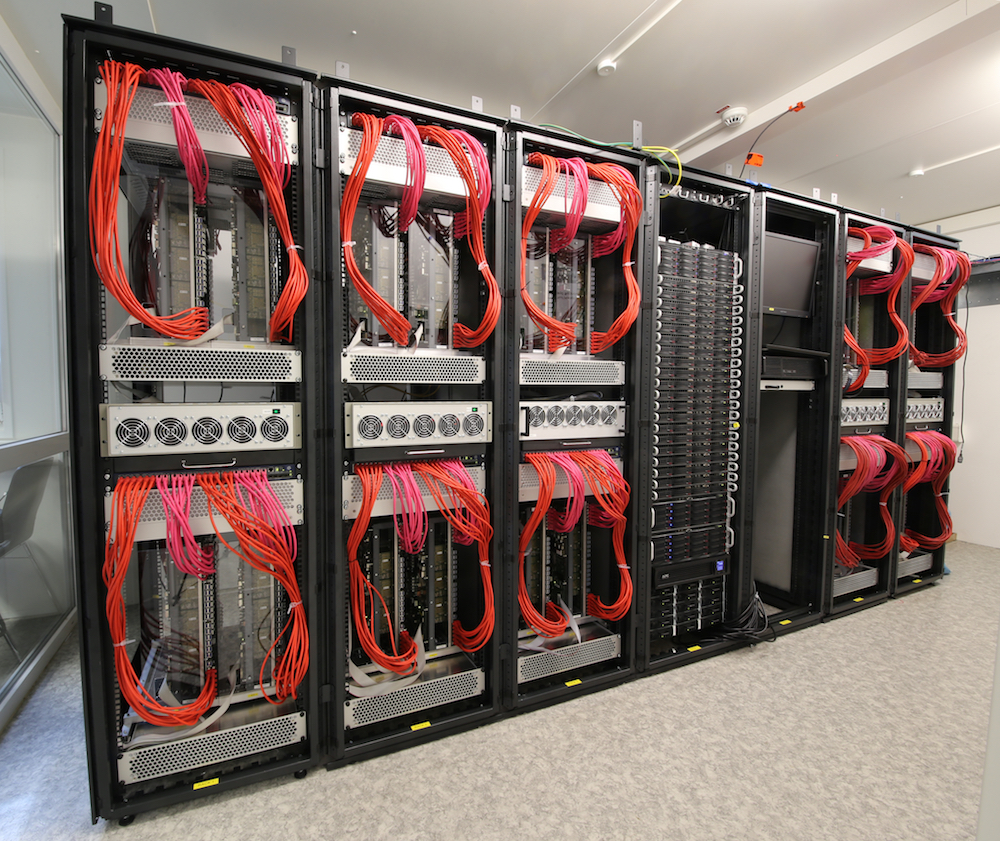}
  \caption{\label{fig:machine_room}The \acrlong{bss1} machine room comprising \num{20} wafer modules organized in \num{5} racks.
  A slot in the middle of each rack hosts the \acrlong{anarm} and the \acrlongpl{macu} of its neighboring wafer modules.
  Gigabit-Ethernet cables connect each wafer module via aggregation switches to the control cluster positioned in the middle rack.
  Taken from~\cite{schmitt2017hwitl}.
  }
\end{figure}

We first illustrate the different components that constitute the system and how they are tested.
Then, we show the steps to assemble the module before it is finally placed in the machine room, as shown in~\cref{fig:machine_room}.
In the second part of the paper, we describe the methods devised to bring such a system into a reliable substrate for neuromorphic experiments: a large number of VLSI analog components inevitably leads to malfunctioning parts and analog variability, for which an underlying fault-tolerant design and suitable handling have to be put in place.
To demonstrate its operation and the successful implementation of these measures, a biologically-motivated network of spiking neurons, a \acrlong{sfc}, is emulated on a fully commissioned \acrlong{bss1} wafer module.

The system belongs to the still-nascent field of neuromorphic computing and remains under continuous development.
Having pioneered a neuromorphic wafer-scale integration of VLSI analog and digital circuits, we also discuss the lessons learned while solving or circumventing the challenges faced along the way.

\section{System Components and Individual Tests}
\label{sec:system_components}

\begin{figure}
  \hfill
  \subfloat[\label{wafer_module:exploded_view}]{\def\svgwidth{.5\linewidth}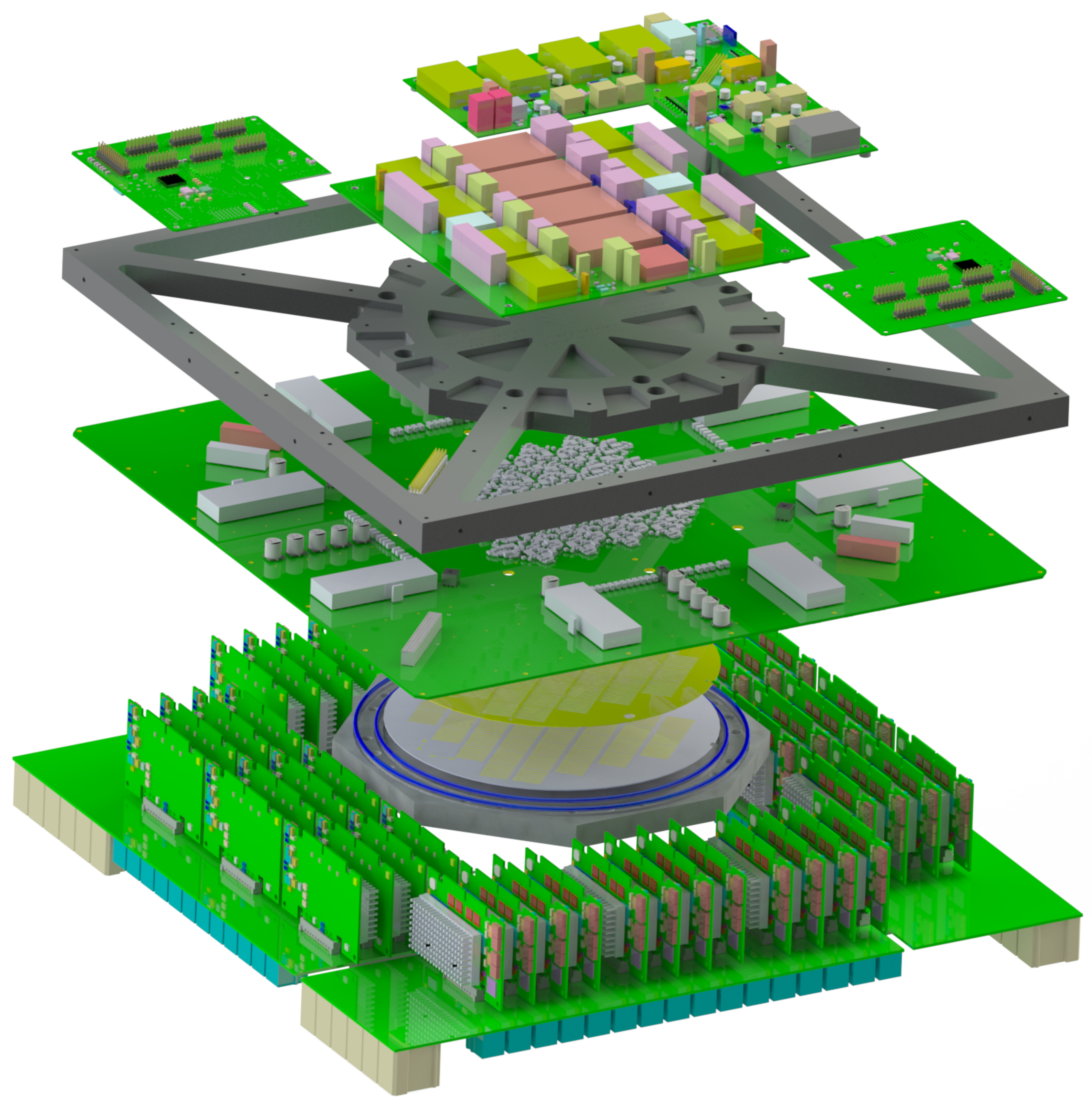}
  \hfill
  \subfloat[\label{wafer_module:picture}]{\includegraphics[height=0.2\textheight]{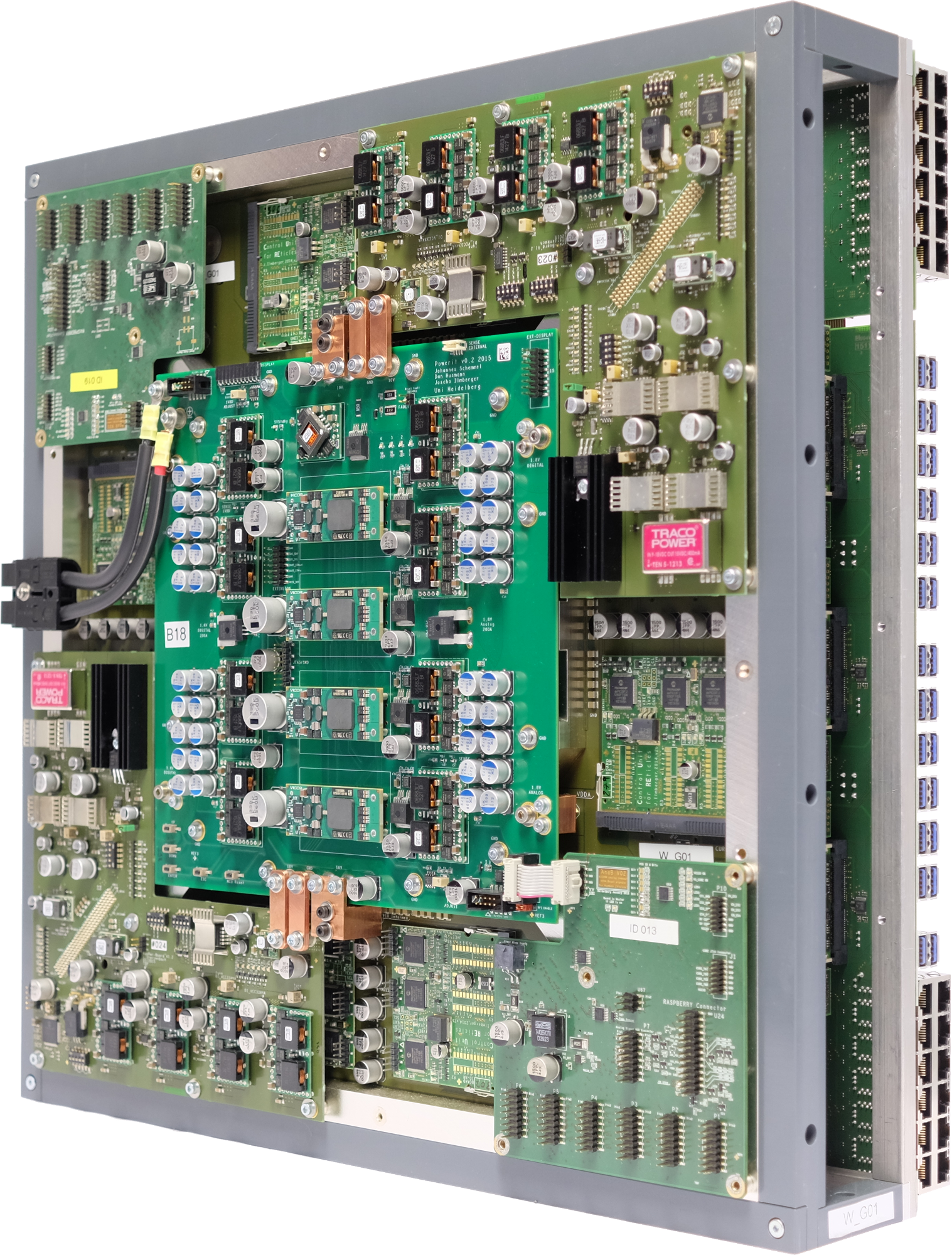}}
  \hspace*{\fill}
  \caption{\label{fig:wafer_module}
    \protect\subref{wafer_module:exploded_view} 3D-schematic of a \acrlong{bss1} wafer module (dimensions: \SI{50}{\centi\meter} \texttimes{} \SI{50}{\centi\meter} \texttimes{} \SI{15}{\centi\meter}) hosting the wafer~(A) and \num{48} communication boards~(B).
    The positioning mask~(C) aligns elastomeric connectors that link the wafer to the large Main PCB~(D).
    Support PCBs provide power supply~(E \& F) for the on-wafer circuits as well as access~(G) to analog dynamic variables such as neuron membrane voltages.
    The connectors for inter-wafer and off-wafer/host connectivity (\num{48} \texttimes{} Gigabit-Ethernet) are distributed over all four edges~(H) of the Main PCB.
    Mechanical stability is provided by an aluminum frame~(I).
    \protect\subref{wafer_module:picture} Photograph of a fully assembled wafer module.
    Taken from~\cite{schmitt2017hwitl}.
  }
\end{figure}

A \acrlong{bss1} wafer module is depicted in~\cref{fig:wafer_module}.
Each of its constituent boards is individually tested before its integration into the system, which permits differentiating errors in the parts from those arising from the assembly.
A short description of each component and the tests it undergoes is given in the following.

\begin{figure}[!ht]
  \subfloat[\label{fig:wafer}]{%
    \begin{overpic}[width=0.48\linewidth]{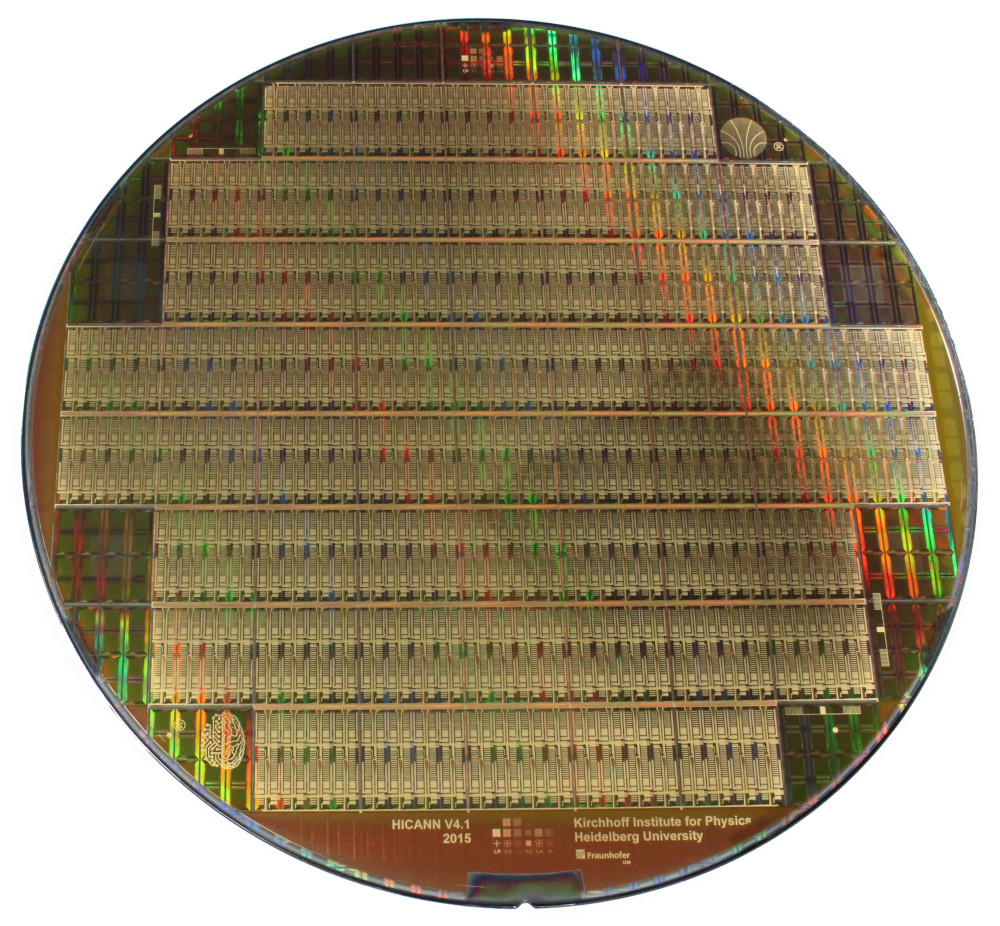}
    \end{overpic}
  }
  \subfloat[\label{fig:main_pcb_bot}]{%
    \begin{overpic}[width=0.48\linewidth]{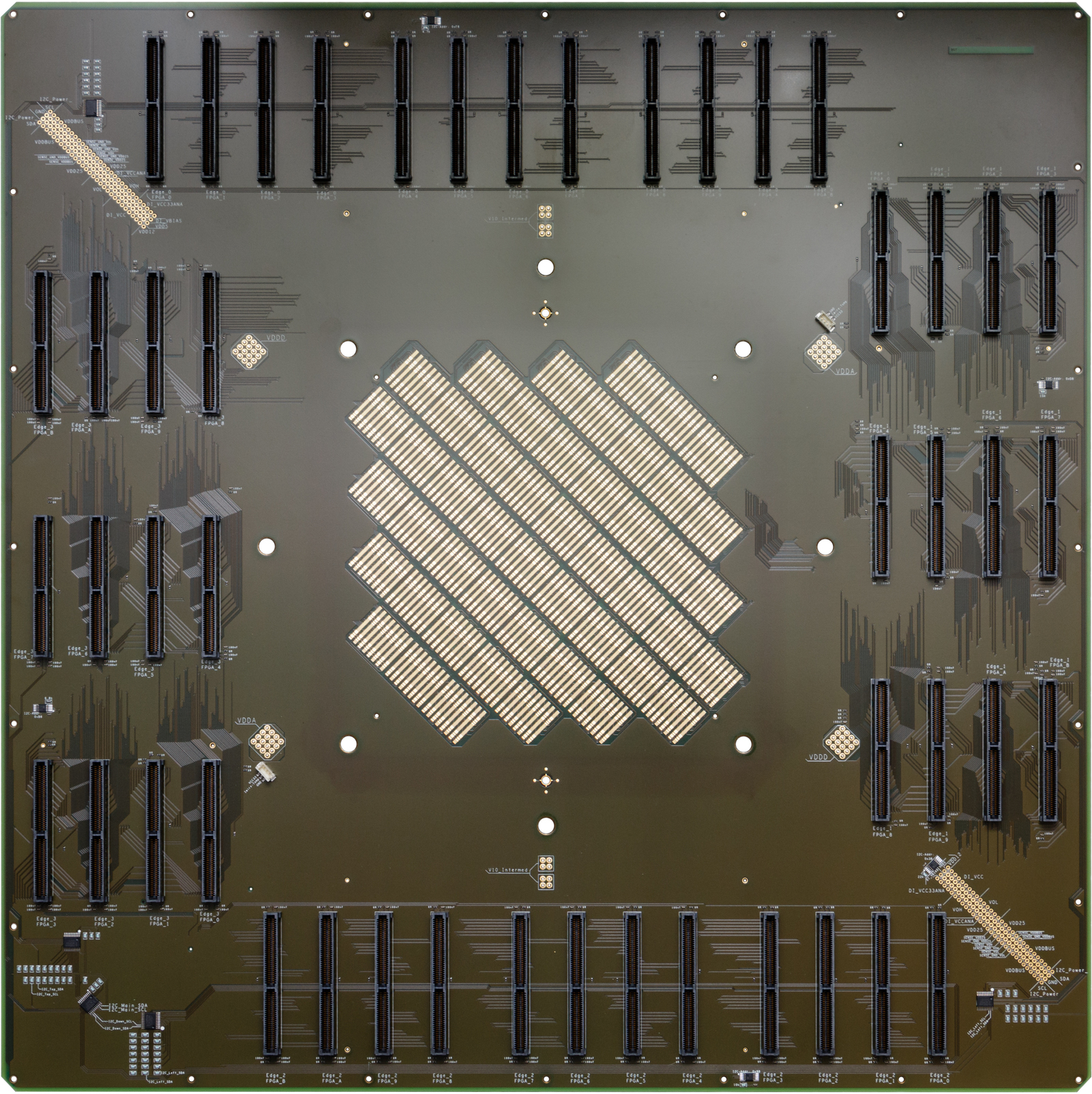}
      \put(45,45){\includegraphics[scale=4]{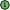}}
      \put(45,5){\includegraphics[scale=4]{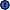}}
      \put(45,80){\includegraphics[scale=4]{img/numbers/num_2.pdf}}
      \put(5,45){\includegraphics[scale=4]{img/numbers/num_2.pdf}}
      \put(80,45){\includegraphics[scale=4]{img/numbers/num_2.pdf}}
    \end{overpic}
  }
  \caption{
    \protect\subref{fig:wafer} The \acrlong{bss1} wafer with applied postprocessing to achieve
      wafer-scale integration and to establish its connection to the \protect\subref{fig:main_pcb_bot} bottom side of the Main PCB.
      There, the wafer connects through elastometic connectors to the center, marked with 1.
      In the borders, \num{48} connectors, marked with 2, accommodate the communication boards.
  }\label{fig:wafer_and_main_pcb}
\end{figure}

\subsection{The \acrlong{bss1} Wafer}
\label{sec:hicann-wafer}

The heart of each module is an uncut $20\,\text{cm}$ wafer, displayed in~\cref{fig:wafer}, fabricated in UMC $\SI{180}{\nano\meter}$ technology comprising 384 \cgls{hicann} ASICs.
Each \cgls{hicann} contains \num{512} analog neuron circuits implementing the \cgls{adex} model~\cite{millner2010vlsi}.
Single neuron circuits receive input from up to \num{220} analog synapses.
Since neuron membranes can interconnect in groups of up to \num{64}, a maximum of $\num{14080}$ synapses can provide input to each of these composite neurons.
Synapse weights are stored with \num{4}-bit resolution in local SRAM at each synapse.

Each \cgls{hicann} stores $\num{12384}$ analog quantities for parameterization of its analog circuits in \cgls{fg} CMOS cells that retain their operation levels according to their isolated gate's accumulated charge~\cite{Millner2012,lande96}.
These \acrshortpl{fg} are written via an onboard \num{10}-bit-resolution DAC, enabling reprogramming via incremental loops with feedback.
Then, the stored values get translated to either a voltage or a current using a source follower or a current mirror, respectively, to set neuron parameters and other onboard circuit operation levels.
While these \cglspl{fg} present a low-power, small-space solution to store analog operation settings, they introduce write-cycle to write-cycle variability, as will be further discussed.\\
Wafer-wide communication is achieved with a custom-developed redistribution layer applied post-wafer-production, creating around $\num{160000}$ lateral connections across chip borders~\cite{zoschkeguettler2017rdlembedding}.
These connections provide the modules with on-wafer spike event communication through \cgls{lvds} buses utilizing an asynchronous serial event transmission protocol.
Furthermore, connections through top-layer pads on the wafer provide the modules with parallel per \cgls{hicann} off-wafer communication, which in conjunction with programmable and redundant components, constitute the system's fault tolerance~\cite{schemmel_ijcnn2008}.

\textit{Testing:} In order to assess the effect of wafer post-processing on the digital yield of an entire wafer, initial needle card tests were carried out on two unprocessed\footnote{"unprocessed" in this context means untested wafers straight from the manufacturer, before the custom redistribution layers have been added.} wafers to determine their yield immediately after production.
Since the wafers undergoing these tests cannot be further processed, comparing results on the same wafers before and after the post-processing is, however, not possible.

\begin{figure}
  \centering
  \subfloat[\label{fig:clean_room_setup_2}]{\includegraphics[width=0.48\linewidth]{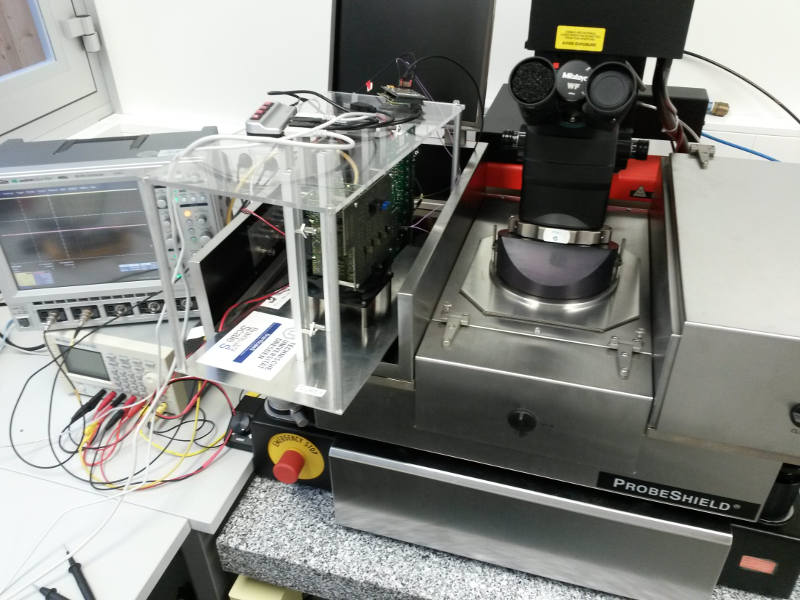}}
  \subfloat[\label{fig:wafer_and_needle_card}]{\includegraphics[width=0.48\linewidth]{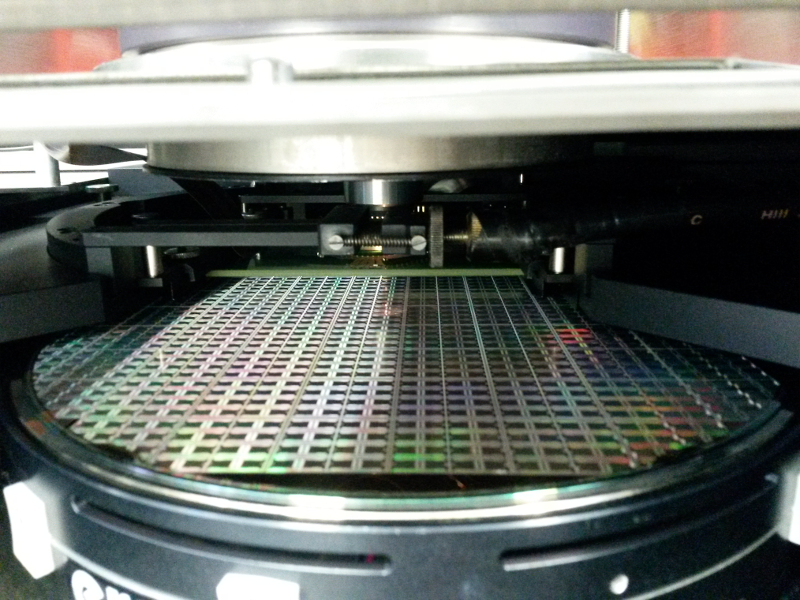}}
  \caption{\protect\subref{fig:clean_room_setup_2} Photograph of the wafer prober and \protect\subref{fig:wafer_and_needle_card}~a close-up of a wafer under test. Different needle cards have been developed and used for tests carried out before wafer post-processing (\cref{sec:hicann-wafer}, visible in this setup) and before wafer module assembly (\cref{sec:tests_installation_steps}), respectively.}
  \label{fig:clean_room_wafer_prober}
\end{figure}

The setup for these tests in the institute's clean room is shown in \cref{fig:clean_room_wafer_prober}, and the procedure is as follows.
The needle card is used to contact each individual ASIC.
Immediately after contacting and powering up, the total current on the used lab supply is measured to detect potential power shorts.
Henceforward, all digital memory cells on the \cgls{hicann} circuits are tested using a built-in \cgls{jtag} access mode.
During these tests, \num{448} \cglspl{hicann} on each of the two wafers were tested, and \SI{93}{\percent} of them showed no single digital error. %
To compare, UMC's calculator estimates a yield of approximately \SI{85}{\percent} by taking into account the process parameters and circuit size.
However, our results are only an estimation:
On the one hand, the tested digital memory cells only cover a fraction of the whole silicon area, which is dominated by analog circuitry.
Therefore, the digital test yield could be assumed to be too optimistic.
On the other hand, perfect power and signal integrity could not be ensured while connecting the circuits through the needles, leading to a possible detection of false negatives, caused for example by slightly underpowered memory cells.
In addition, only wafers from the initial engineering sample production have been available for testing.
No documentation has been available to relate the production yield data from UMC to small batch-size engineering runs.
Nonetheless, the results match the expectations taking the high level of uncertainty into account.
Also, a yield in the order of, e.g., \SI{85}{\percent} would not indicate that \SI{15}{\percent} of the dies cannot be used.
Instead, advantaging from the fault-tolerant design, and depending on the defect type, it could suffice to disable single neuron or synapse circuits, for example, on affected \cglspl{hicann} that are otherwise fully functional and can remain available for experiments.

\subsection{Main PCB}
The Main PCB, displayed in~\cref{fig:main_pcb_bot}, is a \SI{43}{\cm}$\,\times\,$\SI{43}{\cm} passive interconnector board for most parts of the wafer-scale integration system.
Seven of its $\num{14}$ layers are used to distribute $\num{23}$ power rails carrying up to \SI{200}{\ampere} of current.
The rest of the layers are used to route $\num{1152}$ power monitoring, $\num{1472}$ high-speed differential communication, and different sideband signals.
Auxiliary boards, communication infrastructure, and the silicon wafer are connected via various kinds of detachable connectors.
These enable system modularity for development and upgrades, desirable for research and development in dynamic environments over longer timespans.

\textit{Testing:} The manufacturer\footnote{Manufactured by \foreignlanguage{ngerman}{Würth Elektronik}, Germany} performs complete optical inspection and electrical tests of the Main PCB. The \acrlong{bss1} wafer modules are assembled using exclusively fully validated, error-free Main PCBs.

\subsection{Auxiliary Boards}
The wafer module is completed by populating it with \num{48} communication boards and auxiliary boards for power delivery,  control, monitoring, and inter-module communication.

\subsubsection{Communication Boards}
Each communication board\footnote{\label{foot:devAtTud}Developed at the chair of \foreignlanguage{ngerman}{Hochparallele VLSI-Systeme und Neuromikroelektronik} at TU~Dresden}
contains a \cgls{fpga} and connects to one \cgls{hicann} group consisting of \num{8} \cglspl{hicann}.
These boards communicate through separate high-speed \cgls{lvds} interfaces with each of the connected \cglspl{hicann} to configure, monitor, and coordinate the experiment runs; they feed and collect generated spikes into/from the experiments.
Furthermore, they synchronize the start of experiments to allow for wafer wide execution.
Trigger signals generated on these boards also align experiments with analog recordings using the \cgls{anarm}.

\textit{Testing:} The communication boards are tested on a standalone setup that implements loopback connections for the high-speed interfaces.
For this purpose, a test board accommodates and tests four PCBs in parallel, as shown in~\cref{fig:fcp_tud}.
Primarily automated and controlled via software, the tests switch the power supply via \cgls{gpib}. %
Programming is performed via \cgls{jtag} and \cgls{pmbus}.
Tests comprising current consumption measurements, loading and communicating with the \cgls{fpga} design, as well as memory tests are conducted.
In addition, communication with the host computer as well as the links to the wafer and neighboring communication boards are tested.
As per data logs, only \num{18} out of $\num{1404}$ produced PCBs had to be discarded after failed tests.

\subsubsection{Wafer I/O PCB}
Each one of the module's four \cglspl{wio}\footref{foot:devAtTud} attaches to twelve communication boards, aggregating Gbit-Ethernet and connections to other communication boards.

\textit{Testing:} A manual approach is followed as the number of boards is smaller than that of the communication boards.
The board, shown in~\cref{fig:wio_test}, is supplied with power, and the proper functioning of the DC/DC converters is checked with a multimeter.
Individual communication ports are tested.
In addition, the proper transmission of signals using a signal generator and differential probes is measured.
A partial test of the JTAG pins is also carried out.
As per data logs, only \num{2} out of \num{120} produced \cglspl{wio} were discarded after failed tests.

\subsubsection{Main Power Supply}
The \cgls{powerit} has three output channels: Two \SI{1.8}{\volt} outputs as main analog and digital supplies of the wafer with a current limit of \SI{200}{\ampere} each, as well as a \SI{9.6}{\volt} output capable of up to \SI{110}{\ampere} to supply the communication boards.
Multiple custom-milled copper parts ensure a low-resistance screw connection between the \cgls{powerit} and the Main PCB.
Additionally, digital control of the voltages and sensors as near to the wafer as possible allow for compensation of IR-drop.
An integrated microcontroller can measure input and output currents and voltages via shunt resistors, hall sensors, and voltage dividers.

\begin{figure}
  \centering
  \subfloat[\label{fig:fcp_tud}]{\includegraphics[width=0.48\linewidth]{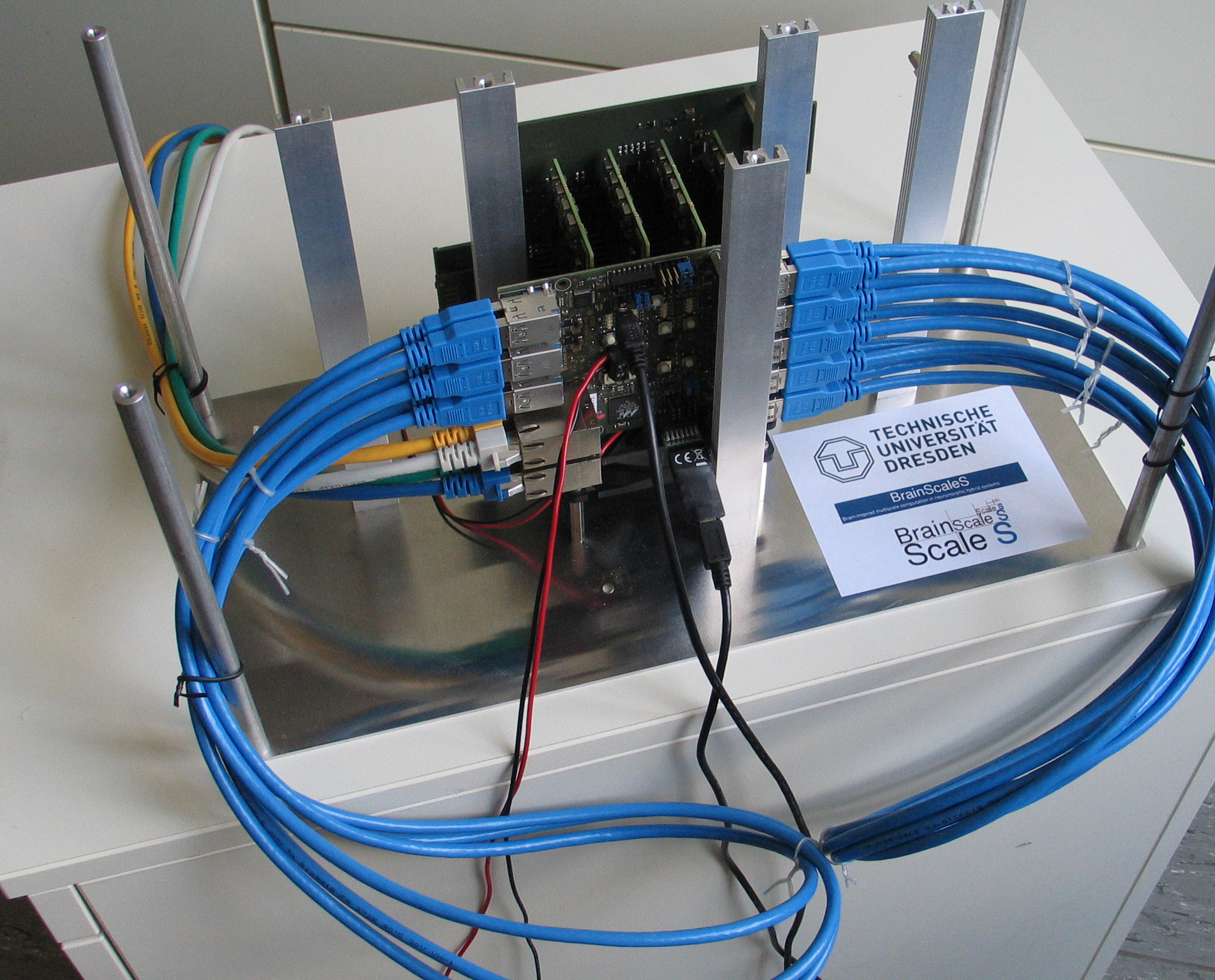}}
  \subfloat[\label{fig:wio_test}]{\includegraphics[width=0.48\linewidth]{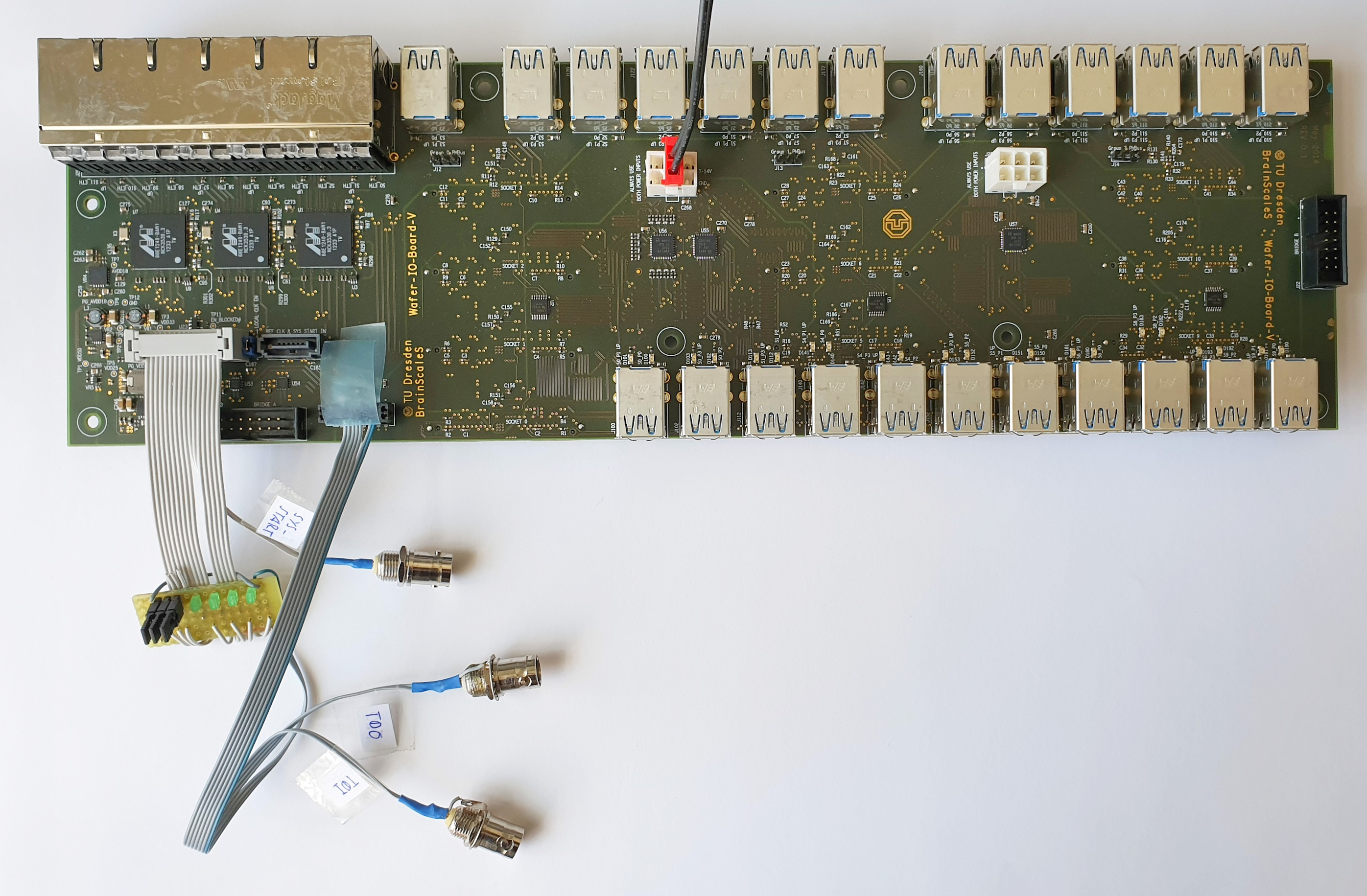}}\\
  \subfloat[\label{fig:prober_powerit}]{\includegraphics[width=0.48\linewidth]{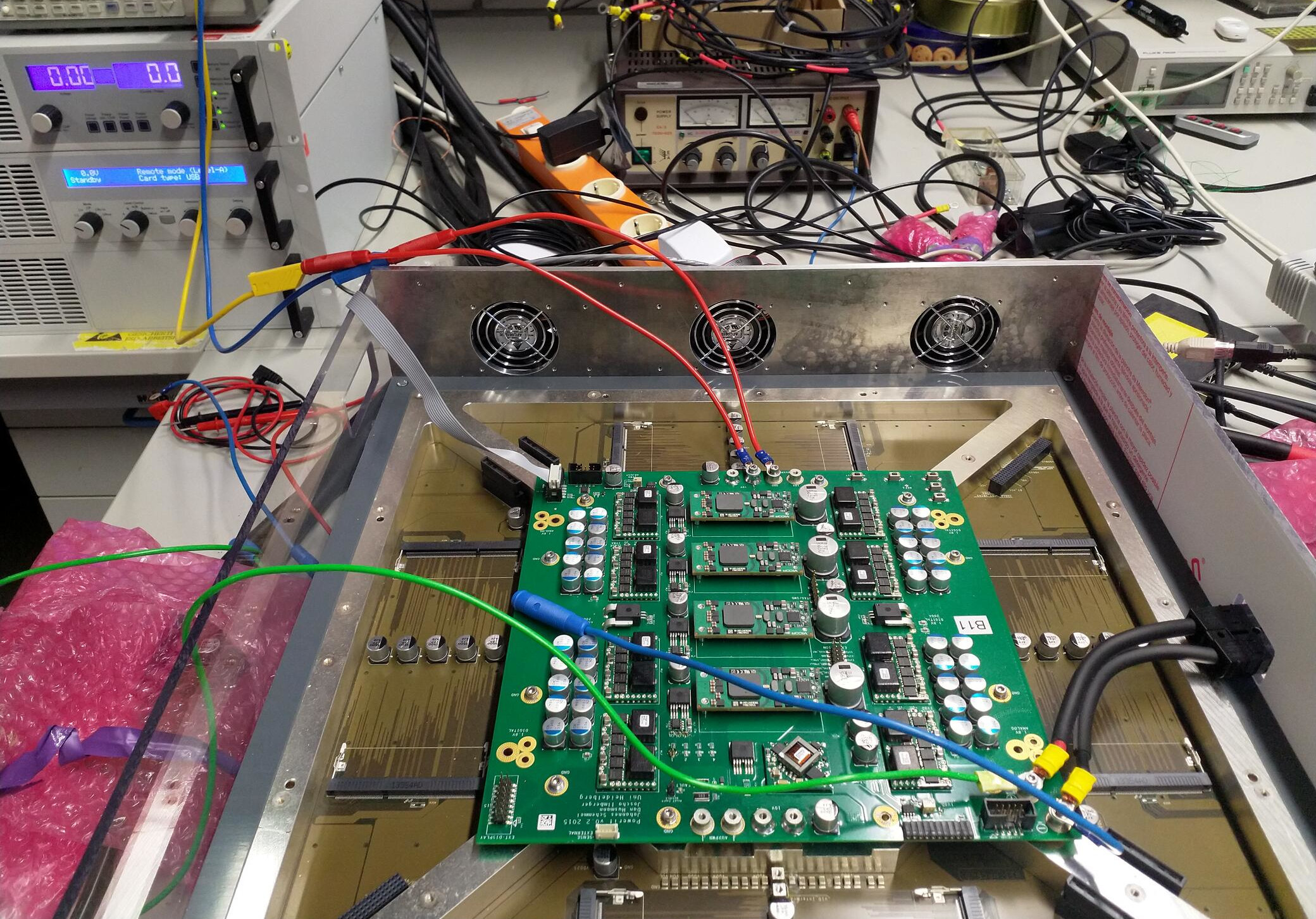}}
  \subfloat[\label{fig:prober_auxpwr}]{\includegraphics[width=0.48\linewidth]{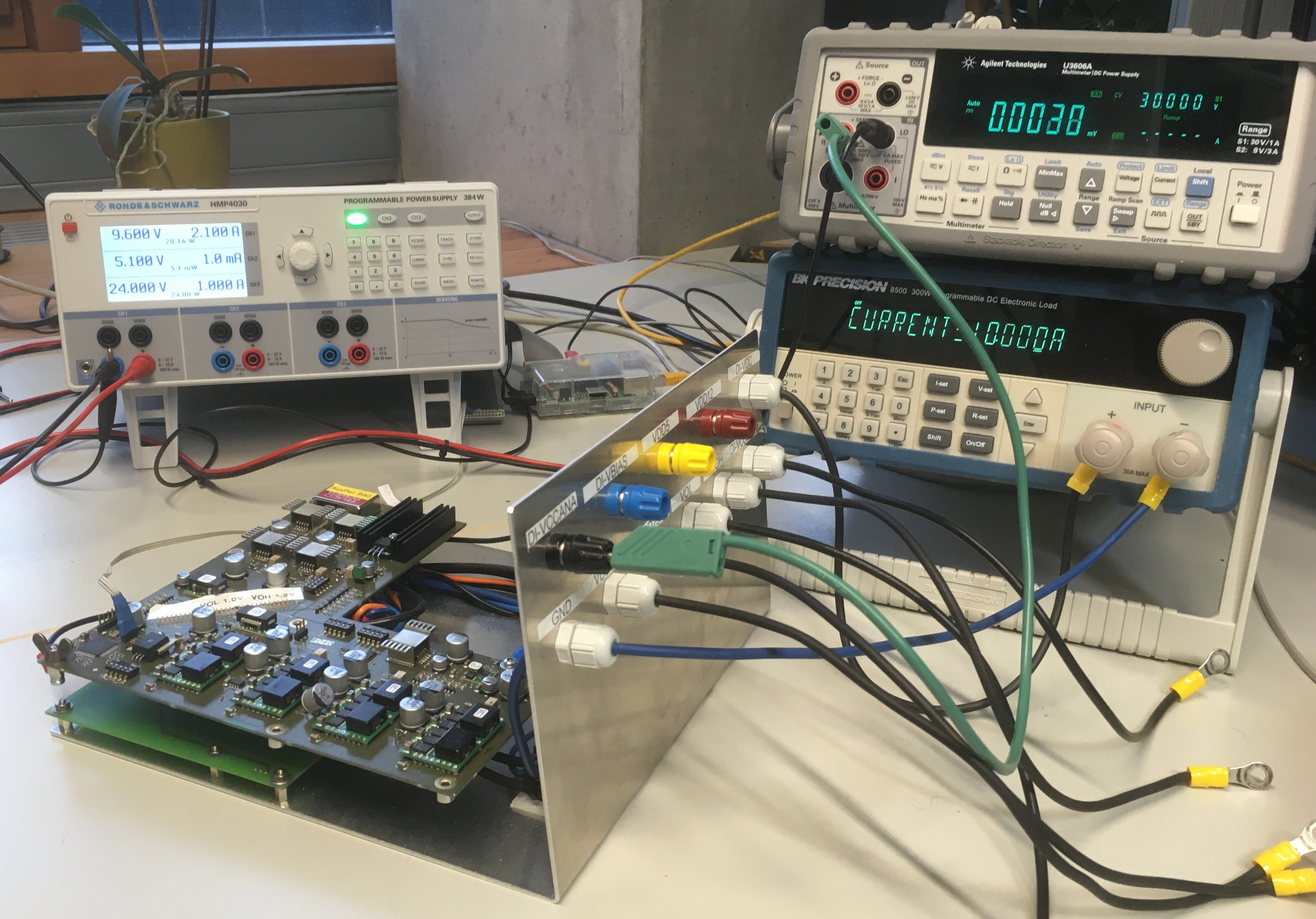}}\\
  \subfloat[\label{fig:prober_cure}]{\includegraphics[width=0.48\linewidth]{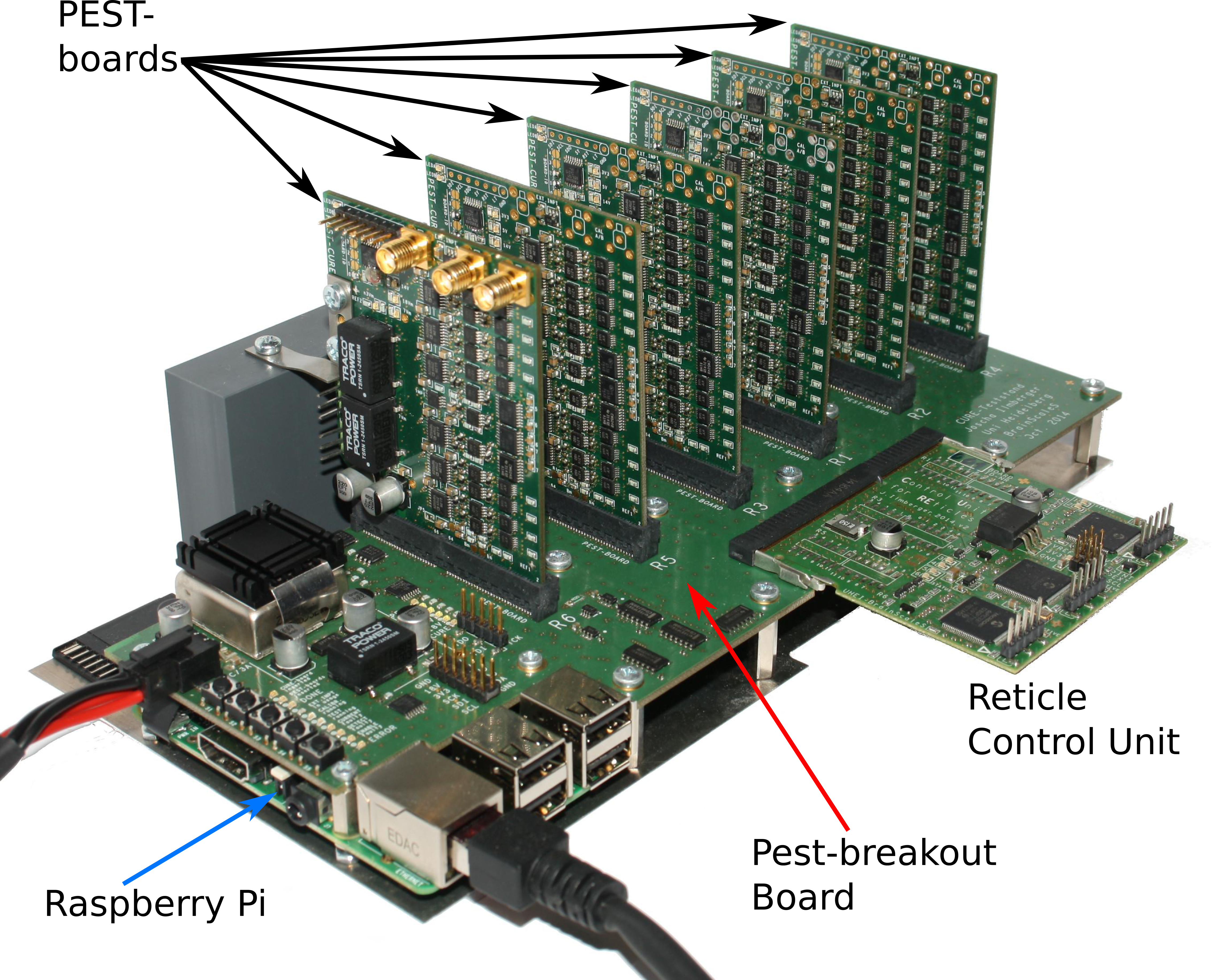}}
  \subfloat[\label{fig:prober_adc}]{\includegraphics[width=0.48\linewidth]{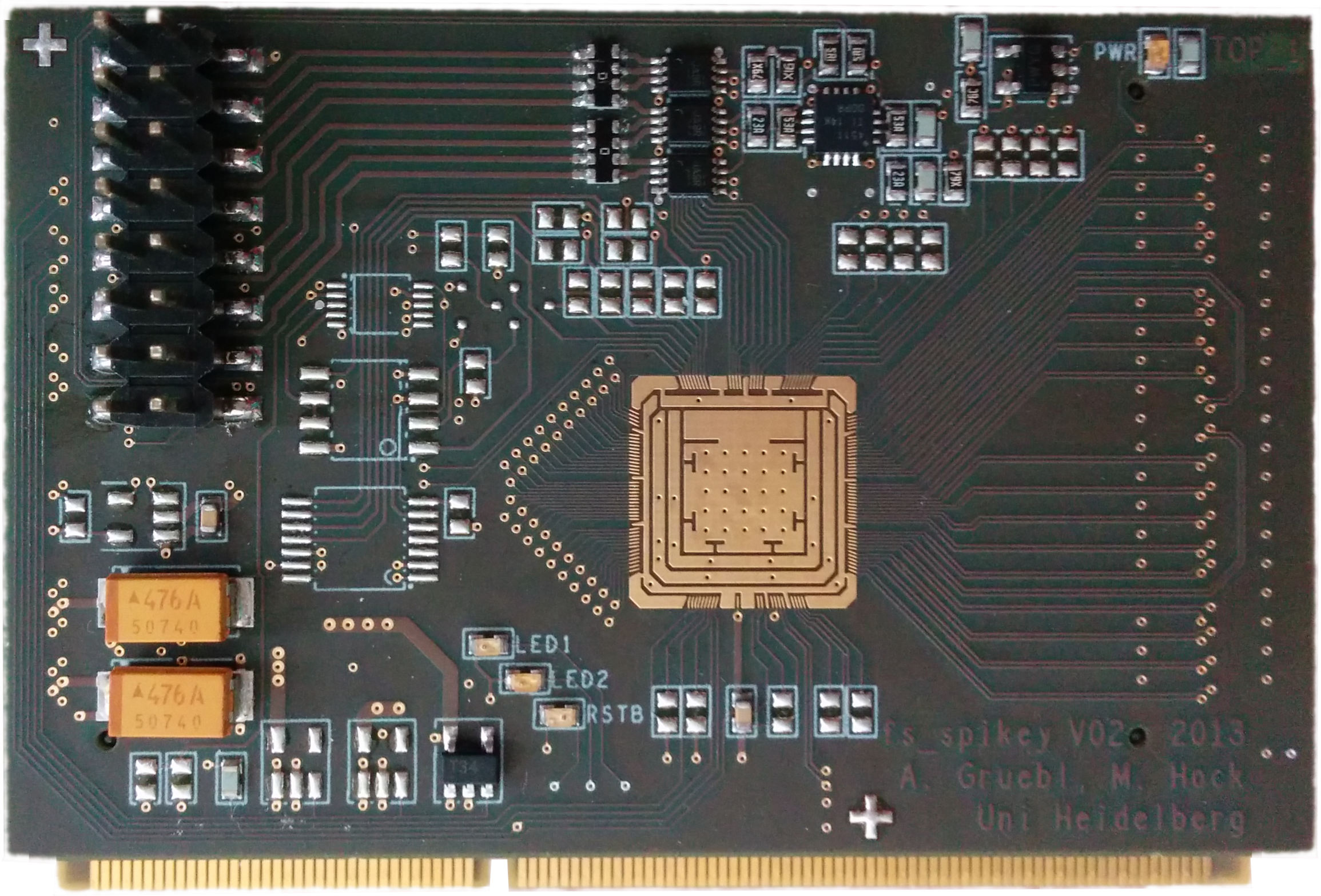}}
  \caption{Auxiliary boards under test.
  \protect\subref{fig:fcp_tud} communication boards test setup and \protect\subref{fig:wio_test} \acrlong{wio} board.
  \protect\subref{fig:prober_powerit} \acrlong{powerit} connected to programmable power supply and electronic load.
  \protect\subref{fig:prober_auxpwr} \acrlong{auxpwr} test stand.
  \protect\subref{fig:prober_cure} \acrlong{cure} test stand.
	Each Power Emulation Systems for Testing (PEST) board emulates the supply voltages of one \cgls{hicann} group.
  \protect\subref{fig:prober_adc} FPGA board of the \acrlong{anarm}. During the calibration, the pins on the top left are connected via a \SI{50}{\ohm} impedance to an external source meter, while the module is connected via USB to the host computer.
  Figures \protect\subref{fig:fcp_tud} and \protect\subref{fig:wio_test} made available by S.~Schiefer, TU-Dresden.
  }
  \label{fig:auxiliary_prober}
\end{figure}

\textit{Testing:} Commissioning of the \cgls{powerit} involves basic functionality tests and calibration of the current and voltage measurement circuits using an external electronic load capable of sinking \SI{4.8}{\kilo\watt} and precision multimeters, see~\cref{fig:prober_powerit}.

\subsubsection{Auxiliary Power Supply}
The \cgls{auxpwr} designed in~\cite{sterzenbach2014auxpwr}, receives \SI{9.6}{\volt} from the \cgls{powerit} and provides ten different voltage outputs for the wafer module.
The currents drawn at the derived voltages vary from \SI{50}{\milli\ampere}, for the common-mode voltage of the \cgls{lvds} on-wafer communication, to \SI{60}{\ampere} for the synapse driver output.
The board has an L-shape with linear and switching regulators placed on different axes to reduce the coils' electromagnetic-noise induction.
In addition, the usage of intermediate voltages reduces the power dissipation for the voltage scaling.
An onboard microcontroller monitors all the voltages and currents.
Four voltages can be controlled digitally through the \cgls{i2c} protocol.

\textit{Testing:} The \cgls{auxpwr} components' functionality is tested during the calibration process of the board, during which an external voltmeter permits adjusting voltage offsets.
A two-point linear calibration under load is performed for the currents.
The test stand can be seen in~\cref{fig:prober_auxpwr}.

\subsubsection{Control Unit for Reticles}
Since the \acrlong{bss1} wafer is not cut into individual chips, the wafer module must be fault-tolerant to individual \cgls{hicann} problems.
For this purpose, the Main PCB features power-FETs for the supply rails of each \cgls{hicann} group of the wafer; overcurrents manifest as a large voltage drop across these power transistors.
The \cgls{cure} controls the gates of these transistors and monitors the supply voltages of the wafer.
Three microcontrollers manage the measured data and react to fault conditions by shutting off the power of the affected \cgls{hicann} groups.
Thus, the \cgls{cure} allows to identify individual fatal faults and to exclude the respective \cgls{hicann} groups from the usable components.
The term reticle refers to the semiconductor manufacturing process and consists of one \cgls{hicann} group.

\textit{Testing:}  The \cgls{cure} is tested using a custom setup producing the voltages expected inside the actual \acrlong{bss1} wafer module, simulating all possible fault conditions while the response time is measured.
Likewise, the drive strength of the control signals for the power transistors on the Main PCB is quantified.
The test setup is displayed in~\cref{fig:prober_cure}.

\subsubsection{Analog Readout Module}
Further insight into the neuron dynamics can be obtained via measurements of its membrane potential, allowing for a better understanding of experiment results and the implementation of calibration routines.
To this end, each neuron contains a switchable analog output amplifier that connects to one of two \SI{50}{\ohm} output buffers per die.
These two outputs are each short-circuited across dies in the same \acrshort{hicann} group. Therefore, each of these groups has two analog outputs, totaling \num{96} independent analog channels available on each wafer module.

The \cgls{anarm} system consists of twelve \cgls{fpga}-controlled \num{12}-bit ADC modules that allow for the digitization of the membrane voltages on one wafer module per \acrlong{bss1} system rack.
Each of the modules in the \cgls{anarm} system connects through a ribbon cable to one of two \cglspl{anab} mounted on the Main PCB, receiving eight analog signals that are multiplexed into the ADC.
An additional digital signal acts as a trigger; four \cgls{hicann} groups share one, allowing synchronization during an experiment between the involved communication boards, \cglspl{hicann} and the \acrshort{anarm} system.
Overall, the \acrshort{anarm} system can simultaneously sample 12 membrane traces per wafer module.

\textit{Testing:} The \acrshort{fpga} board in the \cgls{anarm}, displayed in~\cref{fig:prober_adc}, undergoes DRAM memory tests and basic functional testing of all its peripheral components.
The analog front end is tested during the calibration of the modules.
This calibration is performed using a source meter to generate a series of ground-truth voltages, which are subsequently measured using each input channel.
A \SI{50}{\ohm} series impedance is used at the output of the source meter to match the impedance of the output buffers on the \cgls{hicann}.
This voltage divider formed by the output and input impedances halves the \SI{1.8}{\volt} span of the \cgls{hicann} output to the \SI{0.9}{\volt} maximum input of the \cgls{anarm}.
A linear function fits the recorded signal to the source meter voltages, and the per board offset and gains are stored in a database.

\subsection{Main Control Unit}
The \cgls{macu} consists of a Raspberry Pi powered by the standby voltage of the \cgls{powerit}.
Using the \cgls{i2c} protocol to communicate with all other wafer module components, it controls the start-up sequence of the system.
Additionally, it monitors the multitude of components of a wafer module, which is crucial to ensure robust operation.
With this in mind, the \cgls{macu} aggregates over \num{1800} metrics per wafer, e.g., supply voltages, temperatures, or the active/inactive status of components.
Most data is of a time-series nature and stored via Graphite~\cite{carbon}, with visualization through Grafana dashboards~\cite{grafana}.
These dashboards are hierarchically structured, allowing an intuitive drill-down navigation of the data.
As it is not practical to manually oversee such a large amount of metrics, alerts are set up to check for unexpected events.
For example, supply voltages are checked to be in a valid range and to remain constant over time.
Furthermore, event data, e.g., powering up components, is handled via the ELK stack~\cite{elasticsearch} but also integrated into Grafana and displayed as marks.
These allow easily matching the events with changes in the time-series data.

\textit{Testing:} The Raspberry Pi computers used for the \acrshortpl{macu} are purchased and commissioned without further tests.
However, the maintenance and deployment of the control and monitoring software they run is part of the system's continuous integration development methodology~\cite{mueller2022operating}.

\section{System Assembly and Integration Tests}
\label{sec:system_assembly}

In addition to the tests devised for the individual components, the \acrlong{bss1} wafer module assembly process is carried out along with additional tests that allow pinpointing problems to the individual steps.
In the following, we discuss the module assembly method and the different tests it undergoes during this phase.

\subsection{Wafer to Main PCB Marriage and Module Integration}
\begin{figure}
  \centering
  \subfloat[\label{fig:elastomeric_connectors}]{\includegraphics[width=0.7\linewidth]{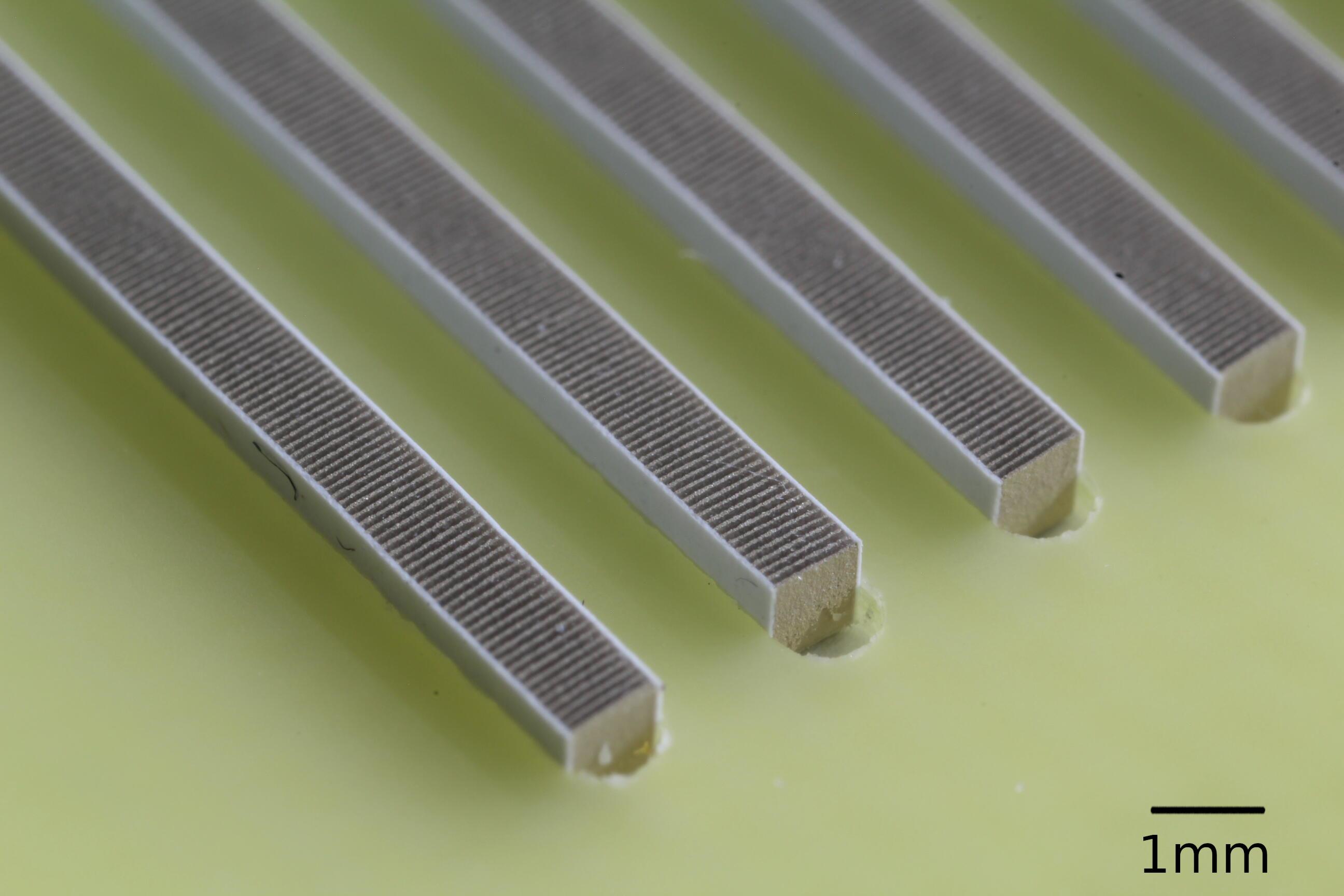}}\\
  \subfloat[\label{fig:wafer_alignment}]{\includegraphics[width=0.53\linewidth]{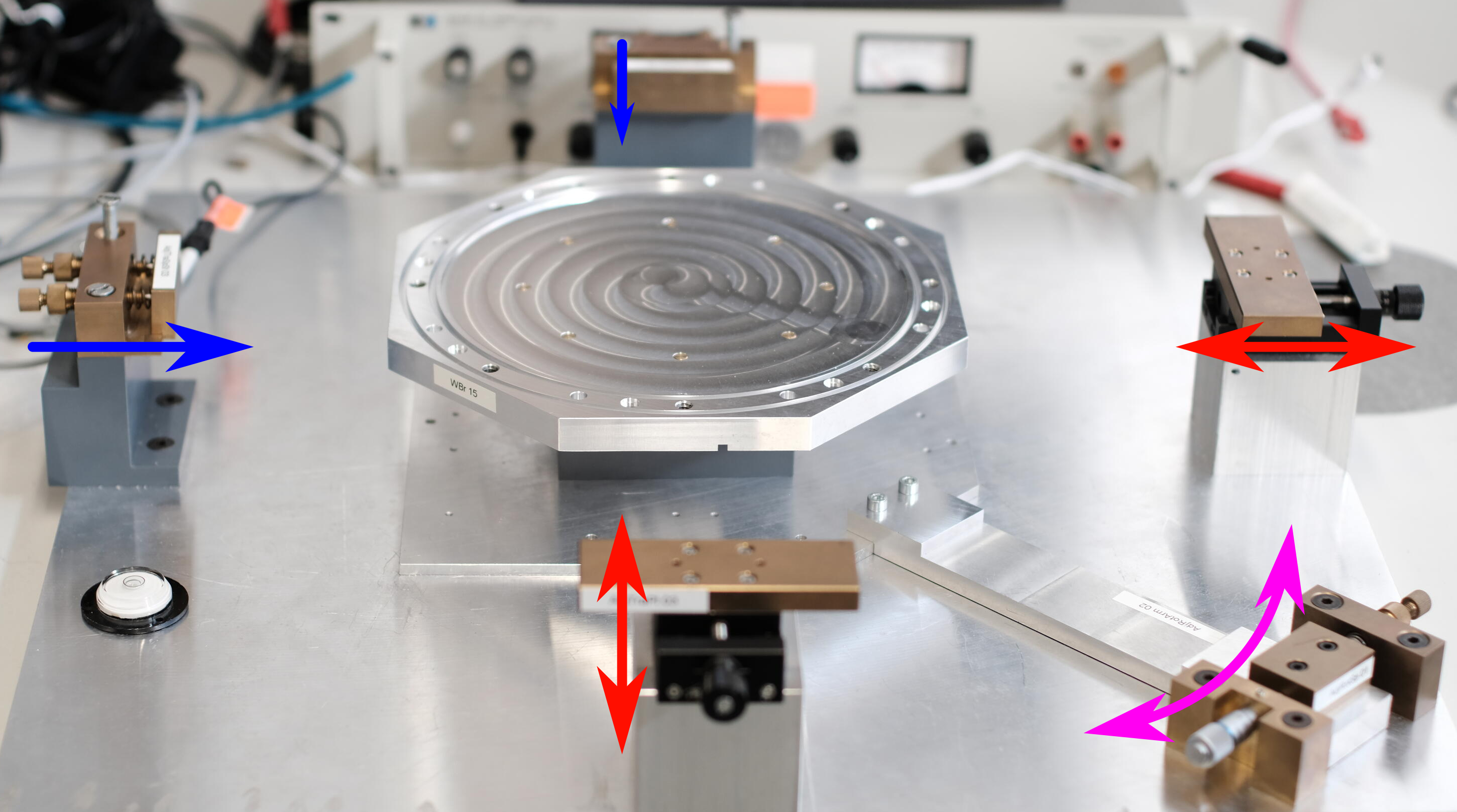}}
  \subfloat[\label{fig:assembly_test_boards}]{\includegraphics[width=0.445\linewidth]{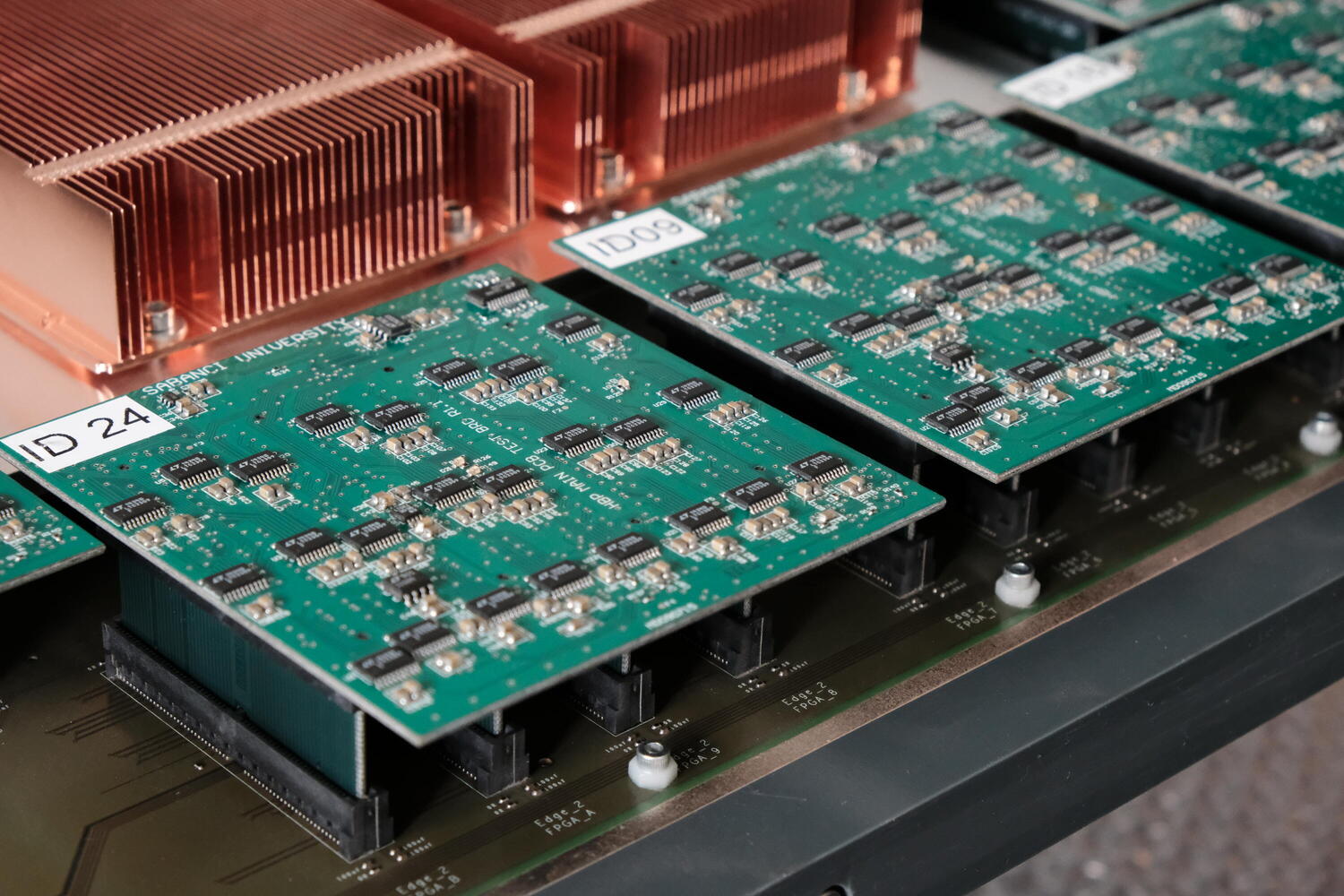}}\\
  \caption{\protect\subref{fig:elastomeric_connectors} Detail view of the elastomeric connectors that connect the pads on the \acrlong{bss1} wafer with the Main PCB.
          \protect\subref{fig:wafer_alignment} Station used to align the Main PCB to the silicon wafer.
           The Main PCB is fixed by springs that apply a constant force (blue arrows).
           Its position is controlled with a micrometer linear stage (red arrows).
           Angular errors can be corrected by rotating the wafer (purple arrows).
           \protect\subref{fig:assembly_test_boards} Test PCBs mounted on the Main PCB to measure the connectivity to the wafer during assembly.}
  \label{fig:alignment}
\end{figure}

The wafer is connected to a total of \num{11904} pads on the Main PCB via \num{384} elastomeric connectors, shown in~\cref{fig:elastomeric_connectors}.
Mounting the Main PCB and the silicon wafer in custom-milled aluminum brackets allows reaching the compression forces required by the connectors.
The station used to align the two components is shown in \cref{fig:wafer_alignment}.
Electrical resistance tests, described in \cref{sec:assembly_tests}, are performed while compressing the elastomeric connectors to ensure correct positioning and even pressure distribution.
Then, the wafer module is populated with the auxiliary boards and, when fully assembled, connected to the \cgls{macu}.
Afterward, it is put on a test stand for initial full-system tests using the same communication chain later used for experiments.
Following this step, the wafer module is placed in a rack in the machine room and attached to the \cgls{anarm} system.

\subsection{Tests at Different Assembly Stages}
\label{sec:tests_installation_steps}

Stage-specific tests allow mapping arising errors to individual assembly steps of the \acrlong{bss1} wafer module, which enables evaluating and improving the procedure.
This section shows the test results obtained for one wafer as an example.

\subsubsection{Pre-Assembly Tests of All \acrshortpl{hicann} on the Wafer}
Before placing a wafer in a module, digital and analog tests are performed on a wafer prober in the institute's clean room, see \cref{fig:clean_room_wafer_prober}.
These tests distinguish production problems from those arising in the wafer module assembly procedure.\\
Similar to the initial needle card tests on the unprocessed wafers, described in \cref{sec:hicann-wafer}, a test system was built using a different needle card connecting to the redistribution layer of a pair of \cglspl{hicann} on a wafer with post-processing.
Extended analog and digital tests are run on the connected dies, a process that is repeated until the entire wafer is analyzed.
These tests serve two purposes: first, to sort out wafers with a high error count that might arise from disrupted connections in the post-processing, and second, to establish a base level for the following assembly tests.
\Cref{fig:prober_switchram} shows the results of a high-level test for all \cglspl{hicann} of one wafer.
The image shows more test results than the number of dies on the picture of the assembled wafer module.
The reason for this was design constraints and limited routing resources on the Main PCB, by which not all \cgls{hicann} groups could be electrically connected and thus used within the module context; those at the edge of the wafer were left out.
For the same reason, the two \cgls{hicann} groups at the center are without high-speed connection.

\subsubsection{Tests During the Assembly Phase}
\label{sec:assembly_tests}
For these additional tests the Main PCB is equipped with test PCBs\footnote{Developed by the group of Yasar Gürbüz at Sabanci University, Istanbul}, shown in \cref{fig:assembly_test_boards}, which measure \acrshort{esd} diode currents and termination resistances between the \cgls{lvds} lines on the wafer.
The tests determine whether a good connection of the wafer to the Main PCB exists.
\Cref{fig:assembly} shows the result of one of these tests, where only the same faulty device on \acrshort{hicann} group 29, also detected in the needle card test, can be seen.
No additional faulty devices validate that the wafer to Main PCB marriage was appropriate.

\subsubsection{Post-Assembly Tests of All \acrshortpl{hicann} on the Wafer}
\begin{figure}
  \centering
  \subfloat[\label{fig:prober_switchram}]{\includegraphics[width=0.5\linewidth]{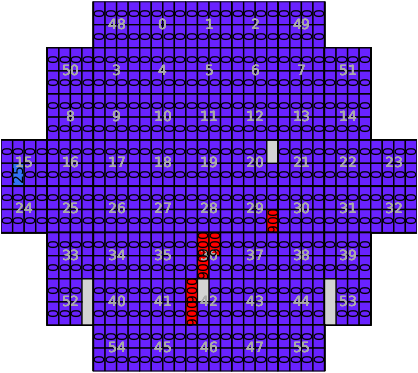}}
  \subfloat[\label{fig:assembly}]{\includegraphics[width=0.5\linewidth]{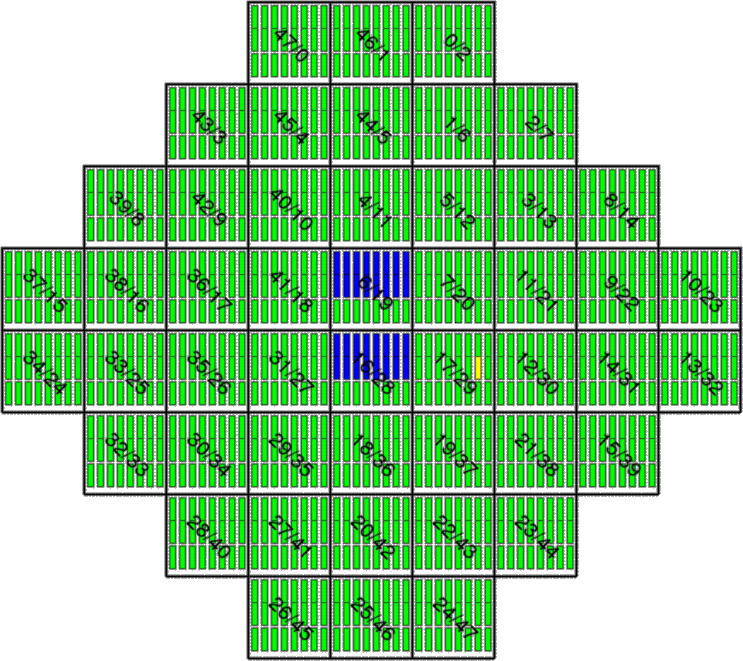}}\\
  \subfloat[\label{fig:module_switchram}]{\includegraphics[width=0.5\linewidth]{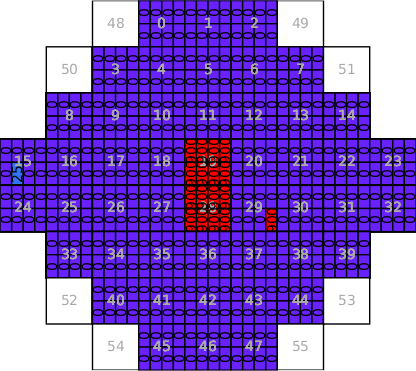}}
  \hspace{0.003\linewidth}
  \subfloat[\label{fig:commtest}]{\includegraphics[width=0.455\linewidth]{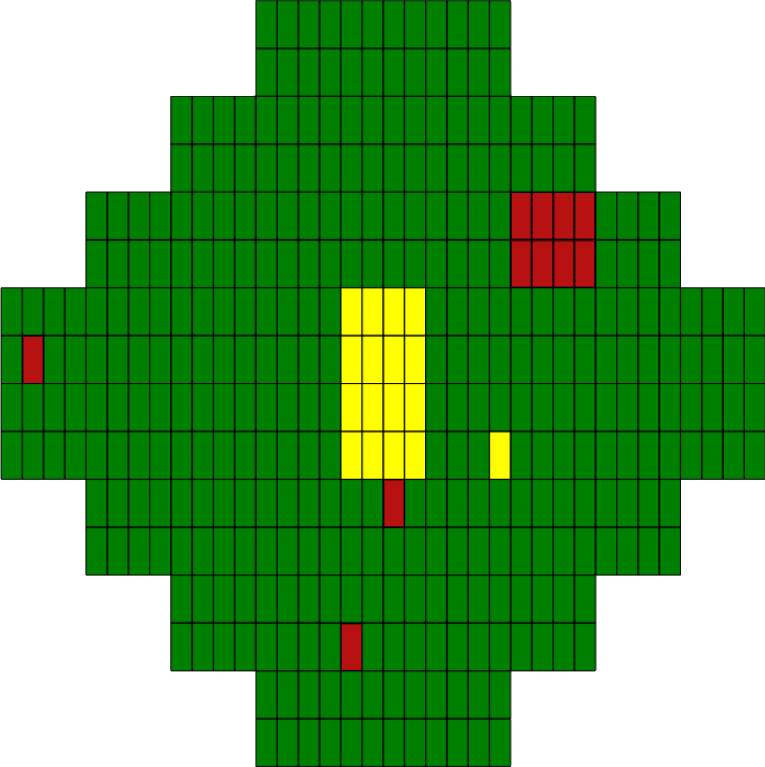}}
  \caption{Test results of one \acrlong{bss1} wafer for the different assembly steps:
	\protect\subref{fig:prober_switchram} Before assembly \protect\subref{fig:assembly} during assembly \protect\subref{fig:module_switchram} after assembly.
	In \protect\subref{fig:prober_switchram} and \protect\subref{fig:module_switchram}, the number in the smallest rectangles shows the amount of errors found on the corresponding \cgls{hicann}.
	Purple or red indicate that all tests were successful or failed, respectively.
	For grey \cglspl{hicann} the test was skipped since no connection could be established using the wafer prober.
	In \protect\subref{fig:assembly}, test results are shown per elastomeric connector and a yellow rectangle indicates a problem in the high-speed communication of one \cgls{hicann}.
	\protect\subref{fig:commtest} Communication test result.
	\cglspl{hicann} without high-speed communication are marked yellow, without JTAG communication red.
	The center two \cgls{hicann} groups have no high-speed interface by design.
	Consequently, they are marked faulty in all tests requiring  high-speed communication to the Main PCB.}
  \label{fig:D9NKP14B6_tests}
\end{figure}

After the assembly of the wafer module is completed, the same tests run on the pre-assembly phase are conducted, and results are compared.
The results for one test are shown in \cref{fig:module_switchram}.
The errors in \acrshort{hicann} groups \num{15} and \num{29} are still present, while the errors in groups \num{36} and \num{42} are not.
Further investigations could trace these last errors to connection problems of the needle card used in the wafer prober.

\section{Commissioning Software}\label{sec:commissioning}
After assembly, additional steps are necessary to bring the \acrlong{bss1} wafer module into readiness for experiments.
These include digital tests to find and exclude malfunctioning components and calibrating the individual neurons to address manufacturing-process-induced circuit mismatches.
Databases store the results from these two steps, allowing serialized data storage to disk.
See~\cite{mueller2022operating} for details.
Furthermore, all steps are fully automated and periodically executed after installation of the module in the machine room to track the systems' current state.

\subsection{Communication Tests}
The first test that is executed on a newly assembled wafer module is the communication test, which is used to find unresponsive \cglspl{hicann}.
Communication problems most likely arise from insufficient connection quality between the Main PCB and the wafer, cf.~\cite{zoschkeguettler2017rdlembedding}, or from scratches or similar defects on the post-processing layers.\\
During the test, an individual connection is established to each of the \num{384} \cglspl{hicann} of one wafer.
The test is split into a high-speed test and a \cgls{jtag} test, which reflects the two possibilities to communicate with the \acrshort{hicann}.
Failures are stored separately in the availability database.
The result of a communication test is shown in \cref{fig:commtest}.
In this example, the result comparison between the test stand and the rack-mounted fully assembled wafer module shows one additional \cgls{hicann} group and \num{3} individual \cglspl{hicann} that cannot communicate via \cgls{jtag}.

\subsection{Memory Tests}\label{sec:blacklisting}
\begin{figure}
  \includegraphics[width=1\linewidth]{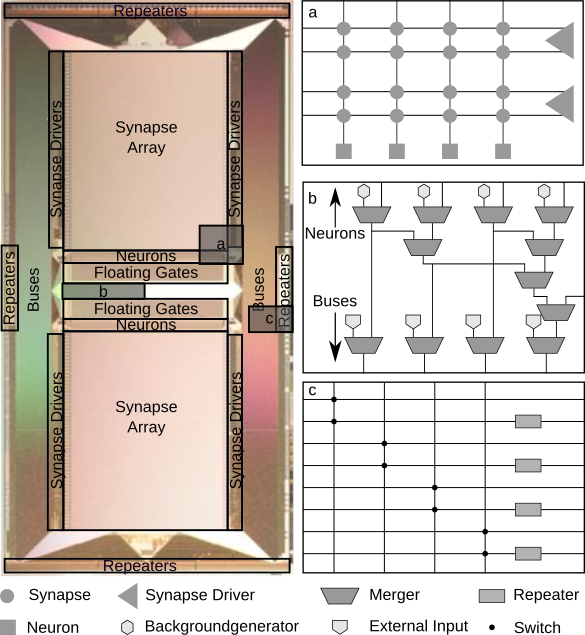}
  \caption{Left: Picture of the \cgls{hicann} with labeled components and marked areas shown on the right side.
  Top right: Detail of the synapse array.
  Two synapse rows are connected to one synapse driver.
  All synapses of the same column are connected to one neuron circuit.
  Middle right: Left half of the merger tree.
  Neuron input from the top gets routed to the buses on the bottom.
  Several inputs can be merged on the same bus.
  Background-generators are used to inject additional signals generated on-chip.
  Right bottom: Sketch of the bus system. Buses are connected by a sparse switch matrix.
  Repeaters, used to regenerate the signals, connect buses of neighboring \cglspl{hicann}.}
  \label{fig:hicann_overview}
\end{figure}

Using a whole uncut wafer, each \acrlong{bss1} wafer module profits from better energy efficiency and higher bandwidth for communication between its ASICs as if these were produced separately and then integrated.
This approach presents a challenge, though, as producing an error-free wafer-scale system in such a way is not possible, as ASICs with manufacturing-induced problems cannot be removed.
The \acrlong{bss1} system addresses this through a digital memory test, which in conjunction with the fault-tolerant system design, enables dynamic handling of malfunctioning components.
Executed after assembly as well as periodically, the test also tracks the state of the systems over time.
Therefore, it allows to operate wafer modules despite a subset of malfunctioning components or connections, consequently increasing the yield of functional systems.\\
The test builds upon the communication test and establishes a connection to a \cgls{hicann} group.
First, it initializes the connected communication board and the \cgls{hicann} under test.
Subsequently, each digital memory is repeatedly write/read-tested using random values.
If a mismatch is found, the largest functional unit that depends on the malfunctioning component is excluded so that it is not utilized in experiments.
\cglspl{hicann} that can communicate only via \cgls{jtag} are exclusively used for spike route-through to and from neighboring \acrshortpl{hicann} on the same wafer.
For these, a routing-specific reduced memory test minimizes the runtime using the slower connection.
In total, more than \SI{42}{\mebi\byte} of digital memory get tested per wafer.
Results for a fully assembled wafer module are shown in \cref{tab:blacklists}.
Tested components and their position on the \acrshort{hicann} are visualized in \cref{fig:hicann_overview}.\\
\begin{table}
  \centering
	\caption{Overview of excluded components extracted from a fully assembled \acrlong{bss1} wafer module.
	"Components" shows the number of components taken into account for the tests and the effective exclusion.
	If two numbers are given, the first one is the number of tested components and the second one is the number of components evaluated for the effective exclusion.
	"Individual" lists the communication and memory test results.
	Buses are marked with \mbox{"-"} because they have no digital memory that could be tested.
	"Effective" shows the results of the effective exclusion of components.
	Here, all components that should not be used during an experiment are included.
	They not necessarily failed a test.}
  \label{tab:blacklists}
  \begin{tabular}{llll}
    \toprule
    Resource & Components & \multicolumn{2}{c}{Excluded}\\ &&Individual & Effective \\
    \midrule
    \cGls{jtag} comm. & 384 & \SI{2.86}{\percent} & \SI{3.39}{\percent} \\
    High-speed comm. & 368/384 & \SI{3.26}{\percent} & \SI{7.81}{\percent} \\
    Synapse drivers & 78320 & \SI{0.04}{\percent} & \SI{0.04}{\percent} \\
    Synapse arrays & 712 & \SI{1.97}{\percent} & \SI{1.97}{\percent} \\
    Synapse rows & 159488 & \SI{0.11}{\percent} & \SI{0.11}{\percent} \\
    Synapses & 40099840 & \SI{0.68}{\percent} & \SI{0.68}{\percent} \\
    \cgls{fg} blocks & 1492 & \SI{0.34}{\percent} & \SI{0.34}{\percent} \\
    External input mergers & 2848/2984 & \SI{0.0}{\percent} & \SI{4.83}{\percent} \\
    Analog outputs & 712 & \SI{0.0}{\percent} & \SI{0.0}{\percent} \\
    Background-generators & 2848 & \SI{0.0}{\percent} & \SI{0.0}{\percent} \\
    Mergers & 5340 & \SI{0.0}{\percent} & \SI{0.0}{\percent} \\
    Buses & -/119360 & - & \SI{2.2}{\percent} \\
    Repeaters & 119360 & \SI{0.21}{\percent} & \SI{0.22}{\percent} \\
    Switches & 2864640 & \SI{0.02}{\percent} & \SI{0.02}{\percent} \\
    \bottomrule
  \end{tabular}
\end{table}

With \SI{110}{\kibi\byte} per \cgls{hicann}, the configuration registers of the synapses make up the largest part of the tested memory.
They are split into two synapse arrays per \cgls{hicann}, each of which is programmed by a custom on-chip SRAM controller described in~\cite{friedmann13phd}.
In the tests, on \SI{1.97}{\percent} of the synapse arrays, unstable behavior is observed.
This means, consecutive write/read operations with fixed values on a single synapse register show varying results.
Since problems in individual synapse registers are very unlikely and could also derive e.g. from the control chain, a special stability test is introduced.
There, each register is tested several times with the same value.
If a single register shows unstable behavior, the whole synapse array is excluded.
Thereby, at the expense of functional components, only stable programmable synapses are used during experiments.\\

A test with ten write/reads of random data per component and a stability test with ten repetitions takes approximately \SI{70}{\second} per \cgls{hicann}.
Since the tests can be executed in parallel for each \cgls{hicann} group, a full wafer test takes approximately \SI{10}{\minute} and can be executed periodically to track the state of the systems.

\subsection{Effective Exclusion of Components}\label{sec:derived_blacklisting}
In special cases, it is not enough to skip malfunctioning components during an experiment, but it is also important to be aware of hardware specific dependencies that can be linked with these components.
This is achieved through an additional step, the effective exclusion of components, where functional but dependent components are excluded.
Several dependencies lead to an effective exclusion.
Some of them are visualized in \cref{fig:hicann_overview}.
\begin{itemize}
  \item Unstable repeater controller:
	To enhance the signal integrity of spike events that have to be routed across several \acrshortpl{hicann}, the signal is regenerated between dies by repeaters.
	These repeaters are organized in blocks where each block has a custom on-chip controller used to program its repeaters.
	Since failures in the digital memory of the repeaters are very unlikely, more than one failing repeater per block indicates that there could also be a problem in the control chain.
	To ensure no unstable components are used, all repeaters connected to the corresponding repeater block are removed from the availability database in such cases.
  \item Buses connected to malfunctioning repeaters:
	Buses are used to route spike events between neuron circuits.
	On boundaries between two \cglspl{hicann}, the buses are connected to repeaters that regenerate the signal.
	Each repeater is connected to a bus on its own \acrshort{hicann} as well as on a neighboring one.
	If a repeater is failing the memory test, there is no possibility to test if it sends wrong signals to its connected buses.
	To circumvent this, all buses connected to such a repeater are excluded and thus not used during an experiment.
	The same holds for repeaters on \cglspl{hicann} without \cgls{jtag} connection.
	As the repeaters cannot be initialized correctly, all neighboring buses connected to repeaters on the problematic \cgls{hicann} are excluded.
  \item Malfunctioning \cgls{fg} controller:
	The \cglspl{fg} are not only used to configure the neurons but also to supply bias voltages to the spike event routing.
	If an error in the controller programming the \cglspl{fg} is found, the whole \cgls{hicann} is excluded from the availability database and, in the following, treated as if there would be no \cgls{jtag} connection.
	Such a \cgls{hicann} is not used at all in experiments.
  \item Without high-speed:
	\cglspl{hicann} that have no high-speed connection are, due to the higher bandwidth requirements, not used to emulate neurons or external inputs but only used to route spike events.
	This is achieved by removing all neurons and external input mergers from the availability database.
  \item No routing options:
	  To improve the placement and prevent lost connections, the algorithm checks that all the components required to establish a route from each neuron and external input merger are available. %
	  If not, the neuron or the external input merger is excluded and therefore skipped in the process of building a network.
  \item Handling hardware versioning:
	In an earlier version of the post-processing, connections were established to \acrshortpl{hicann} on the edges of the wafer that must not be connected.
	To prevent leakage currents from these dies, the connected buses are excluded.
	Therefore, it is unnecessary to distinguish wafer versions in all the following steps.
\end{itemize}
An overview of removed components before and after the effective exclusion of components can be seen in \cref{tab:blacklists}.
The availability database, used to handle the excluded components, allows for storing different states on disk, so malfunctioning components and effective components can be differentiated afterward.
This is for example important during the initialization of the \acrshortpl{hicann}, where only malfunctioning components have to be handled specifically.
\subsection{Analog Readout Tests}
Before usage, the analog recording system gets verified for correct connectivity and configuration by running a series of tests.
Each \cgls{hicann} is set in sequence to generate two different voltage levels, which the \cgls{anarm} measures.
The voltage levels originate from the configuration of one of the \cglspl{fg}.
A recording that agrees with the settings and whose noise levels are within a tolerance threshold indicates that the system is ready for experiments or calibration runs.

\subsection{Calibration}\label{sec:calibration}
VLSI transistors are subject to manufacturing variations translating into differences in signal response.
This problem and the potential impacts have been noted since the first approaches to neuromorphic computing using VLSI~\cite{mead89analog}.
Consequently, the \cgls{hicann}'s microelectronic analog circuits require correction mechanisms to deliver homogeneous responses.

As the manufacturing variability is stationary within the components' operating ranges, thus termed fixed-pattern noise, it can be reduced by suitable calibration.
To this end, a framework has been developed for the \acrlong{bss1} wafer module that performs a one-time circuit characterization through running sequences of experiments that sweep neuron parameters, measure the effect in the observable, and perform appropriate fits on suitable models.
The process creates a database that holds the calibration results and is loaded on routine hardware usage, allowing for automatic translation between biological-space parameters and \cgls{fg}-stored parameters.
Such a conversion is automated and transparent for the users when running an experiment.
See~\cite{mueller2022operating} for details.

The calibration procedure configures all the neuron circuits at once and then processes the individual measurements to allow for programming the \cglspl{fg} in parallel.
In addition, parallelizing the analysis algorithms on the already measured steps further optimizes the time required for calibration.
Regardless, an increase in the number of calibration steps could improve the quality of the fits, while also parameters that are more sensitive to \cgls{fg} parameter variability benefit from an increase in the number of measurement repetitions.
Consequently, calibration time and precision of the results require balancing.

\subsubsection{Calibration Methodology}

In the \acrlong{bss1} system, the only analog neuron property that can be directly recorded is the membrane voltage.
Accordingly, all parameter calibrations are based on membrane recordings under different parameter configurations.
In general, the calibration of one parameter sweeps over its operating range while maintaining the rest of the parameters constant.
The execution order is relevant, as some calibration routines require an already calibrated subset of parameters.
Furthermore, the calibration accounts for analog readout noise, and measurements can be repeated to factor in \cgls{fg} parameter variability.

\begin{figure}
  \centering
  \includegraphics[width=0.8\linewidth]{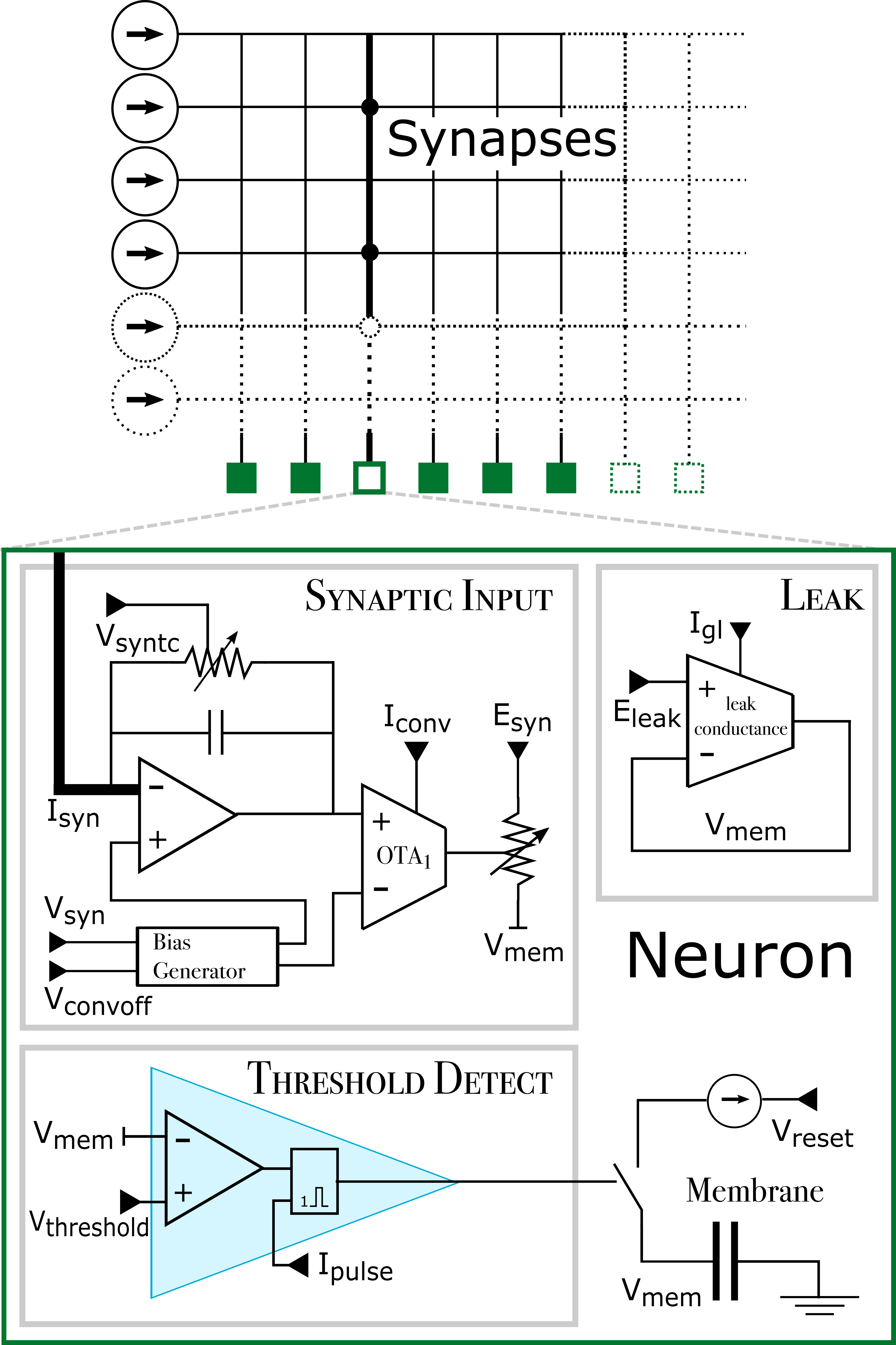}
  \caption{Simplified neuron circuit schematic, displayed on the bottom, with the most relevant calibrated parameters in the \acrlong{bss1} system.
  The leak conductance controlled by $I_\text{gl}$ is constantly driving the membrane potential $V_\text{mem}$ towards the rest voltage $E_\text{leak}$.
  A spike is elicited when the membrane potential reaches the threshold voltage $V_\text{threshold}$.
  After a neuron spikes, its membrane's potential is connected to the reset voltage $V_\text{reset}$ for a period controlled by the parameter $I_\text{pulse}$.
  For simplicity, one synaptic input is displayed out of two through which a neuron integrates excitatory and inhibitory input currents $I_\text{syn}$; this controls a conductance between the reversal potential $E_\text{syn}$ and the membrane with a synaptic time constant controlled by $V_\text{syntc}$.
  Each input receives currents from all the synapses connected to one column in the synaptic array, displayed on top.
  Additional parameters $V_\text{syn}$, $V_\text{convoff}$, and $I_\text{conv}$ provide further control over the synaptic input, as further discussed in the supplementary material.
  }
  \label{fig:neuron_parameters}
\end{figure}

The main neuron calibration parameters are summarized in \cref{fig:neuron_parameters}.
In the following, the calibration procedure is exemplarily shown for the parameter $I_\text{pulse}$, which controls the refractory period $\tau_\text{ref}$, i.e., the time after the emission of a neuron's action potential during which its membrane is clamped to the reset potential and the neuron can elicit no further spike.
The higher $I_\text{pulse}$ is, the shorter the achieved $\tau_\text{ref}$.
Each $I_\text{pulse}$ calibration step sets the resting potential $E_\text{leak}$ above the level at which a spike event is elicited, i.e., $V_\text{threshold}$, which causes the neurons to spike continuously.
The \cgls{isi} is the measurable result.

In the first step, $I_\text{pulse}$ is set to maximum, and the corresponding \cgls{isi} is regarded as \cgls{isi}$_0$, the minimum attainable interval under the current settings.
Larger refractory periods are referenced to \cgls{isi}$_0$ by using

\begin{equation}
  \label{eq:tau_ref_zero}
  \tau_\text{ref}(I_\text{pulse}) = \text{ISI}(I_\text{pulse}) - \text{ISI}_0,
\end{equation}

making the minimum $\tau_\text{ref}$ zero seconds by definition.
Afterward, each step's distinct target \cgls{fg} values of $I_\text{pulse}$ are programmed, causing changes observable in the \cgls{isi} and thus in $\tau_\text{ref}$.
The obtained set of configured parameters and their achieved refractory periods is then fit to a model, which in the case of $\tau_\text{ref}$ corresponds to

\begin{equation}
  \label{eq:Ipl_model}
  I_\text{pulse} = \frac{1}{(c_{0} + c_{1}\cdot \tau_\text{ref})}.
\end{equation}

\begin{figure}
  \includegraphics[width=0.90\linewidth]{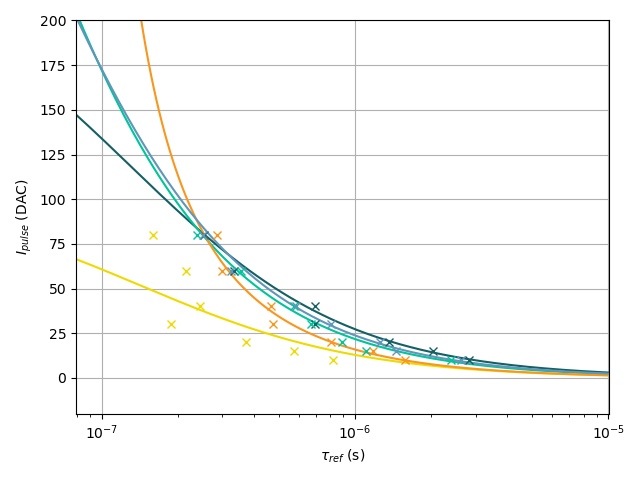}
  \caption{Exemplified calibration procedure for the refractory period.
           Sample fits obtained for a set of neurons, relating the $I_\text{pulse}$ parameter configured with the Floating Gates with the measured $\tau_\text{ref}$.
  Seven values within the dynamic range of $I_\text{pulse}$ were used for the fits.}
  \label{fig:calibration_fit_depiction}
\end{figure}

Such a model derives from transistor-level simulations described in~\cite{schwartz2013diss}.
The resulting fits for five neurons are shown in~\cref{fig:calibration_fit_depiction}.

The pair of constants $c_0$ and $c_1$ corresponding to model~\cref{eq:Ipl_model} is stored in the calibration database for each neuron, which is then used for translation from $\tau_\text{ref}$ in seconds to $I_\text{pulse}$ in digital value.
Further details for each parameter calibration are provided in the supplementary material.

\begin{figure}
  \centering
  \subfloat[
            \label{fig:calibration_I_pl}]{\includegraphics[width=0.5\linewidth]{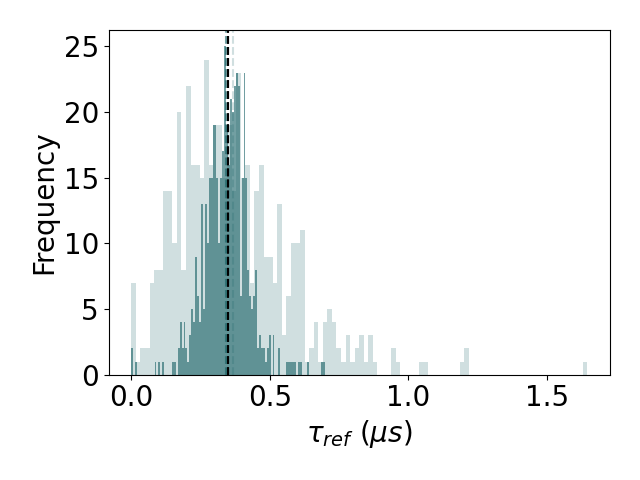}}
  \subfloat[
            \label{fig:calibration_E_leak}]{\includegraphics[width=0.5\linewidth]{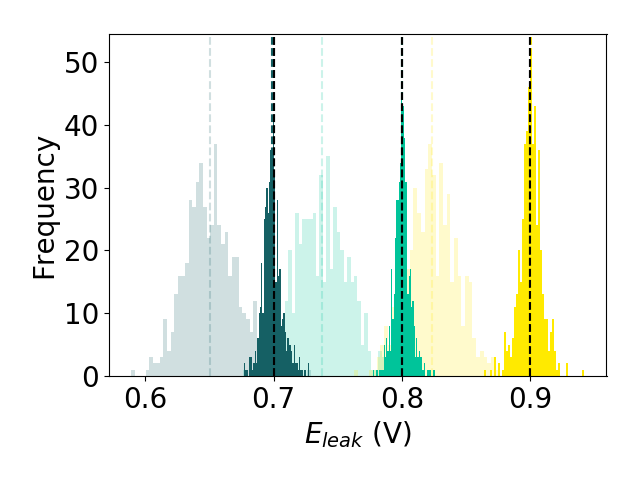}}
    \caption{Histograms of achieved parameter settings on all neurons of one \cgls{hicann} for \protect\subref{fig:calibration_I_pl} the refractory time constant controlled by the parameter $I_\text{pulse}$ and \protect\subref{fig:calibration_E_leak} the leak potential controlled by the parameter $E_\text{leak}$.
            Pale and intense colors correspond to the hardware achieved time constants and voltages for different target values (shown in black-dashed lines), before and after the calibration is applied, respectively.
            The narrowing and centering of the achieved value distributions is better for $E_\text{leak}$ than for $\tau_\text{ref}$.}
  \label{fig:I_pl_E_leak_calib_comparison}
\end{figure}

Depending on each parameter's sensitivity to the programmed \cgls{fg} values, some calibrations enable a more precise setting of parameters than others.
An increased sensitivity due to non-linear hardware dependencies is found where small changes in \cgls{fg} values cause large changes in the observables.
Furthermore, for some \cglspl{fg} only a limited range of their available parameter space is used, reducing the ability to set their corresponding parameters precisely.
As can be seen from the measured values in~\cref{fig:calibration_fit_depiction}, such is the case for $I_\text{pulse}$.
For comparison,~\cref{fig:I_pl_E_leak_calib_comparison} shows how the leak potential $E_\text{leak}$, which is easier to control, obtains a more precise calibration than $I_\text{pulse}$.
For this reason, the control precision of several parameters was improved in the second-generation \acrlong{bss2} chip~\cite{schemmel2020accelerated} partly by enabling digital value storage.

\subsubsection{Synapse Weight Calibration}
\begin{figure}
  \centering
  \subfloat[
            \label{fig:2d_weight_calib}]{\includegraphics[width=0.8\linewidth]{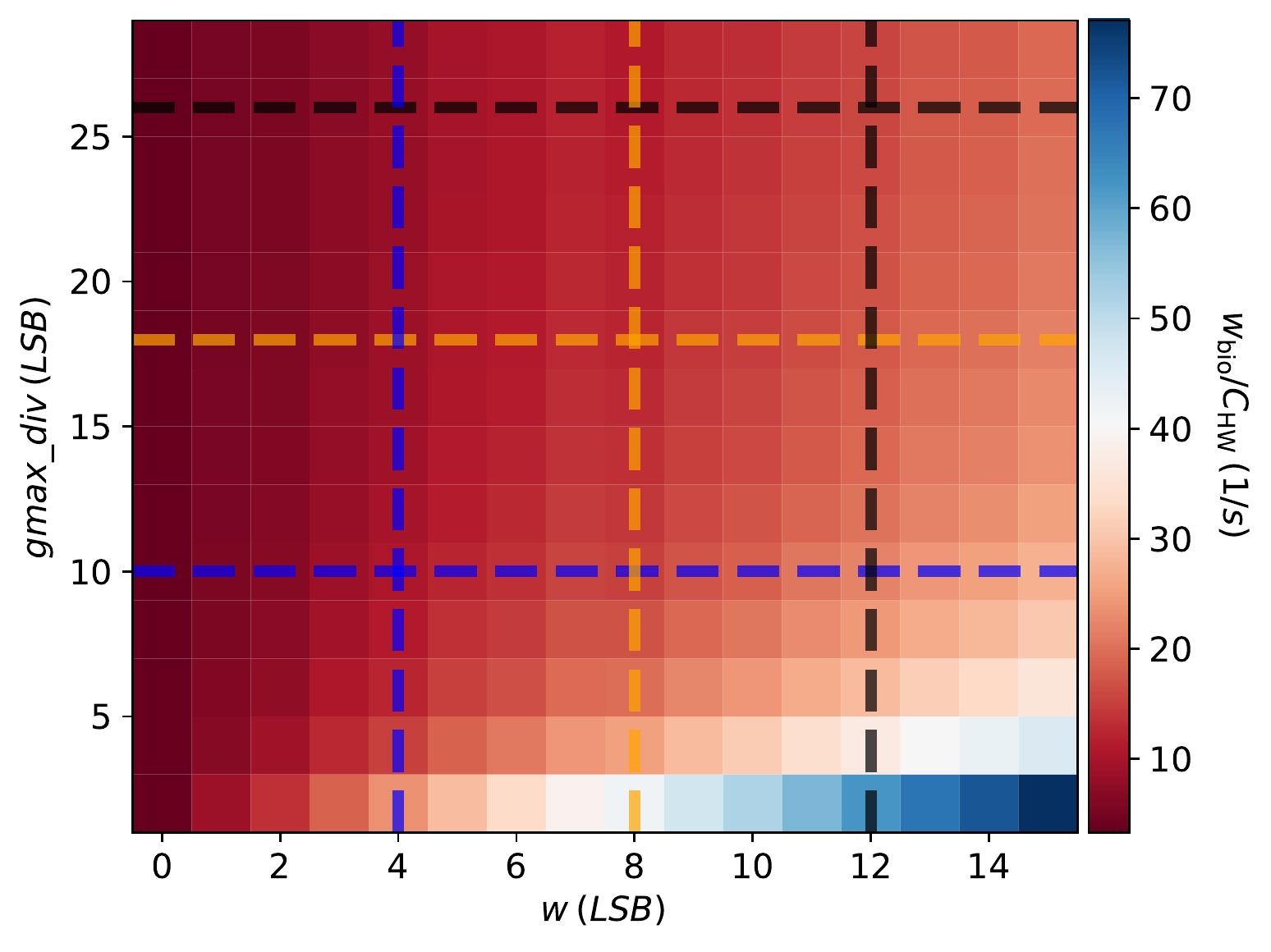}}\\
  \subfloat[
            \label{fig:weight_fit}]{\includegraphics[width=0.49\linewidth]{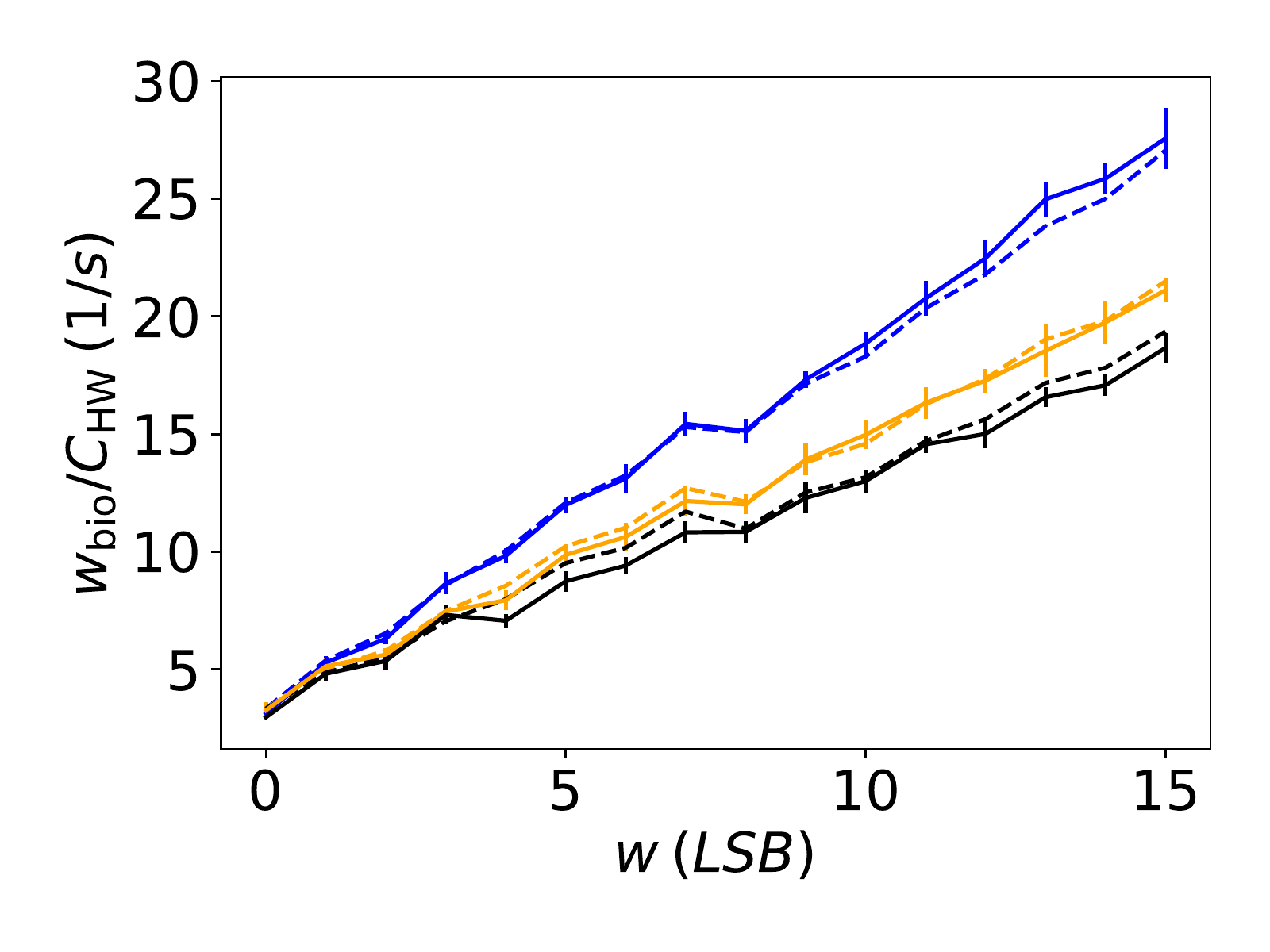}}
  \subfloat[
            \label{fig:gdiv_fit}]{\includegraphics[width=0.49\linewidth]{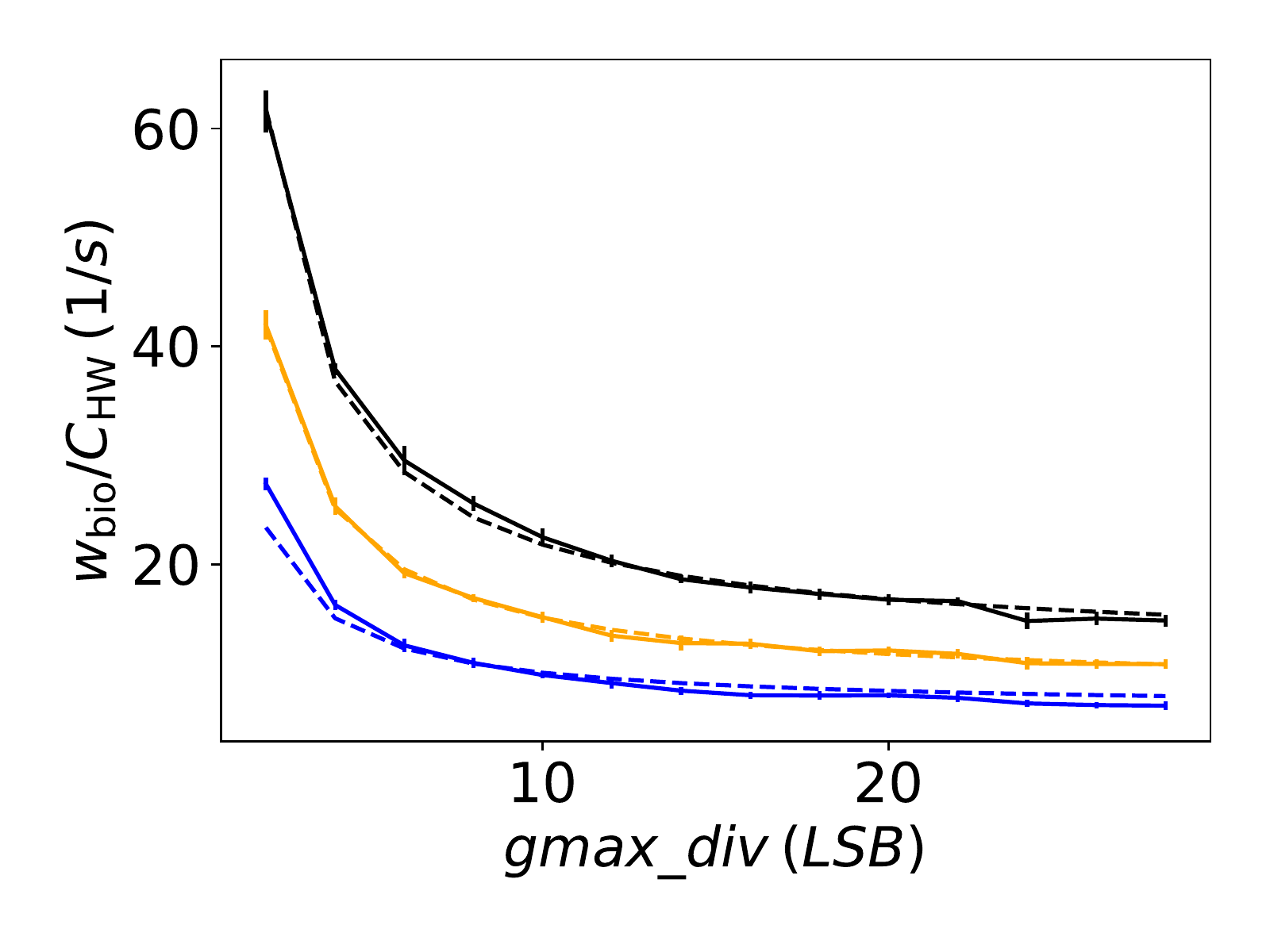}}\\
  \subfloat[
            \label{fig:rewriting_floating_gates}]{\includegraphics[width=0.5\linewidth]{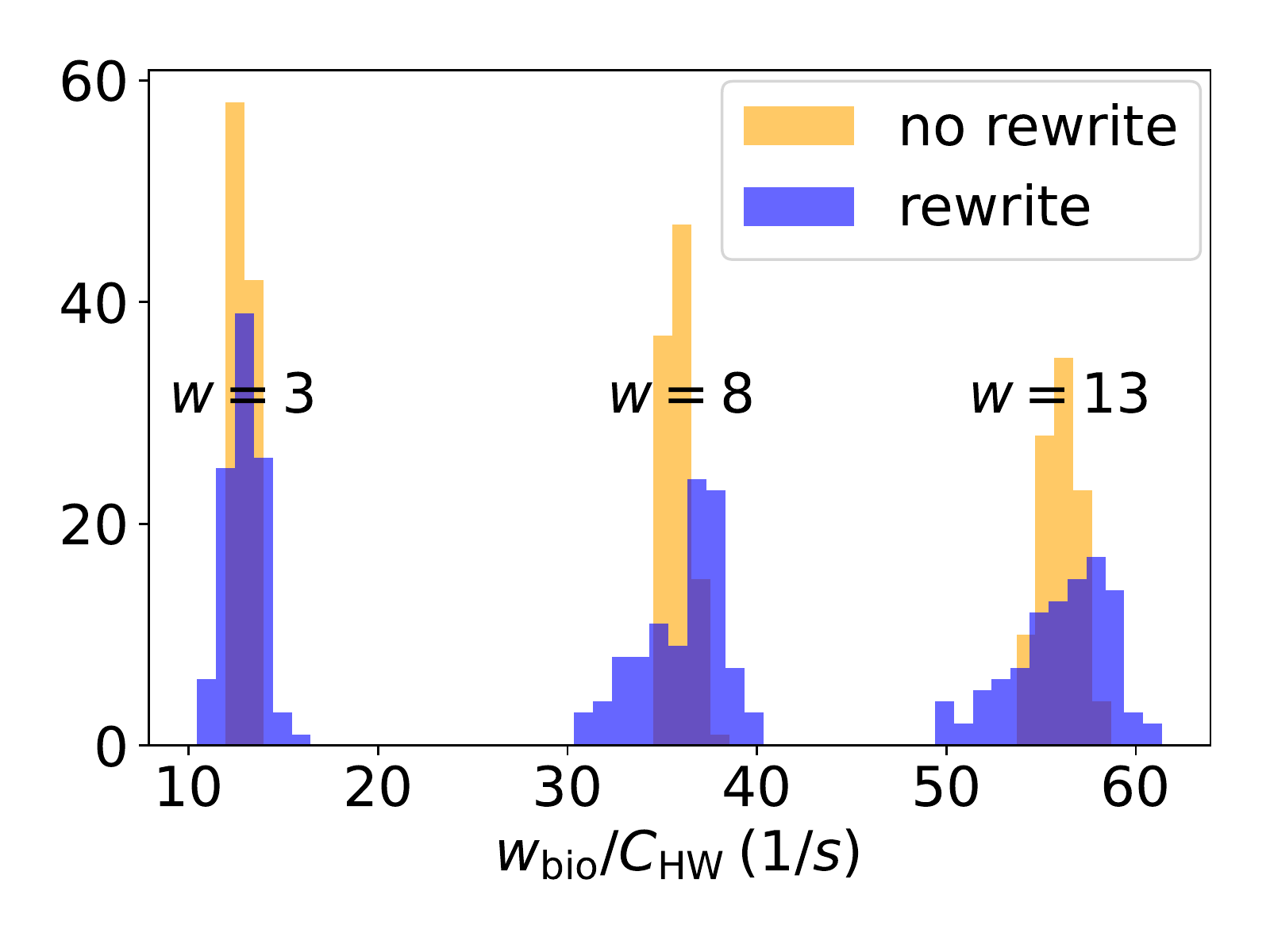}}
  \subfloat[
            \label{fig:mean_error_all_hicanns}]{\includegraphics[width=0.5\linewidth]{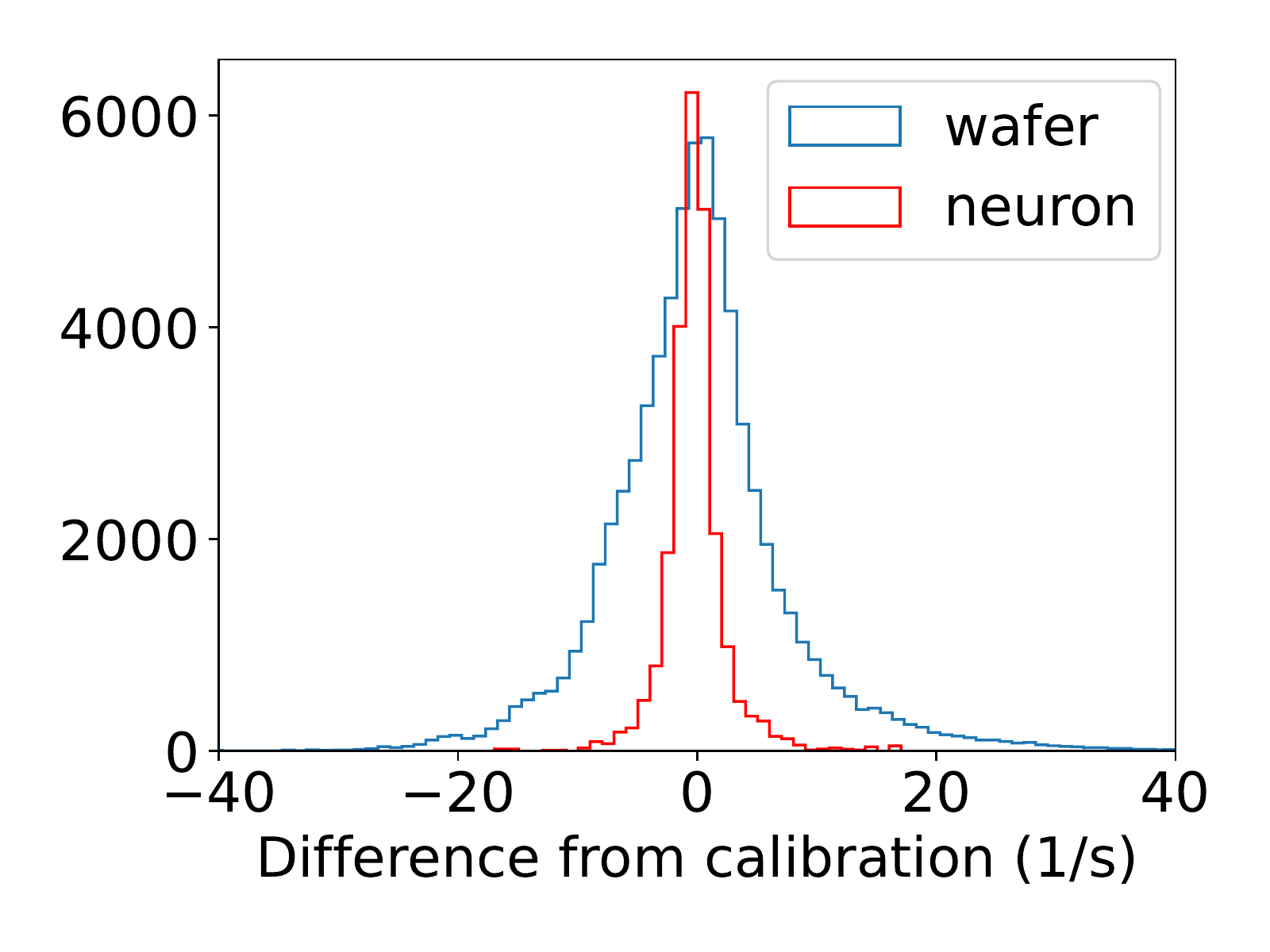}}
    \caption{Results of the synapse weight calibration.
            \label{fig:synapse_weight_calib} \protect\subref{fig:2d_weight_calib} Weight measurement for a fixed neuron circuit for different settings of the digital weight $w$ and hardware parameter $gmax\_div$ with $V_\text{gmax}=\SI{700}{LSB}$.
            Horizontal dashed lines indicate cuts with fixed values of the hardware parameter $gmax\_div$ shown in \protect\subref{fig:weight_fit}, vertical dashed lines indicate cuts with fixed digital weight values $w$ shown in \protect\subref{fig:gdiv_fit}.
            In \protect\subref{fig:weight_fit} and  \protect\subref{fig:gdiv_fit}, solid lines represent measured values, dashed lines the results of the fit of \cref{eq:weight} applied on the whole measured parameter space.
            \protect\subref{fig:rewriting_floating_gates} Variations of weight measurements with and without rewriting of the Floating Gates.
	    Values are extracted for \num{3} digital weight parameters $w$ from a fixed neuron with fixed hardware parameters ($V_\text{gmax} = \SI{700}{LSB}$, $gmax\_div = \SI{2}{LSB}$).
            \protect\subref{fig:mean_error_all_hicanns} Comparison of a per wafer and a per neuron weight calibration.
           Measurements for the entire parameter space are performed on a subset of neurons.
           The calibration is then performed for the whole subset or per individual neuron.
           The histogram shows the difference between the measured and expected values using the obtained calibrations.
         }
\end{figure}

The calibration of the synaptic input differs from the other calibrations due to its additional dependency on the synapse drivers.
The strength of a synapse is configured by three hardware parameters.
The \num{4}-bit digital weight $w$ stored per synapse, a scaling factor $gmax\_div$ stored per synapse row, and the \cgls{fg}-stored reference parameter $V_\text{gmax}$.
This last parameter is set per synapse row and selects one of four possible values shared by blocks of 128 neurons.
Calibrating this large parameter space for each of the \num{512} neurons with \num{110} connected synapse drivers using the analog readout system, which allows for measuring \num{12} membrane traces in parallel, is not possible in a reasonable time frame.
Therefore, a per wafer translation is performed, where only some of the components are taken into account to find the average circuit behavior.
The measurement requires the results of all previous calibrations.
Neurons on different \cglspl{hicann} are stimulated by a single spike for different combinations of the three hardware parameters to cover the whole parameter range.
Subsequently, a fit of the conductance based neuron model is applied to the recorded membrane traces to extract the ratio between biological weight and membrane capacitance $\frac{w_\text{bio}}{C_\text{HW}}$.
Since the membrane capacitance is fixed during experiments, it is unnecessary to separately determine both values.
During the fit, the model parameter of the already calibrated reversal potential is fixed.
The reduced $\chi^2$ value of the fit is used to identify and exclude saturation effects of the involved \cgls{ota}$_{1}$, cf. \cref{fig:neuron_parameters}, which might occur for large weight values.
Finally, the weight translation is found by fitting the expected hardware behavior
\begin{linenomath}
\begin{equation}
  \label{eq:weight}
  A(\frac{w \cdot V_\text{gmax}}{gmax\_div} + i_\text{0} + i_\text{1} \cdot w_\text{1} + i_\text{2} \cdot w_\text{2} + i_\text{4} \cdot w_\text{4}+ i_\text{8} \cdot w_\text{8}),
\end{equation}
\end{linenomath}
adapted from~\cite{Koke2017}, to the results of the first fits.
The fit parameters $i_\text{0-8}$ characterize the effect of parasitic capacitances found in the synaptic circuit for each enabled bit of the \num{4}-bit weight value $w$.
\Cref{fig:2d_weight_calib} demonstrates the large parameter space of the synapse weight calibration.
It shows the measurement of a single neuron, stimulated by a single synapse driver for a single $V_\text{gmax}$ value without rewriting the \cglspl{fg}.
The performance of the fit applied on the whole measured parameter space is shown for fixed values of $gmax\_div$ in \cref{fig:weight_fit} and for fixed digital weight values $w$ in \cref{fig:gdiv_fit}.
Although the whole neuron circuit and consequently the expected noise of each individual component is involved, the error of each measurement does not exceed the variations observed in other calibrations.
However, additional deviations arise from rewriting the \cglspl{fg}, which is demonstrated in \cref{fig:rewriting_floating_gates}; this renders the search for a more precise fit function unbeneficial.
In addition, the per wafer calibration opted over a per neuron circuit calibration introduces a dominant error due to the deviations between neuron circuits, shown in \cref{fig:mean_error_all_hicanns}.
A precise weight calibration within a reasonable runtime would be achievable via a parallel measurement of each neuron circuit.
This would also allow to exclude neurons showing unintended behavior.
However, this is not possible with the currently used analog readout system.
Nonetheless, the lack of a perfect weight calibration can be circumvented via in-the-loop training on the \acrlong{bss1} system, as shown for inference tasks in previous results~\cite{schmitt2017hwitl}.

\subsubsection{Calibration Based Exclusion of Components}
The operation of the \acrshortpl{hicann} during the calibration is similar to the operation during experiments.
All components have to work correctly for the calibration to succeed.
Failing calibrations indicate unintended behavior.
This allows for testing the whole die, especially the analog circuits that cannot be tested directly.
Additionally, thresholds can be defined to exclude outliers.
Consequently, neurons that do not pass all calibration steps are excluded from the availability database.
Numbers of calibration based excluded neurons on a typical wafer are given in \cref{tab:calib_blacklists}.

\begin{table}
  \centering
  \caption{Overview of calibration based excluded neurons of a fully assembled wafer module.
	In the column labeled "Neurons" the first entry shows the number of neurons taken into account for the calibration, the second entry the number of neurons taken into account for the effective exclusion.}
  \label{tab:calib_blacklists}
  \begin{tabular}{lll}
    \toprule
    Neurons & Excluded by Calibration & Effective Exclusion\\
    \midrule
	  182272/190976 & \SI{10.25}{\percent} & \SI{14.59}{\percent} \\
    \bottomrule
  \end{tabular}
\end{table}

\section{Experiment Showcase - Synchronous Firing Chain}

Previous experiments on the \acrlong{bss1} system relied on a small subset of the available neurons~\cite{schmitt2017hwitl,kungl2019accelerated,goeltz2021fast}.
In this section, we use a \cgls{sfc} to utilize a large number of the available wafer module resources.
We start with a relatively short chain to illustrate the behavior of the network and finally present a longer one that utilizes a large part of a single wafer module. %

\cGlspl{sfc} can filter for synchronous activity and propagate the activity along a chain of neuron groups~\cite{aertsen96, gewaltig01_synfire}.
We choose \cglspl{sfc} since they can easily be scaled up to arbitrary sizes by increasing the chain length as well as the number of neurons in a single group and have been studied extensively in previous publications~\cite{diesmann99, diesmann02, kumar08conditions}.
Furthermore, \cglspl{sfc} were used to showcase the functionality of the predecessor of \acrlong{bss1}~\cite{pfeil2013six} and to characterize the behavior of the current system in software simulations~\cite{petrovici2014characterization}.

\begin{figure}
  \centering
  \includegraphics[width=\linewidth]{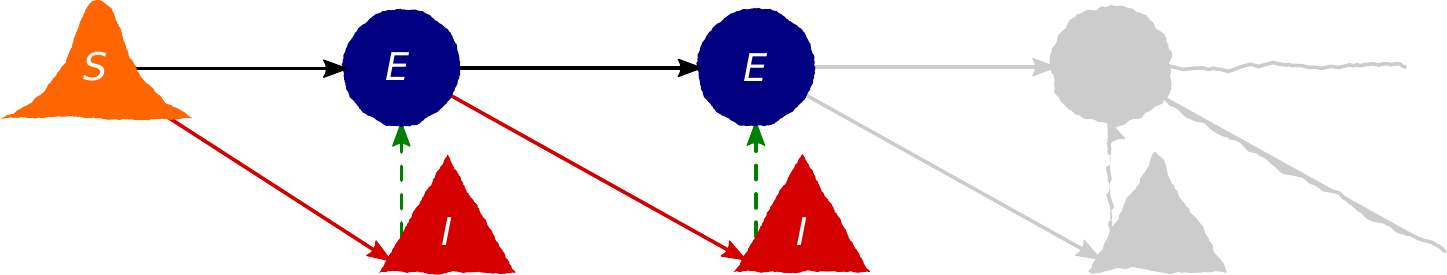}
  \caption{Structure of the \cglspl{sfc} presented in this section.
           The \cgls{sfc} is made up of several groups of excitatory (blue) and inhibitory (red) populations.
           The inhibitory population connects to the excitatory population within the same group and aims to improve the chain's filtering for synchronous input~\cite{kumar08conditions, kremkow2010functional}.
           Each excitatory population is connected to the excitatory and inhibitory population of the next group.
           By repeating this construction schema (grey), chains of arbitrary length can be realized.
           The network is excited by a stimulus population (orange) which projects to the excitatory and inhibitory population of the first group.}
  \label{fig:synfire}
\end{figure}

\Cref{fig:synfire} displays a \cgls{sfc} with feed-forward inhibition.
Each chain link consists of an excitatory and inhibitory population.
The inhibitory populations are connected to the excitatory population within the same group.
This feed-forward inhibition can enhance the filtering properties of the chain~\cite{kremkow2010functional, kumar08conditions}.
The excitatory population forwards its outputs to both populations within the next group.
External stimulus is injected in the form of Gaussian pulse packages~\cite{diesmann99}. %
The strength $a$ denotes the number of input spikes per stimulus neuron and $\sigma$ the standard deviation of the Gaussian from which the spike times are drawn.
We will use $(a, \sigma)$ to refer to specific packages.

\subsection{Network Behavior}

\begin{figure}
  \centering
  \subfloat[\label{fig:sfc_short_propagation}]{\includegraphics[width=\linewidth]{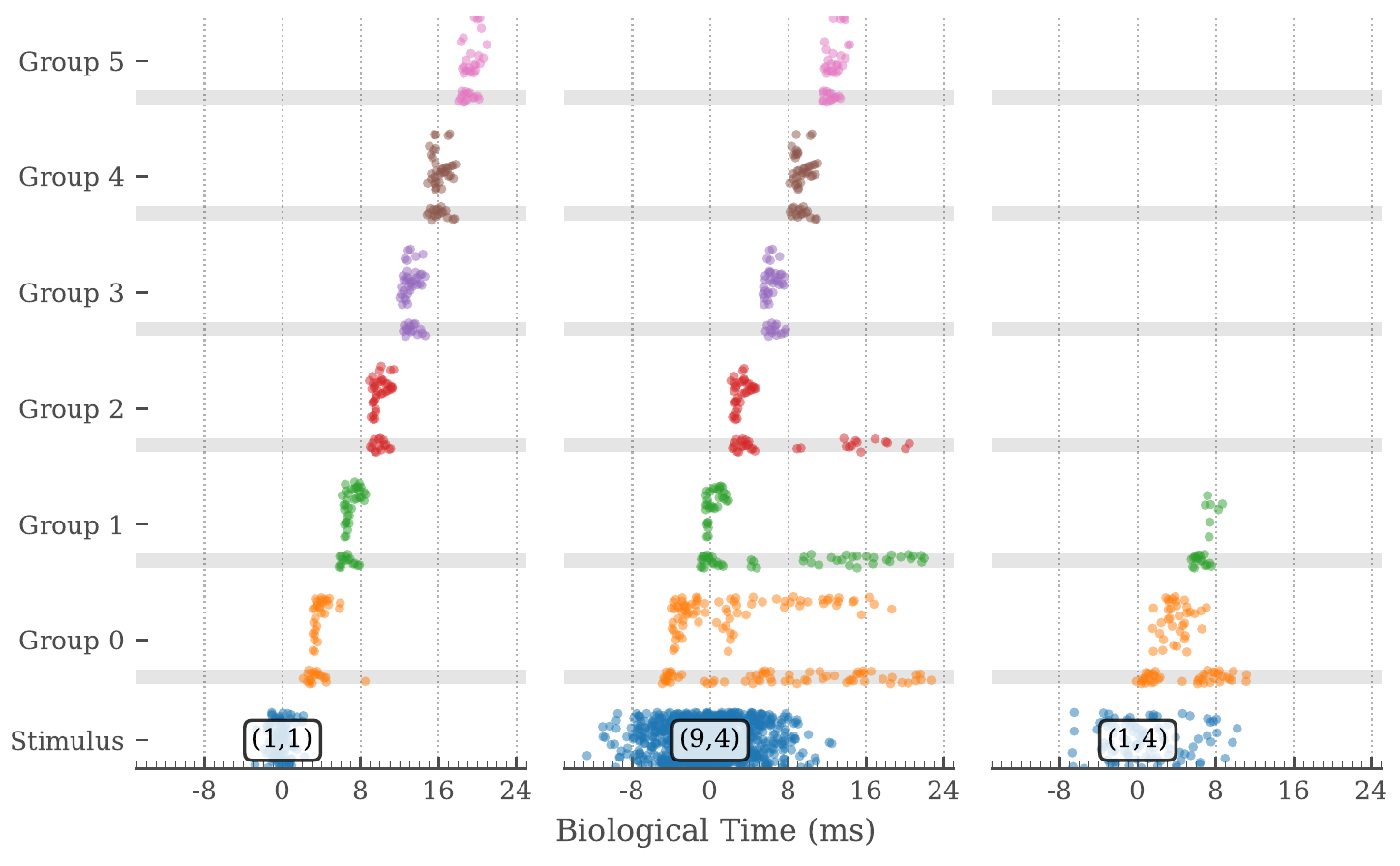}}\\
  \subfloat[\label{fig:sfc_short_phase}]{\includegraphics[width=\linewidth]{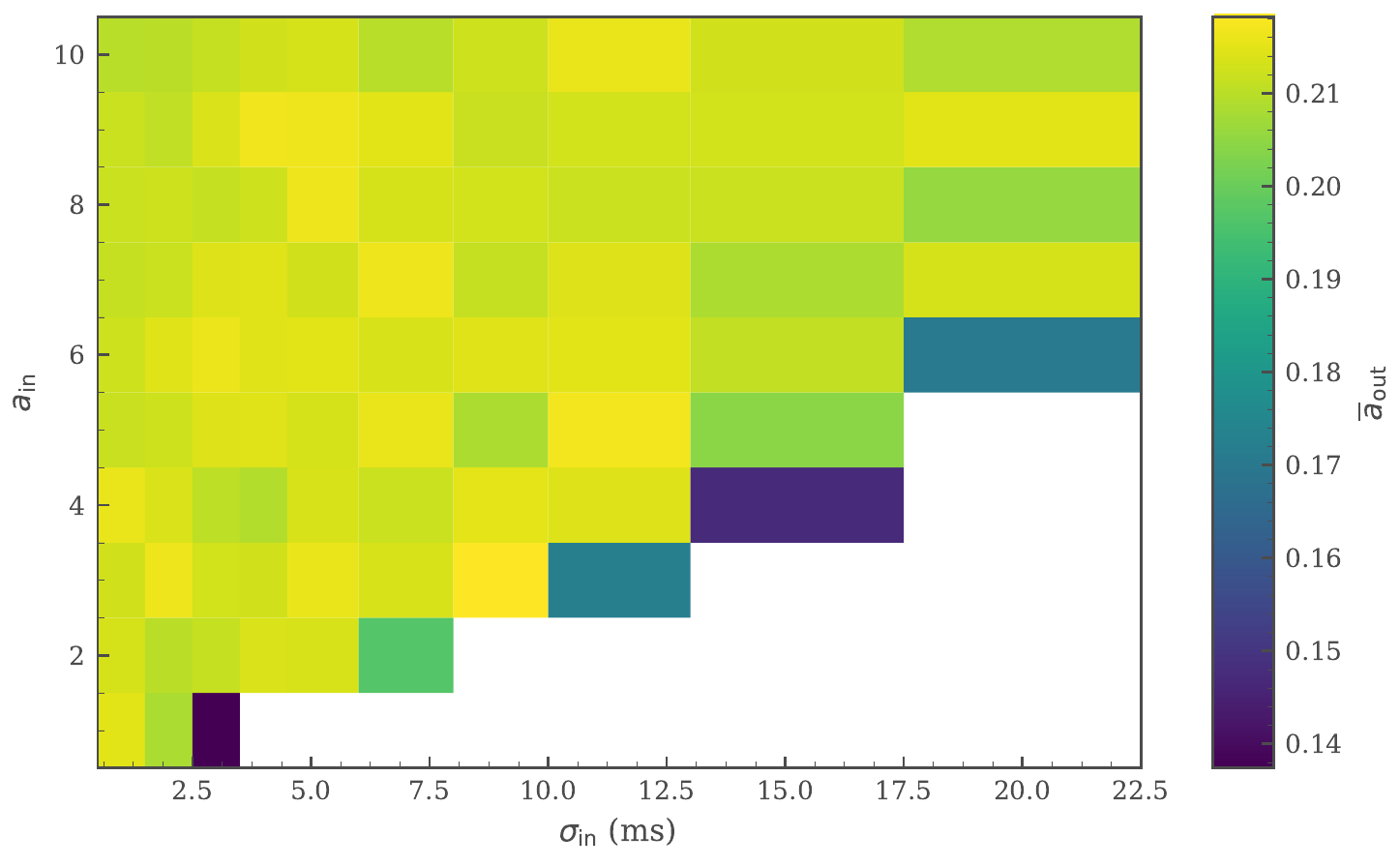}}
  \caption{Hardware emulation of a chain with six chain links.
  		   \protect\subref{fig:sfc_short_propagation} Propagation of pulse packages along the chain.
		   Successful propagation depends both on the strength $a$ as well as the synchronicity $\sigma$ of the initial stimulus, represented by ($a$, $\sigma$).
		   Broad input stimuli synchronize along the chain or do not reach the end of the chain.
		   \protect\subref{fig:sfc_short_phase} Average number of spikes per neuron in the final group $\bar{a}_\text{out}$ of the chain in dependency on the initial strength $a$ and synchronicity $\sigma$.
		   Each input package was presented \num{40} times and the results are averaged over all presentations.
		   The pulse packages propagate if the initial input is strong and synchronous enough.
		   In the region of stable propagation the output strength is almost constant, near the separation of the two regimes the average strength of the final pulse package decreases.
		   This separation line between successful propagation and failure of transmission can be controlled by several parameters such as the synaptic weights.
  }
  \label{fig:sfc_short}
\end{figure}

In a first step we will look at a relatively short chain with six chain links, shown in \cref{fig:sfc_short}, to illustrate how the filtering properties of the chain can be tuned.
\Cref{tab:synfire} summarizes some of the key properties of the network.
We used the manual placement described in~\cite{mueller2022operating} to place the different populations on the wafer.
Specifically, we distribute the external stimulus over several \cglspl{hicann} in order to minimize spike loss due to limited bandwidth.

\begin{table}
  \centering
  \caption{Parameters used for the \cglspl{sfc} presented in this section.}
  \label{tab:synfire}
  \begin{tabular}{lll}
    \toprule
    Parameter                        & Short Chain           & Long Chain            \\
    \midrule
	Chain Length                    & 6                     & 190                   \\
	Stimulus Neurons                & 100                   & 80                    \\
	Excitatory Neurons per Group    & 100                   & 80                    \\
	Inhihibitory Neurons per Group  & 25                    & 20                    \\
	Total Number of Neurons         & 750                   & 19000                 \\
	Total Number of Neuron Circuits & 3000                  & 76000                 \\
	Total Number of Synapses        & $\approx \num{73000}$ & $\approx \num{1.4e6}$ \\
	Used HICANNs                    & 48                    & 230                   \\
    \bottomrule
  \end{tabular}
\end{table}

As mentioned previously, \cglspl{sfc} are able to filter for synchronous input and to synchronize less-synchronous input as it travels along the chain~\cite{diesmann99, gewaltig01_synfire}.
\Cref{fig:sfc_short_propagation} shows the propagation of three different input stimuli along the chain.
In case of a relatively weak and synchronous input $(1,1)$ a single, narrow package travels along the chain.
If the input is stronger and more asynchronous, we observe a broader response in the first groups of the chain which is synchronized as the signal propagates along the chain such that the responses in the final group are comparable.
Too weak and asynchronous input, here $(1,4)$ as an example, dies out and does not cause a response in the final group.
This is in agreement with previous results~\cite{diesmann99, kremkow2010functional, pfeil2013six, petrovici2014characterization}.

\Cref{fig:sfc_short_phase} shows in more detail for which input stimuli the propagation along the chain is successful.
In agreement with the previous observations, weak and asynchronous input is not transmitted to the final group.
The response in the final group is almost uniform.
This indicates that the packages are synchronized as they travel along the chain.
Setting appropriate parameters which reproduce the expected results from simulations relies on the calibration routines, introduced in \cref{sec:calibration}.
The calibration allows to set model parameters in the biological domain and reduces the inherent mismatch between the physical components.

\subsection{Wafer-Scale Network}
The previous section demonstrates the implementation and control of a short \cgls{sfc} on the \acrlong{bss1} system.
This section shows that the commissioning efforts described in \cref{sec:commissioning} also facilitate the implementation of wafer-scale networks.
The properties of this \cgls{sfc} are summarized in \cref{tab:synfire}.

\begin{figure}
  \centering
  \subfloat[\label{fig:sfc_long_mapping}]{\includegraphics[width=0.7\linewidth]{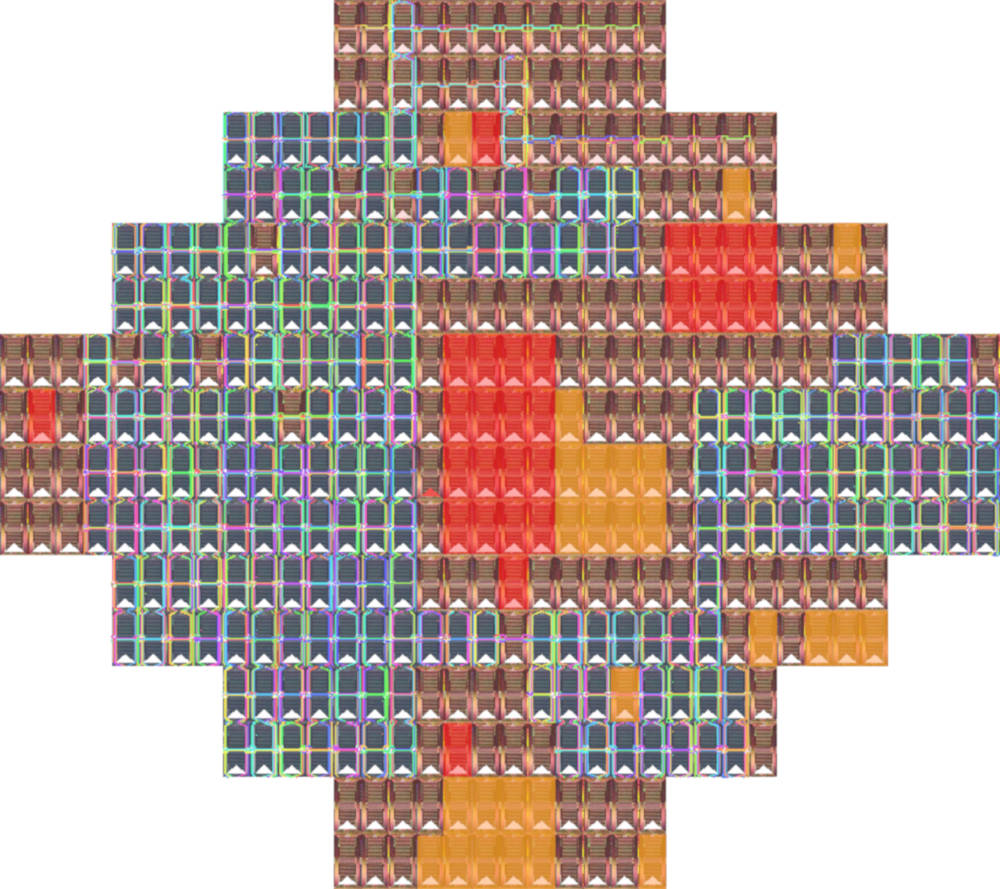}}\\
  \subfloat[\label{fig:sfc_long_propagation}]{\includegraphics[width=\linewidth]{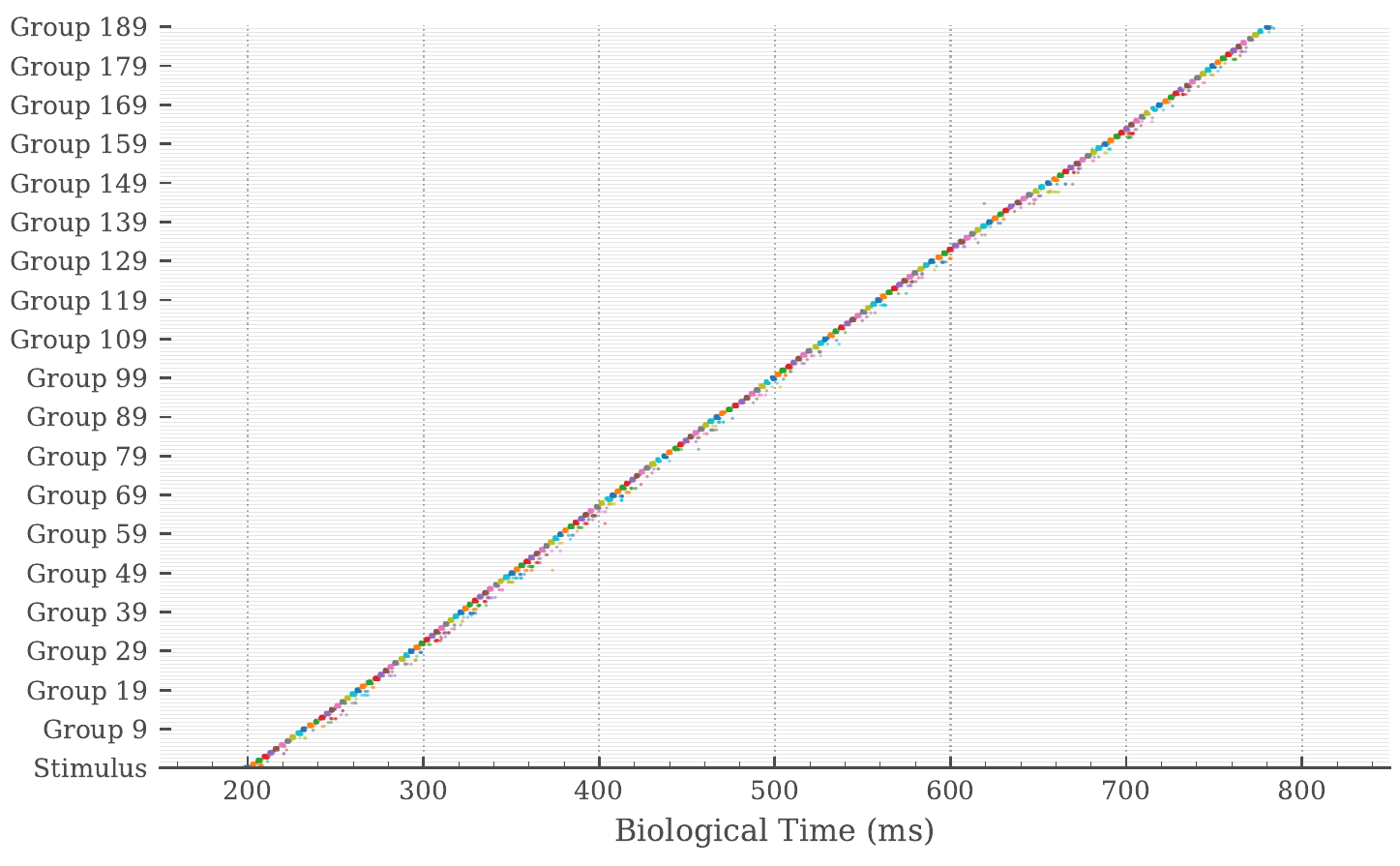}}
  
	\caption{Hardware emulation of a chain with \num{19000} neurons.
			 Further parameters of the network can be found in \cref{tab:synfire}.
			 \protect\subref{fig:sfc_long_mapping} Mapping of the network to a \acrlong{bss1} wafer.
			 \cGlspl{hicann} excluded from the availability database are marked in red, cf.\ \cref{sec:blacklisting}.
			 \cGlspl{hicann} which cannot host an entire group are marked in orange and are not used in the experiment.
			 On each \cgls{hicann} colored in blue an entire group of neurons is placed.
			 Colored lines indicate synaptic connections.
			 \protect\subref{fig:sfc_long_propagation} Response of the chain to an input packet of strength $a=1$ and spread $\sigma=1$.
	   }
  \label{fig:sfc_long}
\end{figure}

The complexity of the emulation increases with the size of the model.
While for a relatively short chain it is possible to investigate the behavior of individual neurons and manually detect malfunctioning and bad calibrated entities, this is not feasible for larger experiments.
Therefore, digital tests described in \cref{sec:blacklisting} are essential to automatically avoid these components during the experiment.

To simplify the automatic routing of the abstract network description to physical entities on the wafer, we once again employ manual mapping, see \cref{fig:sfc_long_mapping}.
We place the different groups in a zig-zag pattern starting from the top-left side towards the bottom of the wafer and then back up towards the top-right side.
This placement schema allows the \acrlong{bss1} operating system~\cite{mueller2022operating} to find appropriate connections between the different populations and minimizes synapse loss, i.e.\ synaptic connections that could not be mapped to the hardware.

We were able to successfully emulate a \cgls{sfc} with \num{190} chain links on the \acrlong{bss1} system.
\Cref{fig:sfc_long_propagation} shows an example of a pulse package that travels along the full length of the chain.
The activity of the individual groups still depends on the exact neuron and synapse properties, but the calibration ensures that the pulse package remains compact.
A synchronous pulse reaches the final group after a signal propagation time of about \SI{600}{\milli\second} in the biological regime, which corresponds to \SI{60}{\micro\second} wall-clock time.

\section{Discussion}

Starting its development more than ten years ago, the first-generation BrainScaleS wafer-scale neuromorphic system represents a milestone toward a large-scale analog neural network emulation platform.
Over years during which several modules have been commissioned and experiments run, we have learned important lessons on building and handling such a complex system.
We discovered drawbacks in our first implementation; some of them could successfully be circumvented via our commissioning software.
Our second-generation neuromorphic \acrlong{bss2} chip~\cite{schemmel2020accelerated} addresses \acrlong{bss1}'s design weaknesses.
Moreover, it enables the application of advanced learning mechanisms by introducing a digital plasticity processor, neuron multi-compartment capabilities, as well as extended analog to digital conversion capacities.

In this paper, we described the individual components of a \acrlong{bss1} wafer module and showed the necessary steps to assemble it.
A wafer-scale analog system is complex and requires many hardware components working concurrently.
Once a wafer module is assembled, it is often not possible to pinpoint defects in individual components.
To alleviate this, each component must get tested on its own; malfunctioning ones must be repaired or replaced before they are added to the system.
Additional tests during the assembly are also crucial to allow for finding and solving errors that arise during that process.
The remaining problems are handled by the exclusion of affected components or circuits from the availability database to ensure the correct operation of the system.

The importance of the tests and monitoring remains after the wafer module gets placed in the rack.
For example, tight monitoring during system operation is necessary to uncover the wear out of system components.
Automated alerts are fundamental for warning in case of values deviating over time.
Furthermore, the tests executed nightly help keep track of the wafer modules' state.

Concerning the wafer in the core of the \acrlong{bss1} system, the probability of fabrication defects in microelectronics is proportional to the circuit area~\cite{werner2016circumvent}.
Thus, it is unfeasible to build such a large analog system without malfunctioning components.
This will most likely further intensify in the future by utilizing novel materials.
With this in mind, the digital tests introduced are executed nightly to identify such malfunctioning components and exclude them from our availability database.
These tests enable storing different states of the database on disk and allow to differentiate actual malfunctioning components from those not usable due to a dependency.
The users can then utilize reliable components, possibly even using a custom availability database.

An additional challenge using analog hardware is the fixed-pattern noise introduced by unavoidable manufacturing process variations.
In the \acrlong{bss1} system, this is worsened by the design decision to use \cglspl{fg} to store the neuron configuration.
These cells allow for long-term storage of analog parameters without storing digital values onboard.
However, the current implementation introduces write-cycle to write-cycle variability. 
Though small, these variations lead to noticeable errors if they are further enlarged by non-linear dependencies between control signal and observable.
To minimize these effects, we presented our calibration framework, which also allows non-expert users to configure experiments in the biological domain without specific knowledge of the hardware.
We demonstrated the narrowing and centering of the achieved value distribution for exemplary parameters after the calibration was applied, limited by thermal noise and the variations caused by the \cglspl{fg}, nonetheless.
Since single-poly floating-gate cells are non-standard devices and not supported by the manufacture, the second-generation \acrlong{bss2} chip reverts to a digital parameter storage scheme employed in a previous neuromorphic architecture~\cite{schemmel_ijcnn04}, thereby vastly improving analog parameter accuracy.
Since the second generation uses a manufacturing process with much smaller geometry, namely \SI{65}{\nano\meter} vs. \SI{180}{\nano\meter}, the area penalty for the digital parameter storage is manageable.
A further advantage of the novel parameter storage is the reduced programming time~\cite{hock13analogmemory}.
In the presented wafer-scale implementation, the single-poly floating-gate parameter storage was the only feasible solution to achieve the required number of analog parameters for the neuron circuits.

On top of explaining the calibration methodology, we demonstrated the necessity for parallel execution of the calibrations.
The large parameter space of the synapse weight calibration exceeds reasonable runtimes using the current readout system.
In order to circumvent this, we introduced a per wafer calibration which, compared to a per circuit calibration, shows larger errors but can be generated in a reasonable time frame.
To improve this, we developed a new readout system, which will replace the external set of ADCs with on-wafer-module boards, increasing the parallel readout capabilities from \num{12} to \num{96} channels~\cite{ilmberger2017masterthesis}.
Moreover, in the \acrlong{bss2} chip, we introduce a per neuron-circuit ADC system, which allows for a massive parallel calibration~\cite{schemmel2020accelerated}.
A per-circuit calibration before each experiment becomes feasible with such a solution.

Finally, we demonstrated the operation of a fully commissioned \acrlong{bss1} wafer module implementing \cglspl{sfc}.
While small chains portray the capability to fine-tune the network parameters, extending to a long chain of \num{190} links illustrates the possibility to scale up networks.
Successfully mapped to an inherently imperfect substrate, it consists of the largest spiking network emulation run with analog components and individual synapses to date.

Our endeavor in developing and maintaining the \acrlong{bss1} system has demonstrated, while illustrating the field's challenges, that building wafer-scale analog neuromorphic hardware is feasible.
Furthermore, the \acrlong{bss1} wafer module with its operating system laid the foundation for the next-generation systems; all lessons learned from the first generation contribute to the success of future large-scale neuromorphic systems.

\section*{Acknowledgments}

The authors wish to thank all present and former members of the Electronic Vision(s) research group contributing to the BrainScaleS-1 platform, development and operation methodologies, as well as software development.
We thank S.~Schiefer and S.~Hartmann from the group \foreignlanguage{ngerman}{Hochparallele VLSI-Systeme und Neuromikroelektronik} at TU-Dresden for the development and before-assembly routine testing of the communication boards and the \cglspl{wio}, as well as for providing test details and images for this writing.
We thank M.~Yaziki and O.~Ceylan from the group of Yasar Gürbüz at Sabanci University, Istanbul for the development of test boards that are being used during assembly of the wafer modules.
We thank Würth Elektronik, Germany for dedicated and detailed support during the development of the Main PCB.
This work has received funding from the EU ([FP7/2007-2013], [H2020/2014-2020]) under grant agreements 604102 (HBP), 269921 (BrainScaleS), 243914 (Brain-i-Nets), 720270 (HBP SGA1), 785907 (HBP SGA2) and 945539 (HBP SGA3), the \foreignlanguage{ngerman}{Deutsche Forschungsgemeinschaft} (DFG, German Research Foundation) under Germany’s Excellence Strategy EXC 2181/1-390900948 (the Heidelberg STRUCTURES Excellence Cluster), the Helmholtz Association Initiative and Networking Fund (ACA, Advanced Computing Architectures) under Project SO-092, as well as from the Manfred Stärk Foundation.

\bibliographystyle{style/IEEEtran}
\bibliography{bib/vision}

\end{document}


\title{Supplemental Material\\From Clean Room to Machine Room: Commissioning of the First-Generation BrainScaleS Wafer-Scale Neuromorphic System}

\DeclareRobustCommand{\enumauthorrefmark}[1]{\smash{\textsuperscript{\footnotesize #1}}}

\newcommand{\contributedSymbol}{\IEEEauthorrefmark{1}}
\newcommand{\uheiSymbol}{\enumauthorrefmark{1}}
\newcommand{\ugoeSymbol}{\enumauthorrefmark{2}}

\author{
	\IEEEauthorblockN{%
		Hartmut Schmidt\contributedSymbol,
		José Montes\contributedSymbol,
		Andreas Grübl,
		Maurice Güttler,
		Dan Husmann,
		Joscha Ilmberger,\\
		Jakob Kaiser,
		Christian Mauch,
		Eric Müller,
		Lars Sterzenbach,
		Johannes Schemmel,
		Sebastian Schmitt\\
	}

	\thanks{
		\IEEEauthorblockA{%
		\contributedSymbol%
		Contributed equally\\
		}
	}
}

\maketitle
\section{Calibration Details for the System's Neuron and Synapse Circuits}\label{sec:appendix}

A calibration procedure is in place for the BrainScaleS-1 system, which compensates for manufacture-induced analog circuit variability.
It accounts for analog readout noise by averaging the features extracted from the membrane traces over time.
In addition, measurements are repeated and then averaged after rewriting the \cglspl{fg}, where stated, to consider \cgls{fg} write-cycle to write-cycle parameter storage variability.

A detailed explanation of each parameter calibration conducted on the wafer module is provided in the following.
We first describe the parameter, explain the calibration approach and the settings used, and show plots illustrating the results.
In addition to the synaptic-weight calibration presented in the main text, these constitute the complete neuron- and synapse-circuit calibrations performed in the system.
Details of the sample points measured, the models utilized, and the average runtimes per parameter are summarized in~\cref{tab:calibration_data}.\\

\begin{figure}
        \centering
        \subfloat[
                  \label{fig:analog_readout_offset}]{\includegraphics[width=0.5\linewidth]{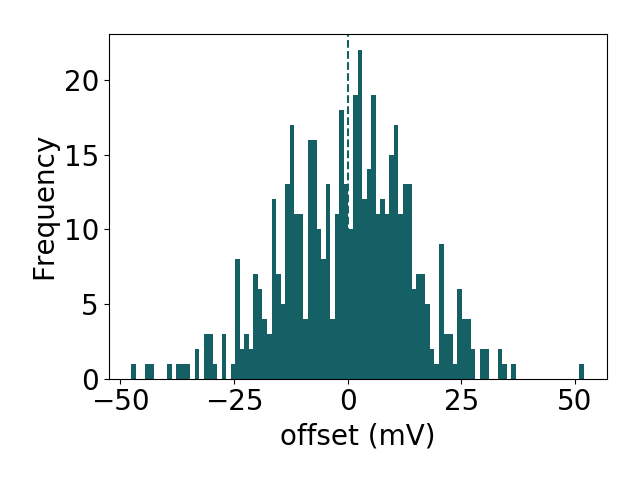}}
        \subfloat[
                  \label{fig:calibration_V_reset}]{\includegraphics[width=0.5\linewidth]{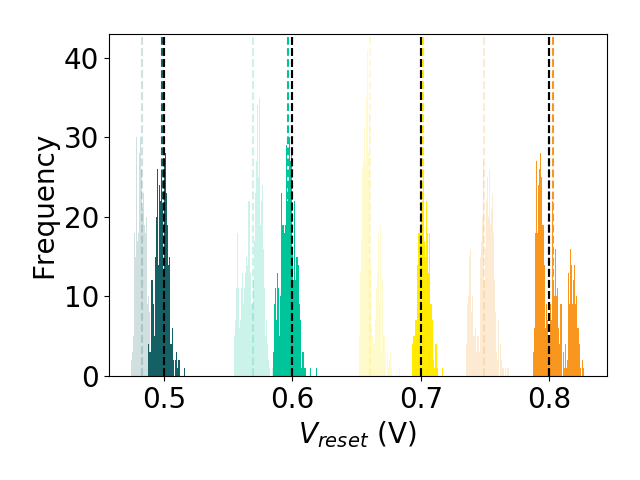}}\\
        \subfloat[
                  \label{fig:calibration_V_threshold}]{\includegraphics[width=0.5\linewidth]{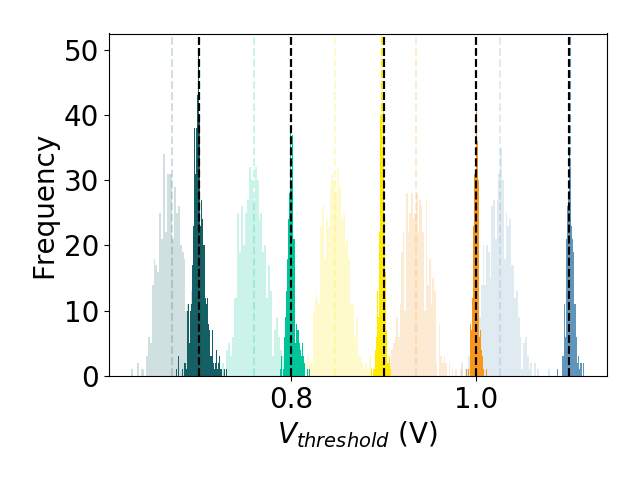}}
        \subfloat[
                  \label{fig:esyni_calib}]{\includegraphics[width=0.5\linewidth]{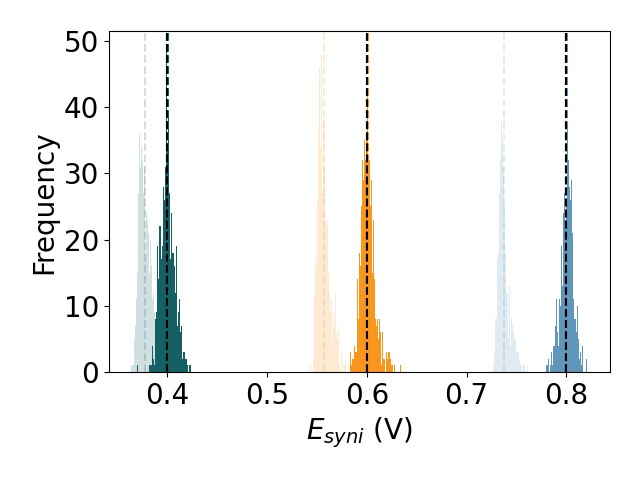}}
	  \caption{\label{fig:calibration_effects_v_fgs} \protect\subref{fig:analog_readout_offset} Analog readout offset distribution for the \num{512} neurons of one \cgls{hicann}.
                  Calibration results for the parameters \protect\subref{fig:calibration_V_reset} $V_\text{reset}$, \protect\subref{fig:calibration_V_threshold} $V_\text{threshold}$ and \protect\subref{fig:esyni_calib} $E_\text{syni}$.
                  Pale and intense colors correspond to the hardware achieved voltages for different target values (shown in black-dashed lines) before and after the calibration is applied, respectively.
		  For $V_\text{reset}$ the correction effect is limited by the parameter being shared by \num{128} neurons.}
      \end{figure}

\textit{Readout shift}: On each \cgls{hicann}, every neuron's membrane trace can be recorded by connecting its switchable analog output amplifier to one of two output buffers.
	     Due to circuit variability, each amplifier adds a constant offset to the recorded traces, the so called readout shift.
	     It has to be determined first, since all further calibrations are influenced by it.
      \begin{itemize}
        \item \textit{How}: Neuron membranes are interconnected in groups of \num{64} (the maximum possible).
        Their individual resting membranes are recorded and every neuron's deviation from the group's mean is stored.
\item \textit{Settings}: $E_\text{leak} = \SI{0.9}{\volt}$, the middle of the range, $V_\text{threshold}$ above resting potential, $I_\text{conv}$ set to \SI{0}{\ampere} to switch off both \cglspl{ota} for the excitatory and inhibitory synaptic input conductances.
        \item \textit{Effects}: The offset is automatically corrected for all subsequent calibrations by loading the calibration backend.
        The distribution of the analog output amplifier offsets of all neurons on one \cgls{hicann} is shown in~\cref{fig:analog_readout_offset}.
      \end{itemize}

  \begin{table}
    \centering
    \caption{Parameter-wise details for the calibration of neurons in one \cgls{hicann}, including the number of steps and repetitions run.
    Extraction refers to the model used to relate and determine variable values from the observables.
    HW-\cgls{fg} refers to the model used to fit parameters between the hardware domain and the programmable \cgls{fg} values.}
    \label{tab:calibration_data}
    \begin{tabular}{l | c | c | S[input-uncertainty-signs=pm,separate-uncertainty=true,table-format = 4.1(1)]}
      \toprule
      Parameter             & \begin{tabular}[x]{@{}c@{}}Target\\values\end{tabular} &  \begin{tabular}[x]{@{}c@{}}Extraction model /\\HW-\cgls{fg} model\end{tabular} & {\begin{tabular}[x]{@{}c@{}}Run time\\(s)\end{tabular}} \\
      \midrule
      Readout shift         & 1                                                      & -                    & 20.5 pm 0.7\\
      $V_\text{reset}$      & 4                                                      & linear / linear      & 87 pm 1\\
      $V_\text{threshold}$  & 5                                                      & linear / linear   & 111 pm 2\\
      $E_\text{syni}$       & 3                                                      & linear / linear      & 59.8 pm 0.8\\
      $I_\text{pulse}$      & 4 rep x 7                                              & \begin{tabular}[x]{@{}c@{}} \acrshort{isi}($I_\text{pulse}$)$-$\acrshort{isi}$_0$ /\\ eq.~(Main-2) \end{tabular} & 920 pm 10\\
      $E_\text{leak}$       & 3                                                      & linear / linear      & 60 pm 1\\
      $V_\text{convoffx,i}$ & 25                                                     & linear / linear      & 313 pm 4\\
      $I_\text{gl}$         & 8                                                      & \begin{tabular}[x]{@{}c@{}}SinglePSP~\cref{eq:psp_model} /\\Softplus~\cref{eq:softplus_function_mem}\end{tabular} & 780 pm 50 \\
      $V_\text{syntcx,i}$   & 10                                                     & \begin{tabular}[x]{@{}c@{}}SinglePSP /\\Softplus~\cref{eq:softplus_function_syn}\end{tabular}   & 810 pm 20\\
      $E_\text{synx}$       & 4 x 4                                                  & linear / linear        & 1300 pm 100\\
      Weight                & 16 x 4 x 4                                             & PSP / eq.~(Main-3) & {$\approx$ 220 x 20000}\\
      \bottomrule
    \end{tabular}
  \end{table}

    $V_\text{reset}$: The potential to which a neuron's membrane is set after a spike is generated.
      It is shared among a group of \num{128} neurons. Each \cgls{hicann} contains four of these groups.
      \begin{itemize}
        \item \textit{How}: Neurons are set to spike continuously by setting their leak potential $E_\text{leak}$ above their threshold potential $V_\text{threshold}$.
         A recording time of \SI{80}{\micro\second} per target value collects an average of \num{39} \cglspl{isi} on each membrane.
        The refractory time $\tau_\text{ref}$ is set to maximum in order to allow for long baseline traces between the spikes.
        The reset voltage is calculated as the average over all the interspike baseline samples to account for readout noise.
    \item \textit{Settings}: $I_\text{conv} = \SI{0}{\ampere}$ for both excitatory and inhibitory synaptic inputs, shutting off the \cgls{ota} of their synaptic conductance.
        $I_\text{gl} = \SI{1.1}{\micro\ampere}$, $I_\text{pulse} = \SI{20}{\nano\ampere}$ to set the refractory time to a high value.
        \item \textit{Sweep}: $V_\text{reset}$, with $E_\text{leak} = V_\text{reset} + \SI{0.4}{\volt}$,
        $V_\text{threshold} = V_\text{reset} + \SI{0.2}{\volt}$
        \item \textit{Effects}: The achieved hardware voltage distribution is shifted towards the correct target value from its original mean, as can be seen in~\cref{fig:calibration_V_reset}.
        The standard deviation does not improve for all targets since the shared nature of the parameter limits the action of the correction over the individual neurons. 
      \end{itemize}
  
    $V_\text{threshold}$: The threshold potential of the \cgls{lif} model, at which an action potential is elicited and the membrane's voltage is forced into the reset potential for the refractory period.
     \begin{itemize}
        \item \textit{How}: Synaptic inputs are minimized in order to isolate the membrane and the threshold detect circuits.
        The threshold potential $V_\text{threshold}$ is set below the leak potential $E_\text{leak}$ to elicit constant spiking.
        The maximum membrane voltage at several spike peaks is averaged and considered the true threshold voltage.
        \item \textit{Settings}: $I_\text{conv} = \SI{0}{\ampere}$, $I_\text{gl} = \SI{1.5}{\micro\ampere}$
        \item \textit{Sweep}: $V_\text{threshold}$, $V_\text{reset} = V_\text{threshold} - \SI{200}{\milli\volt}$, $E_\text{leak} = V_\text{threshold} + \SI{200}{\milli\volt}$
        \item \textit{Effects}: The corrected hardware voltage distribution is centered around the correct target value.
        The standard deviation decreases, as can be seen in \cref{fig:calibration_V_threshold}.
      \end{itemize}
  
    $E_\textit{syni}$: The inhibitory reversal potential towards which the \cgls{ota} in the inhibitory synaptic input drives the membrane when processing synaptic input.
      \begin{itemize}
        \item \textit{How}:
        $V_\text{convoffi}$ of the inhibitory synaptic input is set to a small value so the bias generator forces the membrane potential to the inhibitory reversal potential.
        No spikes are elicted since the threshold voltage is never reached.
        Once the neuron is at rest the averaged membrane voltage characterizes the reversal potential.
        \item \textit{Settings}: $E_\text{leak} = \SI{0.8}{\volt}$, $I_\text{convx} = \SI{0}{\ampere}$, $I_\text{gl} = \SI{0}{\ampere}$, $V_\text{convoffi} = \SI{0.1}{\volt}$, $V_\text{syntcx,y} = \SI{1.8}{\volt}$, $V_\text{threshold} = \SI{1.2}{\volt}$
        \item \textit{Sweep}: $E_\text{syni}$
        \item \textit{Effects}: The achieved inhibitory reversal potential voltages before and after calibration are shown in~\cref{fig:esyni_calib}.
      \end{itemize}
  
    $I_\text{pulse}$: Bias current that controls how fast the neuron's timing mechanism recovers from the reset state after a spike is generated.
      \begin{itemize}
        \item \textit{How}: Neurons are set to spike continuously by setting $E_\text{leak}$ above $V_\text{threshold}$.
        For the refractory time constant measurements, the baseline traces corresponding to the reset-state of the membranes are extracted.
        $I_\text{pulse}$ is first set to its maximum and the effective refractory period is measured and recorded; this constitutes the minimum achievable period denoted thus $\tau_{0}$.
		      The subsequent measured refractory periods are referenced to $\tau_{0}$ by substracting $\tau_{0}$ from them, and fitting eq.~(Main-2) from the main text.
        \item \textit{Settings}: $E_\text{leak} = \SI{1.2}{\volt}$, $V_\text{threshold} = \SI{0.8}{\volt}$, $E_\text{synx} = \SI{1.2}{\volt}$, $E_\text{syni} = \SI{0.8}{\volt}$,
        $V_\text{reset} = \SI{0.5}{\volt}$
        \item \textit{Sweep}: $I_\text{pulse}$
	\item \textit{Effects}: The achieved refractory time constants' mean is closer to the target value after the calibration is obtained and applied, as can be observed in fig.~11a in the main text.
        The standard deviations reduce.
        In fig.~10 in the main text the limited precision to configure the refractory time constant is demonstrated, as only a fraction of the possible parameter range of $I_\text{pulse}$ results in reasonable configurations.
      \end{itemize}
  
    $E_\text{leak}$: The reference voltage towards which the membrane potential is constantly driven through the leak conductance.
      \begin{itemize}
        \item \textit{How}: Synaptic inputs are minimized and the membranes are read on a resting state.
        \item \textit{Settings}: $I_\text{conv} = \SI{0}{\ampere}$, $V_\text{t} = \SI{1.2}{\volt}$, $V_\text{reset} = \SI{0.9}{\volt}$
        \item \textit{Sweep}: $E_\text{leak}$
        \item \textit{Effects}: The corrected hardware voltage distribution is centered around the correct target value.
        The standard deviation decreases, as can be seen in fig.~11b in the main text.
      \end{itemize}
  
\begin{figure}
        \centering
        \subfloat[
                  \label{fig:V_convoff_effects_uncalibrated}]{\includegraphics[width=0.51\linewidth]{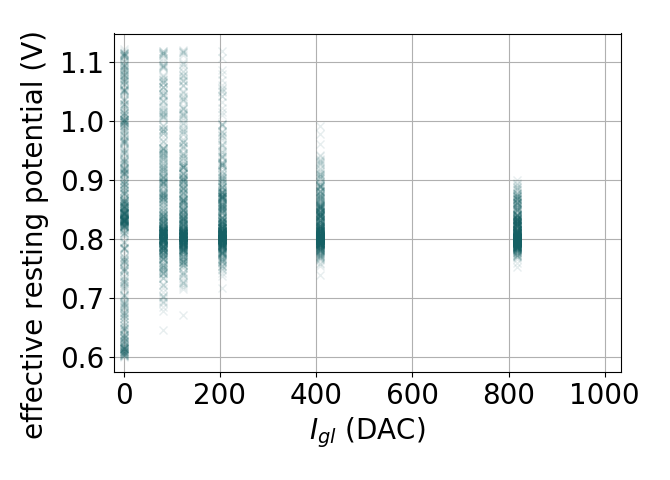}}
        \subfloat[
                  \label{fig:V_convoff_effects_calibrated}]{\includegraphics[width=0.49\linewidth]{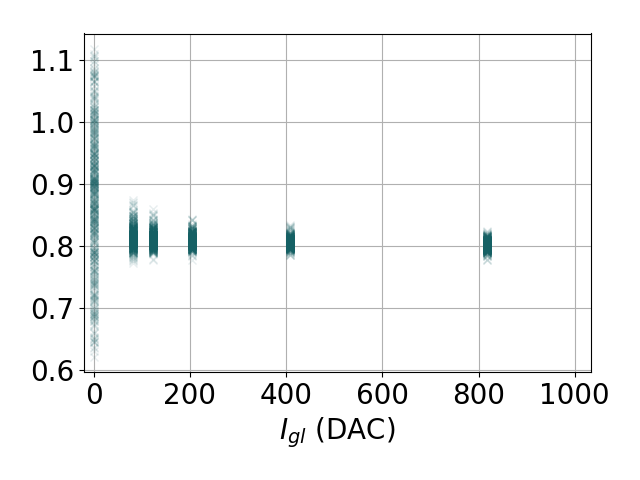}}
        \caption{\label{fig:V_convoff_effects} Effective resting potential of the neurons on one \cgls{hicann} \protect\subref{fig:V_convoff_effects_uncalibrated} before and \protect\subref{fig:V_convoff_effects_calibrated} after calibration of parameter $V_\text{convoffx}$.}
      \end{figure}

      $V_\text{convoffx}$: Offset voltage for the integrator on the excitatory synaptic input.
      The voltage parameter is used by a bias generator that controls the reference of \cgls{ota}$_{1}$, compensating for mismatches.
      The offset should balance two effects: minimize an undesired permanent current flowing to the membrane, which shifts the neuron's resting potential, against the weakening of the synaptic input caused by a too substantial compensation.
      Consequently, the goal of the calibration is to find the sweet spot in between, where the bias generator compensates precisely for the mismatch of \cgls{ota}$_{1}$.
      \begin{itemize}
        \item \textit{How}:
        The point of interest is the transition from a zero to a non-zero conductance on \cgls{ota}$_{1}$.
        It is measured by the shift of the resting potential arising for different values of $V_\text{convoffx}$.
        The calibrated value of $V_\text{convoffx}$ corresponds to the first value where the resting potential is no longer shifted.
        In addition, the linear range of the relation between the membrane rest-voltage shift and $V_\text{convoffx}$ is characterized.
        Effects from the inhibitory synaptic input are minimized by using low values for $E_\text{syni}$, $I_\text{convi}$ and a high $V_\text{convoffi}$.
        Furthermore, the effect is more pronounced for lower values of $E_\text{leak}$.
        \item \textit{Settings}: $E_\text{leak} = \SI{0.8}{\volt}$, $E_\text{syni} = \SI{0.4}{\volt}$, $E_\text{synx} = \SI{1.2}{\volt}$, $I_\text{convi} = \SI{0}{\ampere}$,
        $I_\text{gl} = \SI{0.2}{\micro\ampere}$ a low value that limits the leakage current from the synapse onto the membrane,
        $V_\text{convoffi} = \SI{1.8}{\volt}$, $V_\text{threshold} = \SI{1.2}{\volt}$, $V_\text{reset} = \SI{0.3}{\volt}$
        \item \textit{Sweep}: $V_\text{convoffx}$
        \item \textit{Effects}: A calibrated $V_\text{convoffx}$ parameter limits the deviations in the effective resting potential arising from leaks through the excitatory synaptic input, as shown in~\cref{fig:V_convoff_effects}.
        Nevertheless, a minimal $I_\text{gl}$ is required to allow the neuron membranes to exhibit uniform effective resting potentials.
      \end{itemize}

      $V_\text{convoffi}$: Offset voltage for the integrator on the inhibitory synaptic input conductance.
      The calibration principle is the same as for $V_\text{convoffx}$, but it should be performed independently as both inputs introduce leak currents into the membrane.
      \begin{itemize}
        \item \textit{How}: A low $E_\text{synx}$, $I_\text{convx}$ and high $V_\text{convoffx}$ minimize effects from the excitatory synaptic input.
        \item \textit{Settings}: $E_\text{leak} = \SI{0.8}{\volt}$, $E_\text{syni} = \SI{0.4}{\volt}$, $E_\text{synx} = \SI{1.2}{\volt}$, $I_\text{convx} = \SI{0}{\ampere}$,
        $I_\text{gl} = \SI{0.2}{\micro\ampere}$ a low value that limits the leakage current from the synapse onto the membrane,
        $V_\text{convoffx} = \SI{1.8}{\volt}$, $V_\text{threshold} = \SI{1.2}{\volt}$, $V_\text{reset} = \SI{0.3}{\volt}$
        \item \textit{Sweep}: $V_\text{convoffi}$
      \end{itemize}

    The following parameter calibrations use input spikes to generate \cglspl{psp} on the membrane.
    From the shape of the voltage traces, it is possible to approximate parameters related to the time constants of synaptic inputs ($\tau_\text{syn}$) and the membrane ($\tau_\text{mem}$).
    For a single input spike arriving while the membrane of a LIF neuron is in a steady state, the \cgls{psp} shape can be either described by an $\alpha$-function, if both time constants are the same, or by a difference of exponentials if one of the time constants is smaller~\cite{petrovici2016form}.
    This behavior is described by $V(t)\approx$
    \begin{equation}
      \label{eq:psp_model}
      \begin{cases}
          \begin{split}
            &E_{\text{leak}}+
            \theta(t_{\scaleto{0\mathstrut}{4pt}})A\left(\text{exp}\left(\frac{t_{\scaleto{0\mathstrut}{4pt}}-t}{\tau_{\scaleto{1\mathstrut}{4pt}}}\right)-\text{exp}\left(\frac{t_{\scaleto{0\mathstrut}{4pt}}-t}{\tau_{\scaleto{2\mathstrut}{4pt}}}\right)\right)\\
            &\qquad\qquad\qquad\qquad\qquad\qquad\qquad\qquad\quad\text{if }\tau_{\scaleto{1\mathstrut}{4pt}}\neq\tau_{\scaleto{2\mathstrut}{4pt}}\\
            &\\
            &E_{\text{leak}}+\theta(t_{\scaleto{0\mathstrut}{4pt}})\text{exp}\left(1-\frac{t-t_{\scaleto{0\mathstrut}{4pt}}}{\tau_{\scaleto{1\mathstrut}{4pt}}}\right)\frac{t-t_{\scaleto{0\mathstrut}{4pt}}}{\tau_{\scaleto{1\mathstrut}{4pt}}}\\
            &\qquad\qquad\qquad\qquad\qquad\qquad\qquad\qquad\quad\text{if }\tau_{\scaleto{1\mathstrut}{4pt}}=\tau_{\scaleto{2\mathstrut}{4pt}},\\
        \end{split}
      \end{cases}
    \end{equation}
    with
    \begin{equation}
    	A=\frac{h}{\tau^{\frac{1}{1-\tau}}-\tau^{\frac{\tau}{1-\tau}}}
    \end{equation}
    and $\tau=\frac{\tau_{2}}{\tau_{1}}$ a ratio between $\tau_\mathrm{mem}$ and $\tau_\mathrm{syn}$,
    derived in~\cite{bytschok2011shared} and further developed in~\cite{Koke2017}.
    It relates the membrane's voltage course with both relevant time constants and the height $h$ of the \cgls{psp}.
    The fitting algorithm fixes one of the time constants and varies the other.
    Although the \cglspl{psp} are symmetric in $\tau_\text{mem}$ and $\tau_\text{syn}$, the fact that typically $\tau_\text{mem}>\tau_\text{syn}$ is considered.
    Once the parameters are determined from the measurements through fitting the model, a linear fit is used to obtain a calibration relating parameters with \cgls{fg} values, as with the previously treated calibrations.
  
\begin{figure}
        \centering
        \subfloat[
                  \label{fig:I_gl_fits}]{\includegraphics[width=0.5\linewidth]{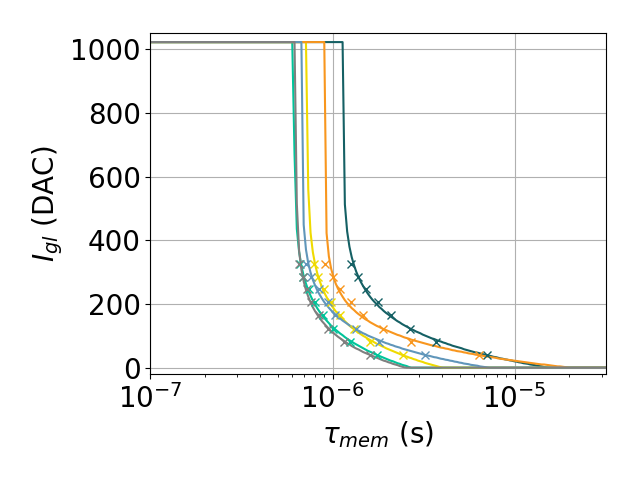}}
        \subfloat[
                  \label{fig:calibration_I_gl}]{\includegraphics[width=0.5\linewidth]{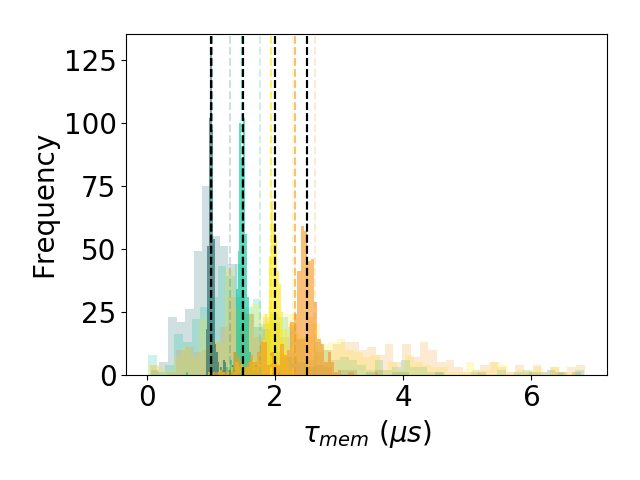}}
        \caption{\protect\subref{fig:I_gl_fits} Fits for the parameter $I_\text{gl}$ against the achieved membrane time constant on five neurons, using a softplus function model and eight measurement steps.
        \protect\subref{fig:calibration_I_gl} Distribution of membrane time constants before and after the $I_\text{gl}$ calibration is applied for all neurons, with pale and intense colors, respectively.}
        \label{fig:I_gl_figures}
      \end{figure}

    $I_\text{gl}$: Bias current that controls the membrane's leakage conductance.
      This parameter and the chosen membrane capacitance, which can be set to two different values, determines the membrane time constant.
      \begin{itemize}
        \item \textit{How}: The input spikes should arrive with enough space to allow the membrane to return to a steady-state after each perturbation.
        A strong excitatory synaptic input is set to achieve a better signal-to-noise ratio.
        Fitting~\cref{eq:psp_model} returns both the membrane and the synaptic input time constant from the \cgls{psp} shape.
        A fit of the softplus function
        \begin{equation}
          \label{eq:softplus_function_mem}
           \tau_\text{mem} = \frac{a \cdot \text{log}(1 + \text{exp}(c \cdot (b - I_\text{gl})))}{c} + \text{offset}
        \end{equation}
        is subsequently used to translate between biological-space parameters and \cgls{fg}-stored parameters.
        \item \textit{Settings}: $E_\text{leak} = \SI{0.8}{\volt}$, $E_\text{synx} = \SI{1.3}{\volt}$, $E_\text{syni} = \SI{0.6}{\volt}$, $V_\text{syntcx} = \SI{1.6}{\volt}$,
        $V_\text{convoffi} = \SI{0.9}{\volt}$, $V_\text{convoffx} = \SI{0.9}{\volt}$,
        $V_\text{threshold} = \SI{1.2}{\volt}$, $V_\text{reset} = \SI{0.3}{\volt}$,
        $V_\text{gmax0} = \SI{1}{\volt}$, $\text{gmax\_div} = \SI{30}{LSB}$,
        \textit{big capacitor}, $I_\text{gl}$ speedup normal.
        \item \textit{Sweep}: $I_\text{gl}$
        \item \textit{Effects}: The achieved membrane time constants' mean is closer to the target value after the calibration is obtained and applied, as can be observed in~\cref{fig:calibration_I_gl}.
        The standard deviations are reduced.
        However, as seen in~\cref{fig:I_gl_fits}, the precision to configure $\tau_\text{mem}$ is limited as only a fraction of the possible parameter range of $I_\text{gl}$ results in reasonable configurations.

      \end{itemize}

      \begin{figure}
        \centering
        \subfloat[
                  \label{fig:vsyntcx_calib}]{\includegraphics[width=0.5\linewidth]{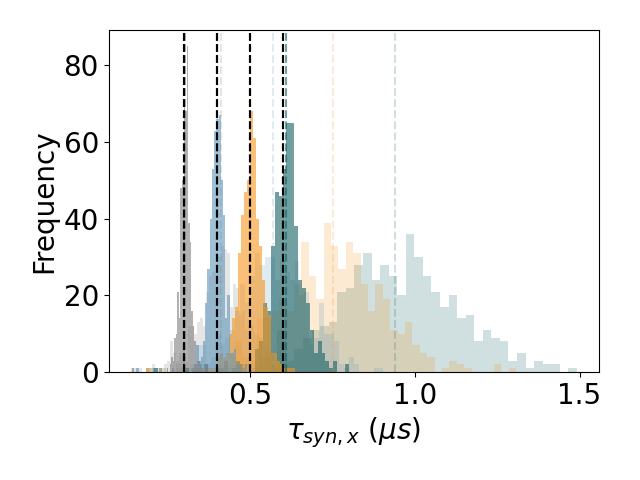}}
        \subfloat[
                  \label{fig:vsyntci_calib}]{\includegraphics[width=0.5\linewidth]{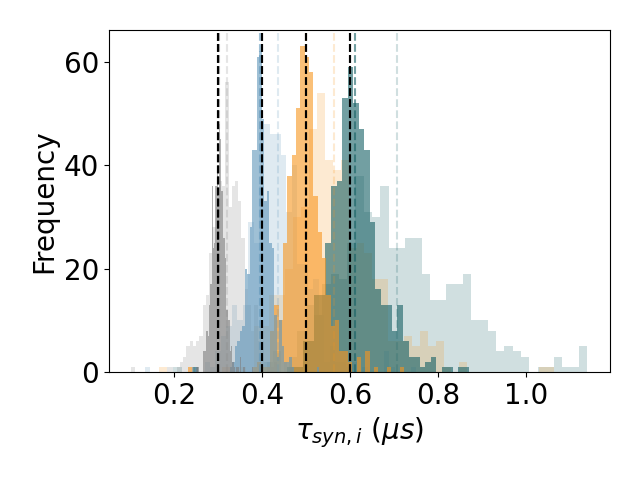}}
        \caption{Distribution of the achieved \protect\subref{fig:vsyntcx_calib} excitatory and \protect\subref{fig:vsyntci_calib} inhibitory synaptic time constants before and after calibration in pale and intense colors, respectively,
        for four different target values (black dashed lines).}
      \end{figure}
  
    $V_\text{syntcx}$: Voltage controlling the excitatory synapse time constant, $\tau_{syn,x}$, by varying the voltage integrator's resistive element.
      Large values of $V_\text{syntcx}$ shift $E_\text{leak}$ towards the reversal potential, since leak currents in the synaptic input integrator inrease for higher voltages.
      \begin{itemize}
        \item \textit{How}: Similar to the $I_\text{gl}$ calibration, input spikes that arrive with enough separation are used.
        \Cref{eq:psp_model} is fitted to extract the time constants.
	Afterwards, a fit of the softplus function
        \begin{equation}
          \label{eq:softplus_function_syn}
           \tau_\text{syn} = \frac{a \cdot \text{log}(1 + \text{exp}(c \cdot (b - V_\text{syntc})))}{c} + \text{offset}
        \end{equation}
        to the extracted values is used to translate between biological-space parameters and \cgls{fg}-stored parameters.
        \item \textit{Settings}: $E_\text{leak} = \SI{0.8}{\volt}$, $E_\text{syni} = \SI{0.6}{\volt}$, $E_\text{synx} = \SI{1.3}{\volt}$, $I_\text{gl} = \SI{0.3}{\micro\ampere}$,
        $V_\text{convoffx} = \SI{1.8}{\volt}$, $V_\text{convoffi} = \SI{1.8}{\volt}$, $V_\text{threshold} = \SI{1.2}{\volt}$, $V_\text{reset} = \SI{0.3}{\volt}$,
        $V_\text{gmax0} = \SI{0.05}{\volt}$, $\text{gmax\_div} = \SI{30}{LSB}$
        \item \textit{Sweep}: $V_\text{syntcx}$
        \item \textit{Effects}: The distribution of the achieved excitatory synaptic time constants, $\tau_{syn,x}$ before and after calibration of $V_{syntcx}$ is shown in~\cref{fig:vsyntcx_calib}.
      \end{itemize}
  
      $V_\text{syntci}$: Voltage controlling the inhibitory synapse time constant, $\tau_{syn,i}$, by varying the voltage integrator's resistive element.
    \begin{itemize}
        \item \textit{How}: Similar to the $V_\text{syntcx}$ calibration.
        \item \textit{Settings}: $E_\text{leak} = \SI{0.8}{\volt}$, $E_\text{syni} = \SI{0.3}{\volt}$, $E_\text{synx} = \SI{1.3}{\volt}$, $I_\text{gl} = \SI{0.3}{\micro\ampere}$,
        $V_\text{convoffx} = \SI{1.8}{\volt}$, $V_\text{convoffi} = \SI{1.8}{\volt}$, $V_\text{threshold} = \SI{1.2}{\volt}$, $V_\text{reset} = \SI{0.3}{\volt}$,
        $V_\text{gmax0} = \SI{0.05}{\volt}$, $\text{gmax\_div} = \SI{30}{LSB}$
        \item \textit{Sweep}: $V_\text{syntci}$
        \item \textit{Effects}: The distribution of the achieved inhibitory synaptic time constants, $\tau_{syn,i}$ before and after calibration of $V_{syntcx}$ is shown in~\cref{fig:vsyntci_calib}.
      \end{itemize}

\begin{figure}
        \subfloat[
                  \label{fig:exc_rev_pot}]{\includegraphics[width=0.49\linewidth]{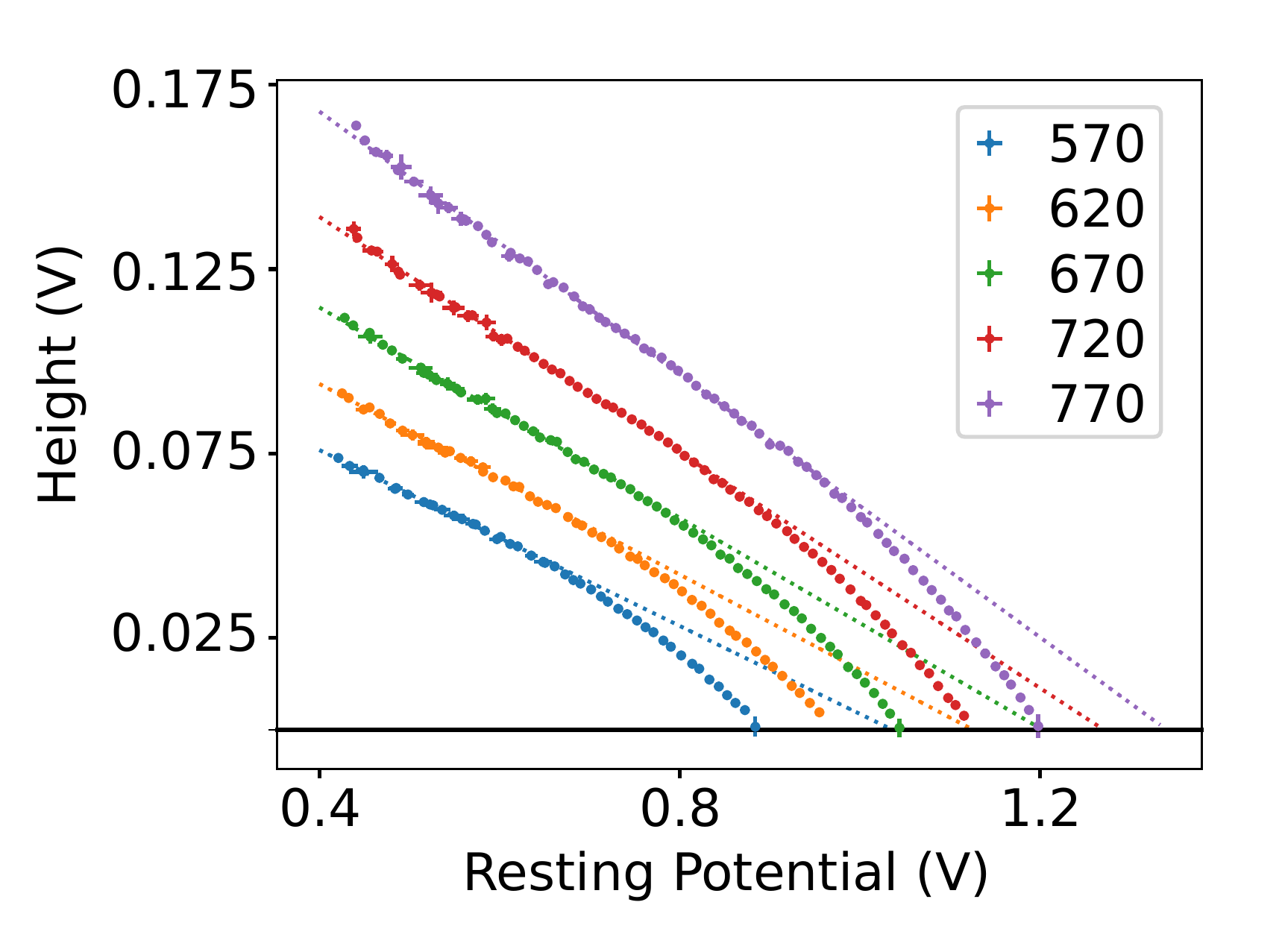}}
        \subfloat[
                  \label{fig:result_exc_rev_pot}]{\includegraphics[width=0.49\linewidth]{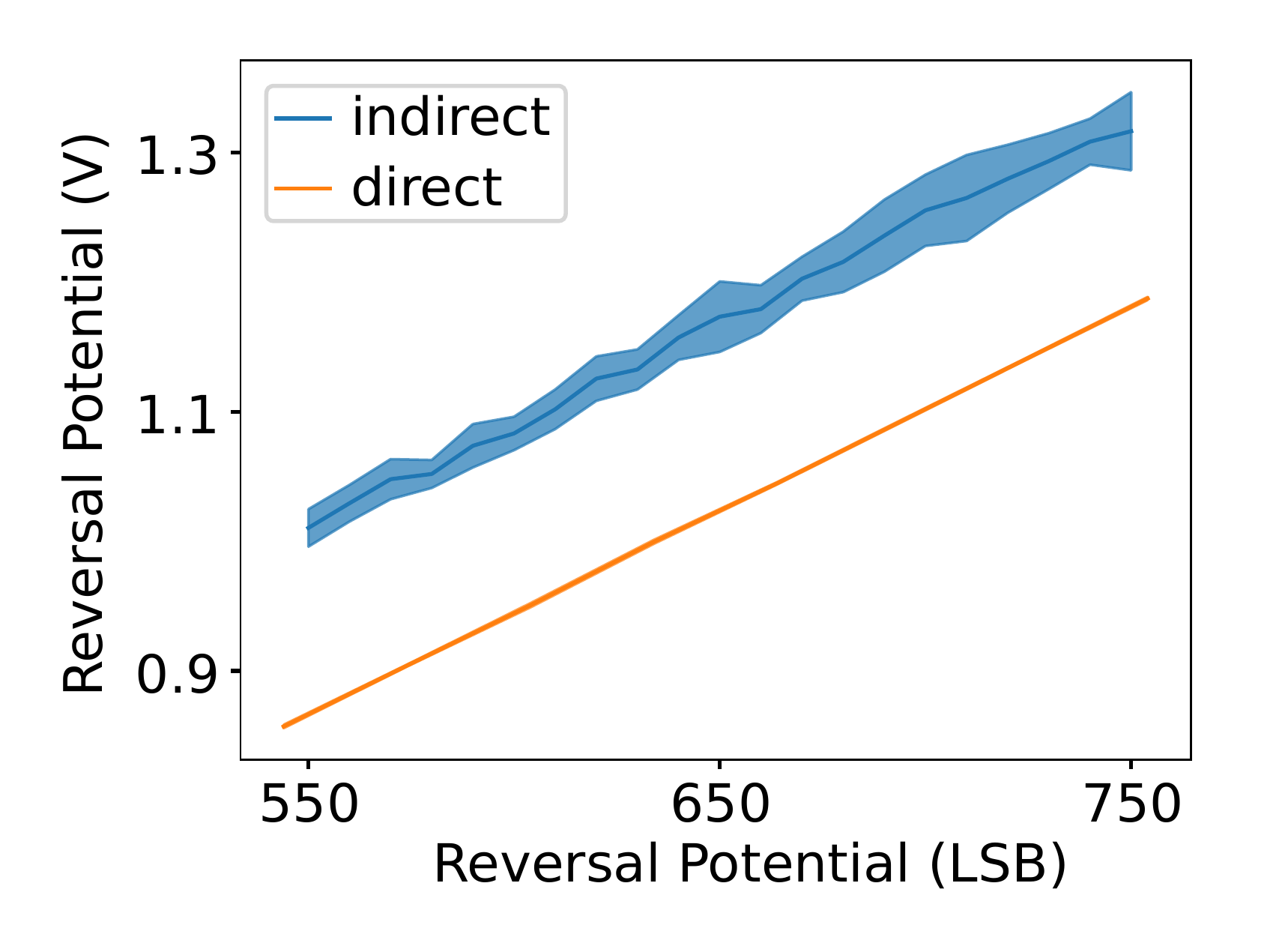}}
        \caption{Excitatory reversal potential calibration. \protect\subref{fig:exc_rev_pot} Indirect measurement of the excitatory reversal potential.
             The height of the \cgls{psp} of a stimulated neuron is extracted for different resting potentials.
             Different colors indicate different hardware settings of the reversal potential in LSB.
             Since the circuits are not designed to reach the excitatory reversal potential, non linear behavior is observed for small distances between membrane potential and reversal potential.
             A linear extrapolation of the linear region (dotted line) is used to extract the correct reversal potential.
             During experiments the neuron is exclusively operated in the linear regime.
             \protect\subref{fig:result_exc_rev_pot} Comparison of direct and indirect measurement of the excitatory reversal potential.
             Because of the non-linear behavior of the circuits close to the reversal potential, the direct measurement provides too small values.
             The indirect measurement has larger errors due to its dependency on the whole neuron circuit and the enlargement by the linear extrapolation.}
  \label{fig:exc_rev_pot_both}
  \end{figure}

  \textit{$E_\textit{synx}$}:
  In biologically plausible networks, the excitatory reversal potential is above the threshold and thus never reached by the membrane potential.
  Its calibration is a good showcase for pitfalls during the operation of analog circuits.
  Intuitively, a direct measurement using the membrane potential would be used for both reversal potentials.
  However, similar to their biological counterparts, the circuits of the \cgls{hicann} chip are not designed for the membrane potential to get close to the excitatory reversal potential.
  Thus, the circuits show a non-linear behavior when approaching the reversal potential, as they deviate from the center of their design ranges.
  This can be observed in~\cref{fig:exc_rev_pot}.
  Therefore, the excitatory reversal potential is measured indirectly.
    \begin{itemize}
      \item \textit{How}:
      The height of the \cgls{psp} of a stimulated neuron is measured for different resting potentials in the linear regime of the circuits.
      A linear extrapolation is used to extract the resting potential where the height reaches zero, shown in~\cref{fig:exc_rev_pot}.
      In the conductance based synapse model this resting potential is equal to the reversal potential.
      The measurements are repeated for different reversal potentials to extract the linear dependency between hardware value and applied voltage.
      \item \textit{Settings}: $I_\text{convi} = \SI{0}{\ampere}$, $I_\text{gl} = \SI{E-7}{\second}$, $V_\text{convoffx,i} = \SI{0.9}{\volt}$, $V_\text{syntcx,i} = \SI{2E-7}{\second}$, $V_\text{threshold} = \SI{1.8}{\volt}$, $V_\text{gmax} = \SI{0.9}{\volt}$, $gmax\_div = \SI{2}{LSB}$, $w = \SI{15}{LSB}$
      \item \textit{Sweep}:
      $E_\text{leak}$, $E_\textit{synx}$
      \item \textit{Results}:
      Results of the calibration compared to a direct measurement can be seen in~\cref{fig:result_exc_rev_pot}.
      The disadvantage of the indirect measurement is the increased runtime and the dependency on the shape of the \cgls{psp}.
      Small variations of hardware parameters, most likely due to the necessity to rewrite the \cgls{fg} value of the resting potential, are enlarged by the linear extrapolation performed to find the reversal potential.
      As a result,~\cref{fig:result_exc_rev_pot} shows larger variations for the indirect calibration than the direct measurement.
      Nevertheless, the technique allows for correctly calibrating the excitatory reversal potential without directly measuring it.
    \end{itemize}

\bibliographystyle{style/IEEEtran}
\bibliography{bib/vision}